\documentclass[12pt]{iopart}

\expandafter\let\csname equation*\endcsname\relax
\expandafter\let\csname endequation*\endcsname\relax
\usepackage{amsmath}  
\usepackage{mathtools} 
\usepackage{hyperref}
\usepackage{booktabs}
\usepackage{graphicx}
\usepackage{enumitem}
\usepackage{subcaption}
\usepackage{amssymb}
\usepackage{cite}
\hypersetup{
    colorlinks=true,
        linktoc=all,
    linkcolor=blue,
    citecolor=blue,
    urlcolor=blue
}
\usepackage{datetime}
\usepackage{xcolor}

\newcommand{\nh}[1]{{\color{black} {#1}}}

\begin{document}
\title{Robust Bayesian inference with gapped LISA data using all-in-one TDI-$\infty$}

\author{Niklas Houba$^{1}$, Jean-Baptiste Bayle$^2$, and Michele Vallisneri$^{3}$} 
\address{$^1$ ETH Zurich, Institute of Geophysics, Department of Earth and Planetary Sciences, Switzerland}
\address{$^2$ University of Glasgow, Glasgow G12 8QQ, United Kingdom}
\address{$^3$ Jet Propulsion Laboratory, California Institute of Technology, Pasadena CA 91109 USA}

\ead{niklas.houba@eaps.ethz.ch}

\vspace{10pt}

\newdateformat{monthyear}{\monthname[\the\month] \the\year}
\begin{indented}
\item[]\monthyear\today
\end{indented}

\begin{abstract}
The Laser Interferometer Space Antenna (LISA), an ESA L-class mission, is designed to detect gravitational waves in the millihertz frequency band, with operations expected to begin in the next decade. LISA will enable groundbreaking studies of astrophysical phenomena such as massive black hole mergers, extreme mass ratio inspirals, and compact binary systems. A key challenge in analyzing LISA's data is the significant laser frequency noise, which must be suppressed using time-delay interferometry (TDI) during on-ground processing. Classical TDI mitigates this noise by algebraically combining phase measurements taken at different times and spacecraft. However, data gaps caused by instrumental issues or operational interruptions complicate the process. These gaps affect multiple TDI samples due to the time delays inherent to the algorithm, rendering surrounding measurements unusable for parameter inference and degrading overall data quality. In this paper, we apply the recently proposed variant of TDI known as TDI-$\infty$ to astrophysical parameter inference, focusing on the challenge posed by data gaps. TDI-$\infty$ frames the LISA likelihood numerically in terms of raw measurements, marginalizing over laser phase noises under the assumption of infinite noise variance. Additionally, TDI-$\infty$ is set up to incorporate and cancel other noise sources beyond laser noise, including optical bench motion, clock noise, and modulation noise, establishing it as an all-in-one TDI solution. The method gracefully handles measurement interruptions, removing the need to explicitly address discontinuities during gravitational-wave template matching. We integrate TDI-$\infty$ into a Bayesian framework, demonstrating its superior performance in scenarios involving data gaps. Compared to classical TDI, the method preserves signal integrity more effectively and is particularly interesting for low-latency applications, where the limited amount of available data makes data gaps particularly disruptive. The study's results highlight the potential of TDI-$\infty$ to enhance LISA's scientific capabilities, paving the way for more robust data analysis pipelines.
\end{abstract}
\vspace{-5pt}

\section{Introduction}
The Laser Interferometer Space Antenna (LISA) is a pioneering space mission recently adopted by the European Space Agency (ESA) to detect gravitational waves, with its launch planned for the mid-2030s \cite{colpi2024lisa}. 
Expected to be the first spaceborne gravitational-wave observatory, the mission builds on the groundbreaking observations of ground-based detectors like LIGO and Virgo, which recently opened a new gravitational-wave window to the Universe \cite{PhysRevLett.116.061102, PhysRevLett.116.241103, PhysRevLett.118.221101, Abbott_2017, PhysRevLett.119.141101, PhysRevLett.119.161101, PhysRevX.9.031040, Abbott_2020, PhysRevD.102.043015, Abbott_2020_2, PhysRevLett.125.101102}. Operating in a frequency range beyond the reach of these observatories, LISA will offer an unparalleled view of the cosmos in the millihertz band \cite{SciReqDoc, Amaro-Seoane_2012} enabling the study of astrophysical phenomena such as the mergers of massive black hole binaries (MBHBs), extreme mass ratio inspirals (EMRIs), and compact binary systems in the Galaxy \cite{PhysRevD.81.104014,Mangiagli2022}. Moreover, the mission will provide valuable insights into the early Universe and test the limits of Einstein's theory of general relativity \cite{amaroseoane2017laser}.

LISA will comprise a triangular constellation of three spacecraft, nominally separated by 2.5 million kilometers, forming  {large-scale Michelson interferometers} in a heliocentric orbit \cite{Martens2021}. The spacecraft will continuously exchange laser beams, measuring the extremely small changes in distance caused by passing gravitational waves. These distance variations are on the order of picometers, requiring precision measurement techniques to observe them \cite{Schuldt_2009}. 
To achieve the sensitivity required for detecting LISA's astronomical targets, the mission employs a measurement principle based on the use of drag-free controlled test masses \cite{PhysRevD.99.082001}. These test masses are designed to be in free fall, ideally isolated from all external forces except gravity by the surrounding spacecraft. The test masses are cubes of a gold--platinum alloy, each measuring 46 millimeters on a side and weighing approximately 2 kilograms \cite{Carbone2006-kz}. These properties are carefully selected to ensure minimal interaction with external forces. Each spacecraft maneuvers around its test masses using an advanced drag-free control system, minimizing non-gravitational forces like solar radiation pressure and interplanetary dust impacts \cite{BenderArt}. Without drag-free control, these non-gravitational forces would disrupt the test masses' free fall along the sensitive measurement axes.  {Although drag-free technology allows the test masses to closely follow geodesic paths dictated by spacetime curvature, residual acceleration noise still constrains LISA's measurement precision. Test-mass acceleration noise cannot be mitigated in post-processing. Therefore, stringent stability requirements are imposed on the drag-free control system and on the test masses to achieve the necessary accuracy for tracking their gravitational motion.}

Another challenge LISA faces is due to the unequal lengths of its science interferometer arms \cite{Dhurandhar2010}. In equal-arm interferometers, the noise induced by the intrinsic frequency instabilities of the laser sources cancels out automatically when recombining the laser beams at the detector. In LISA, the arm-length imbalance causes the intrinsic laser frequency instability to dominate the measurements, masking the faint gravitational-wave signals by seven to eight orders of magnitude \cite{Bayle2018a, Bayle2018b}. As a consequence, LISA needs to synthesize an equal-arm interferometer
in post-processing, using \emph{time-delay interferometry} (TDI) \cite{Tinto2002TDI1stGen, TDITINTO2005}. TDI is a complex algorithm that cancels out laser noise by combining phase measurements taken at different times and spacecraft.
When the time delays between these measurements are adjusted carefully, TDI effectively suppresses the laser frequency noise, allowing the much weaker gravitational-wave signals to be isolated and analyzed. The classical TDI approach relies on constructing these time-delayed combinations algebraically based on the varying relative distances between the spacecraft. The laser-canceling combinations are then utilized for gravitational-wave
data analysis \cite{PhysRevD.103.042006}.

 {One of the significant challenges in astrophysical data analysis is the presence of data gaps \cite{PhysRevD.104.044035, PhysRevD.100.022003, castelli2024extractiongravitationalwavesignals}. These gaps can arise for various reasons, including unplanned interruptions, such as instrument downtime or communication losses, as well as planned operations, like antenna re-pointing, where gaps are expected and can be managed accordingly.} The impact of data gaps on TDI is profound: because TDI relies on precise timing and on the use of
delayed data points to cancel laser noise at each observation, a gap in the data affects not only the immediate TDI sample but also subsequent samples that are part of the same noise-cancellation sequence, thus rendering surrounding measurements unusable for astrophysical analysis \cite{Pollack2004, carre2010, PhysRevD.103.082001}.  {Furthermore, TDI requires high-precision fractional delay filters with long interpolation kernels to achieve the necessary timing accuracy, which amplifies the sensitivity of the TDI data streams to interruptions. Consequently, even short gaps can disrupt the continuity of the noise-cancellation process, compromising the integrity of the data and, thus, the accuracy of gravitational-wave parameter estimation via classical template matching
} 
Template matching, a standard technique in gravitational-wave data analysis, involves comparing the observed data with precomputed waveforms or templates representing different possible sources  \cite{IMPRSStatsLecture8, BabakLISAPresentation}. The likelihood function, which quantifies how well a given template matches the observed data, plays a central role in this process. Incorporating data gaps into the likelihood calculation is not straightforward. The missing data points disrupt the continuous signal model assumed by frequency-domain template matching, making it challenging to correctly define a likelihood function.  {In classical TDI, data gaps
result in the loss of additional samples around each gap. Because Bayesian inference with classical TDI is traditionally conducted in the frequency domain, accurately accounting for the missing information and the discontinuities in the frequency-based likelihood becomes non-trivial.} Inaccuracies in the likelihood can lead to biased or inaccurate parameter estimates, further complicating the interpretation of detected gravitational-wave signals.

Given these challenges, there is a pressing need for more robust methods that can handle data gaps without compromising the integrity of the data. This paper introduces and explores TDI Infinity (TDI-$\infty$) for LISA, an extension of classical TDI, focusing on its ability to tackle challenges related to data gaps. TDI-$\infty$ was proposed for a toy model in \cite{PhysRevD.103.082001} and reframes parameter inference by formulating the LISA likelihood numerically in terms of the raw phase measurements rather than relying on their algebraic combinations. The approach involves marginalizing over the laser phase noises under the assumption of infinite noise variance, which allows TDI-$\infty$ to gracefully manage measurement dropouts and reduce their impact on parameter-estimation accuracy. 

The primary goal of this paper is to evaluate the application of TDI-$\infty$ in the context of LISA with a particular focus on its performance in scenarios involving measurement interruptions and discontinuities. We aim to demonstrate how TDI-$\infty$ can preserve signal integrity more effectively than classical TDI.  {This is particularly important in scenarios where rapid alerts and updates to the astronomical community are essential and rely on limited data, such as detecting impending transient events. For example, in the case of merging MBHBs, the accumulation of signal-to-noise ratio is often concentrated in short timespans that must be analyzed as soon as each data package is received. In low-latency settings, where timely response is critical, there is limited tolerance for data package loss -- requesting retransmission of lost data may take several hours, a delay we can't afford in these scenarios. This sensitivity to data gaps strongly motivates our study, as it underscores the need for robust methods to manage incomplete data streams without compromising the reliability of the analysis.}
By integrating TDI-$\infty$ into a Bayesian framework, we seek to show its superiority in handling incomplete data and its potential to enhance LISA's scientific robustness. 
Section \ref{sec:Telemetered} introduces the model of the telemetered beat-note signals recorded by LISA, which is used in Section \ref{sec:TDI} to set up classical TDI and TDI-$\infty$. Section \ref{sec:Bayesian} introduces Bayesian parameter inference and details how TDI-$\infty$ is integrated into this methodology with its specific likelihood function. Section \ref{sec:Simulation}  presents simulation results, comparing the performance of TDI-$\infty$ based parameter inference with classical TDI in the presence of data gaps. Finally, Section \ref{sec:Conclusion} summarizes the results and discusses the implications of the paper's findings for future data-analysis pipelines.

\section{Telemetered beat-note signals in LISA}\label{sec:Telemetered}

LISA employs a measurement concept known as \emph{split interferometry} \cite{Otto_2012, PhysRevD.99.084023}. In the split-interferometry setup, the distance between two test masses on distant spacecraft is reconstructed from three individual beat-note measurements. The inter-spacecraft interferometer measures the distance between the optical benches of the local and distant spacecraft. The test-mass interferometers on each spacecraft then measure the distance between the optical bench and the associated test mass on that spacecraft. These measurements are combined to determine distance variations between the test masses related to gravitational-wave events \cite{https://doi.org/10.15488/8545}. Each optical bench (there are six in total) is equipped with a third interferometer, known as the reference interferometer. The reference interferometer
measures the relative phase noise of the laser sources between adjacent optical benches. This information is essential for TDI
on-ground processing.

\begin{figure}[]
	\centering
  \includegraphics[trim=7 10 380 0, clip,width=0.7\textwidth]{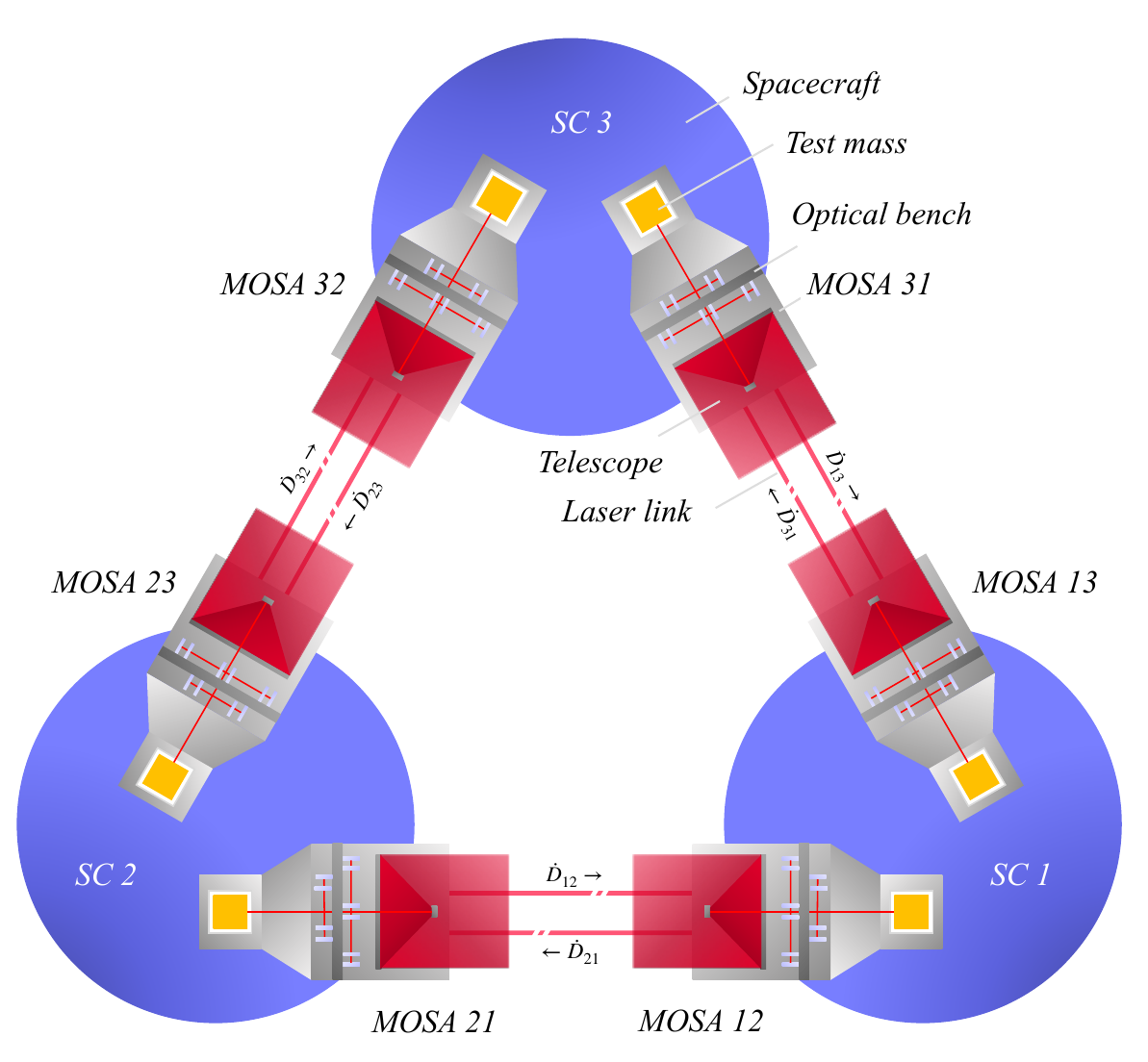}
	\caption{Configuration of the LISA constellation, including the index convention for this paper, which follows Ref.\ \cite{PhysRevD.104.023006}. 
The constellation consists of three spacecraft, each connected by bi-directional near-infrared laser links covering a distance of $2.5 \times 10^6$ km. The acronym MOSA refers to the \emph{moving optical subassembly}, consisting primarily of an optical bench, a telescope, and a test mass. This illustration is highly simplified.}
	\label{im:LISAConstellation}
\end{figure}

Below we summarize the mathematical model for the telemetered beat-note signals \cite{PhysRevD.107.083019}, considering LISA's active transponder technology.
In Section \ \ref{sec:TDI} we will use this model to extend the simplified TDI-$\infty$ framework of Ref.\ \cite{PhysRevD.103.082001} and adapt it for realistic application. In our formalism, each measurement is decomposed into two parts: large frequency offsets and drifts (superscript \(o\)) and smaller in-band fluctuations (superscript \(\epsilon\)). The index notation follows the convention illustrated in Fig.\ \ref{im:LISAConstellation}.

\subsection{Inter-spacecraft interferometer}
The inter-spacecraft interferometer (isi) measures the phase difference between a laser beam received from a distant spacecraft and the local laser. Taking optical bench 12 of spacecraft 1 as an example, the beat-note signal of the inter-spacecraft interferometer at the carrier frequency (index $\text{c}$) can be expressed as
\begin{align}
    \text{isi}_{12,\text{c}}^{o} &= \mathbf{F} \dot{\mathbf{T}}_1 \left\{\dot{\mathcal{D}}_{12} O_{21} - \nu_0 d_{12}^o - O_{12}\right\},
    \label{eq:isi_o_12} 
    \end{align}\\
    \begin{align}
    \text{isi}_{12,\text{c}}^{\epsilon} &= \mathbf{F} \dot{\mathbf{T}}_1 \bigg\{ \dot{\mathcal{D}}_{12} \left(\dot{p}_{21} - \dot{N}_{21}^{\Delta} \right) - \left( \nu_0 + \dot{\mathcal{D}}_{12} O_{21} \right) \dot{H}_{12}  \nonumber \\
    &\quad + \frac{\nu_0}{c} \dot{N}_{\text{isi}12 \leftarrow 21}^{\text{ob}} - \left( \dot{p}_{12} - \dot{N}_{12}^{\Delta} + \frac{\nu_0}{c} \dot{N}_{\text{isi}12 \leftarrow 12}^{\text{ob}} \right) \nonumber \\
    &\quad + \dot{N}_{\text{isi}12,\text{c}}^{\text{ro}} - (\dot{\mathcal{D}}_{12} O_{21} - \nu_0 d_{12}^o - O_{12})\frac{ q_1^{\epsilon}}{1 + \dot{q}_1^{o}}\bigg\},
    \label{eq:isi_eps_12} \\
    \text{isi}_{12,\text{c}} &= \text{isi}_{12,\text{c}}^{o} + \text{isi}_{12,\text{c}}^{\epsilon}.
    \label{eq:isi_12}
\end{align}
\noindent
In equations \eqref{eq:isi_o_12} and \eqref{eq:isi_eps_12}, $\nu_0$ is the nominal laser frequency (around 281 THz \cite{PhysRevD.104.023006});
$d_{12}^o$ represents the slowly varying proper pseudorange offsets, which include the light time of flight and frame conversions between the proper times of spacecraft 1 and 2;
$O_{12}$ and $O_{21}$ represent frequency offsets related to the special-relativistic Doppler shift between spacecraft;
the Doppler-delay operator $\dot{\mathcal{D}}_{12}$ represents the light travel time from spacecraft 2 to spacecraft 1, as defined in 
\cite{PhysRevD.107.083019}; the timestamping operator \(\dot{\mathbf{T}}_1\) in equations \eqref{eq:isi_o_12} and \eqref{eq:isi_eps_12} translates signals from (idealized) spacecraft proper time to the actual clock-implemented on-board time of spacecraft 1; and the continuous linear filter operator \(\mathbf{F}\) represents the effects of the filtering and downsampling process (see more below).

Equation \eqref{eq:isi_eps_12} for in-band beat-note fluctuations includes the frequency variations of the local laser and (delayed) distant laser, denoted by $\dot{p}_{12}$ and $\dot{p}_{21}$ respectively, along with local and distant optical-bench path-length noise contributions, which manifest as Doppler shifts $\dot{N}^{\text{ob}}_{\text{isi}{12 \leftarrow 12}}$ and $\dot{N}^{\text{ob}}_{\text{isi}{12 \leftarrow 21}}$. Longitudinal motion of the MOSA (see Fig.\ \ref{im:LISAConstellation}) along the line of sight associated with the local and delayed distant optical benches introduce additional optical noises $\dot{N}^{\Delta}_{12}$ and $\dot{N}^{\Delta}_{21}$.  The gravitational-wave signal $\dot{H}_{12}$ adds further Doppler shifts. Finally, equation  \eqref{eq:isi_eps_12} includes the readout noise $\dot{N}^{\text{ro}}_{\text{isi12},\text{c}}$ and clock noise, represented by the term $\dot{q}_1^{\epsilon}/(1 + q_1^{o})$. Here, \(q_1^{\epsilon}\) and \(\dot{q}_1^{o}\) are noise sources related to the timing deviations between the spacecraft's measurement time scale and its true relativistic proper time.

Spacecraft proper time is defined as the time shown by perfect clocks co-moving with the spacecraft's center of mass, and it is a theoretical construct unavailable in practice. Instead, the recorded measurements rely on an imperfect onboard timer,
referred to as the onboard clock time frame. The timestamping operator \(\dot{\mathbf{T}}_1\) translates from the former to the latter.
The LISA mission design currently indicates that the raw 80 MHz phasemeter beat-note signals will be processed by filtering and downsampling to progressively lower rates, ultimately reaching the final measurement rate of 4 Hz. The filtering effect is represented by the filter operator \(\mathbf{F}\).

In addition to the carrier beat note, each inter-spacecraft interferometer records a sideband measurement. While the sideband--sideband beat-note signal shares  similarities with the carrier measurement, it differs in two key aspects. Firstly, both the distant and local laser frequency instabilities are influenced by the interaction between the modulation frequency, clock jitter, and modulation noise, represented by \( \nu_{21}^{m}(\dot{q}_2^\epsilon + \dot{M}_{21}) \) and \( \nu_{12}^{m}(\dot{q}_1^\epsilon + \dot{M}_{12}) \). Secondly, the modulation frequency introduces shifts in the distant-sideband beat-note frequency offsets, which are impacted by out-of-band clock errors, described as \( \nu_{21}^{m}(1+\dot{q}_2^o) \).
The model for the sideband--sideband beat-note frequency measurement (index $\text{sb}$) of the inter-spacecraft interferometer is

\begin{align}
    \text{isi}_{12,\text{sb}}^{o} &= \mathbf{F} \dot{\mathbf{T}}_1 \left\{\dot{\mathcal{D}}_{12} O_{21} - \nu_0 d_{12}^o - O_{12} + \dot{\mathcal{D}}_{12} \left[ \nu_{21}^m \left( 1 + \dot{q}_2^o \right) \right] - \nu_{12}^m \left( 1 + \dot{q}_1^o \right)\right\},
    \label{eq:isi_o_12_sb} 
    \\ \text{isi}_{12,\text{sb}}^{\epsilon} &= \mathbf{F} \dot{\mathbf{T}}_1 \bigg\{ \dot{\mathcal{D}}_{12} \left( \dot{p}_{21} + \nu_{21}^{m} \left( \dot{q}_2^{\epsilon} + \dot{M}_{21} \right)-\dot{N}_{21}^{\Delta} \right) \nonumber \\
    &\quad - \left( \nu_0 + \dot{\mathcal{D}}_{12} \left[ O_{21} + \nu_{21}^{m} \left( 1 + q_2^o \right) \right] \right) \dot{H}_{12} + \frac{\nu_0}{c} \dot{N}_{\text{isi}12 \leftarrow 21}^{\text{ob}} \nonumber \\
    &\quad - \left( \dot{p}_{12} - \dot{N}_{12}^{\Delta}+ \nu_{12}^{m} \left( q_1^{\epsilon} + \dot{M}_{12} \right) + \frac{\nu_0}{c} \dot{N}_{\text{isi}12 \leftarrow 12}^{\text{ob}} \right) \nonumber \\ &\quad + \dot{N}_{\text{isi}12, sb}^{\text{ro}} \nonumber  \\ &\quad - (\dot{\mathcal{D}}_{12} O_{21} - \nu_0 d_{12}^o - O_{12} + \dot{\mathcal{D}}_{12} \left[ \nu_{21}^m \left( 1 + \dot{q}_2^o \right) \right] - \nu_{12}^m \left( 1 + \dot{q}_1^o \right))\frac{ \dot{q}_1^{\epsilon}}{1 + \dot{q}_1^{o}}\bigg\},
    \label{eq:isi_eps_12_sb} \\
    \text{isi}_{12,\text{sb}} &= \text{isi}_{12,\text{sb}}^{o} + \text{isi}_{12,\text{sb}}^{\epsilon},
    \label{eq:isi_12_sb}
\end{align}
where
\(\nu_{12}^m\) and \(\nu_{21}^m\) are the modulation frequencies,
\(\dot{M}_{12}\)  and \(\dot{M}_{21}\) represent the modulation noise,
and $\dot{N}^{\text{ro}}_{\text{isi12},\text{sb}}$ is the readout noise in the sideband-sideband measurement.

\subsection{Test-mass interferometer}
The carrier--carrier beat-note frequency measurement in the test-mass interferometer includes the frequency variations from the local and neighboring laser beams originating from the adjacent optical bench on the same spacecraft. These variations are denoted as \( \dot{p}_{12} \) and \( \dot{p}_{13} \). Additionally, it accounts for the related path length noises on the optical bench, \( \dot{N}^{\text{ob}}_{\text{tmi}12 \leftarrow 12} \) and \( \dot{N}^{\text{ob}}_{\text{tmi}12 \leftarrow 13} \). As the adjacent beam passes through the optical fiber, it is contaminated by backlink noise \( \dot{N}^{\text{bl}}_{12} \). The measurement incorporates also the readout noise \( \dot{N}^{\text{ro}}_{\text{tmi}12,\text{c}} \) and the clock noise term \( \dot{q}_1^\epsilon / (1 + q_1^0) \).
Overall, the measurement is modeled as
\begin{align}
    \text{tmi}_{12,\text{c}}^{o} &= \mathbf{F} \dot{\mathbf{T}}_1 \left\{ O_{13} - O_{12}\right\},
    \label{eq:tmi_o_12} \\
    \text{tmi}_{12,\text{c}}^{\epsilon} &= \mathbf{F} \dot{\mathbf{T}}_1 \Big\{ \dot{p}_{13} + \frac{\nu_0}{c} \left( \dot{N}_{\text{tmi}12 \leftarrow 13}^{\text{ob}} + \dot{N}_{12}^{\text{bl}} \right)  \nonumber \\
    &\quad - \left( \dot{p}_{12} + \frac{\nu_0}{c} \left( \dot{N}_{\text{tmi}12 \leftarrow 12}^{\text{ob}} + 2\left(\dot{N}^{\delta}_{12}-\dot{N}_{12}^{\Delta} \right) \right)\right) \nonumber \\
    &\quad + \dot{N}_{\text{tmi}12,\text{c}}^{\text{ro}} -  \left( O_{13} - O_{12}\right)\frac{\dot{q}_1^{\epsilon}}{1 + \dot{q}_1^{o}}\Big\},
    \label{eq:tmi_eps_12} \\
    \text{tmi}_{12,\text{c}} &= \text{tmi}_{12,\text{c}}^{o} + \text{tmi}_{12,\text{c}}^{\epsilon},
    \label{eq:tmi_12}
\end{align}
where
\(\dot{N}^{\delta}_{12}\) is the additional local test-mass noise, and
$\dot{N}^{\Delta}_{12}$ is the local optical-bench noise as also measured in the inter-spacecraft interferometer.

\subsection{Reference interferometer}
The carrier--carrier beat-note frequency measurement in the reference interferometer is equal to the measurement of the test-mass interferometer, except that it does not include additional optical-bench displacement noise and test-mass noise since the laser beam is not reflected off the test mass in this case:
\begin{align}
    \text{rfi}_{12,\text{c}}^{o} &= \mathbf{F} \dot{\mathbf{T}}_1 \left\{O_{13} - O_{12}\right\},
    \label{eq:rfi_o_12} \\
     \text{rfi}_{12,\text{c}}^{\epsilon} &= \mathbf{F} \dot{\mathbf{T}}_1 \Big\{ \dot{p}_{13} + \frac{\nu_0}{c} \left( \dot{N}_{\text{rfi}12 \leftarrow 13}^{\text{ob}} + \dot{N}_{12}^{\text{bl}} \right) \nonumber \\
    &\quad - \left( \dot{p}_{12} + \frac{\nu_0}{c}  \dot{N}_{\text{rfi}12 \leftarrow 12}^{\text{ob}}  \right) \nonumber \\
    &\quad + \dot{N}_{\text{rfi}12,\text{c}}^{\text{ro}} -  \left( O_{13} - O_{12}\right)\frac{\dot{q}_1^{\epsilon}}{1 + \dot{q}_1^{o}} \Big\},
    \label{eq:rfi_eps_12} \\
    \text{rfi}_{12,\text{c}} &= \text{rfi}_{12,\text{c}}^{o} + \text{rfi}_{12,\text{c}}^{\epsilon}.
    \label{eq:rfi_12}
\end{align}
The sideband--sideband beat-note frequency measurement follows a similar principle, where the adjacent and local laser frequency variations are influenced by the in-band clock and modulation noise, represented as \( \nu_{13}^{m}(\dot{q}_{1} + \dot{M}_{13}) \) and \( \nu_{12}^{m}(\dot{q}_{1} + \dot{M}_{12}) \):
\begin{align}
    \text{rfi}_{12,\text{sb}}^{o}       &= \mathbf{F} \dot{\mathbf{T}}_1 \left\{O_{13} - O_{12}+ \left( \nu_{13}^m - \nu_{12}^m \right) \left( 1 + \dot{q}_1^o \right)\right\},
    \label{eq:rfi_o_12_sb} \\
    \text{rfi}_{12,\text{sb}}^{\epsilon} &= \mathbf{F} \dot{\mathbf{T}}_1 \Big\{ \dot{p}_{13} + \nu_{13}^{m} \left( \dot{q}_1 + \dot{M}_{13} \right)  + \frac{\nu_0}{c} \left(\dot{N}_{\text{rfi}12 \leftarrow 13}^{\text{ob}}+\dot{N}^{\text{bl}}_{12}\right) \nonumber \\
    &\quad - \left( \dot{p}_{12} + \nu_{12}^{m} \left( \dot{q}_1 + \dot{M}_{12} \right)  + \frac{\nu_0}{c} \dot{N}_{\text{rfi}12 \leftarrow 12}^{\text{ob}}\right) \nonumber \\
    &\quad + \dot{N}_{\text{rfi}12,\text{sb}}^{\text{ro}} - \left(O_{13} - O_{12}+ \left( \nu_{13}^m - \nu_{12}^m \right) \left( 1 + \dot{q}_1^o \right)\right)\frac{\dot{q}_1^{\epsilon}}{1 + \dot{q}_1^{o}}\Big\},
    \label{eq:rfi_eps_12_sb} \\
    \text{rfi}_{12,\text{sb}}           &= \text{rfi}_{12,\text{sb}}^{o} + \text{rfi}_{12,\text{sb}}^{\epsilon}.
    \label{eq:rfi_12_sb}
\end{align}
Equations \eqref{eq:isi_o_12}--\eqref{eq:rfi_12_sb} serve as the foundation for developing the extended TDI-$\infty$ framework in the following section. We will introduce two key extensions to the model of Ref.\ \cite{PhysRevD.103.082001}. The first extension, akin to classical TDI, eliminates laser noise \( \dot{p}_{ij} \) with \( i,j \in \{1,2,3\} \) and \( i \neq j \), see Ref. \cite{Houba2023}. The second extension addresses all suppressible noise sources, including optical-bench displacement noise $\dot{N}_{ij}^{\Delta}$, clock  noise $\dot{q}_i$, and modulation noise $\dot{M}_{ij}$, in addition to laser noise.

\section{Time-delay interferometry}\label{sec:TDI}
TDI is a fundamental component of the data-processing strategy for LISA, with origins tracing back to the 1990s. The development of TDI has been the subject of extensive research, contributing to its theoretical underpinnings and experimental demonstrations \cite{Tinto_Massimo_1999, Armstrong_John_1999, Tinto_Massimo_2004, PhysRevD.70.062002, GeomTDIVallisneri2005, SyntheticLISA, Tinto_Massimo_2021, Bayle_Jean-Baptiste_2021, Muratore_Martina_2020, Page_Jessica_2021, de_Vine_Glenn_2010, Mitryk_Shawn_2010, Cruz_Rachel_2006}. The primary function of TDI is to account for the unequal and dynamically changing arm lengths of the LISA constellation by constructing linear combinations of time-shifted interferometric signals. Without TDI, laser frequency noise would dominate the measurements, effectively obscuring the gravitational-wave signals that LISA aims to detect. TDI is essential for spaceborne gravitational-wave detectors like LISA because, unlike terrestrial interferometers where arm lengths can be equalized, LISA's arms span millions of kilometers and vary due to the spacecraft's orbital motion. Consequently, laser noise does not cancel out automatically when the two interferometer arms are recombined at the detector.

\subsection{First- and second-generation TDI}
Classical TDI addresses the challenge of laser noise by algebraically combining the telemetered beat-note measurements from multiple spacecraft with appropriate time delays, effectively creating a synthetic interferometer that suppresses laser noise.
Since its inception, TDI has evolved through two main generations, each improving upon the previous approach to address challenges in the realistic LISA setting. The first generation of TDI,  {TDI-1 in the following}, is thoroughly described in \cite{Tinto2002TDI1stGen} and is based on the assumption that arm lengths are unequal but remain constant over time. 

\begin{figure}[]
    \centering
  \includegraphics[trim=270 60 50 100, clip,width=0.8\textwidth]{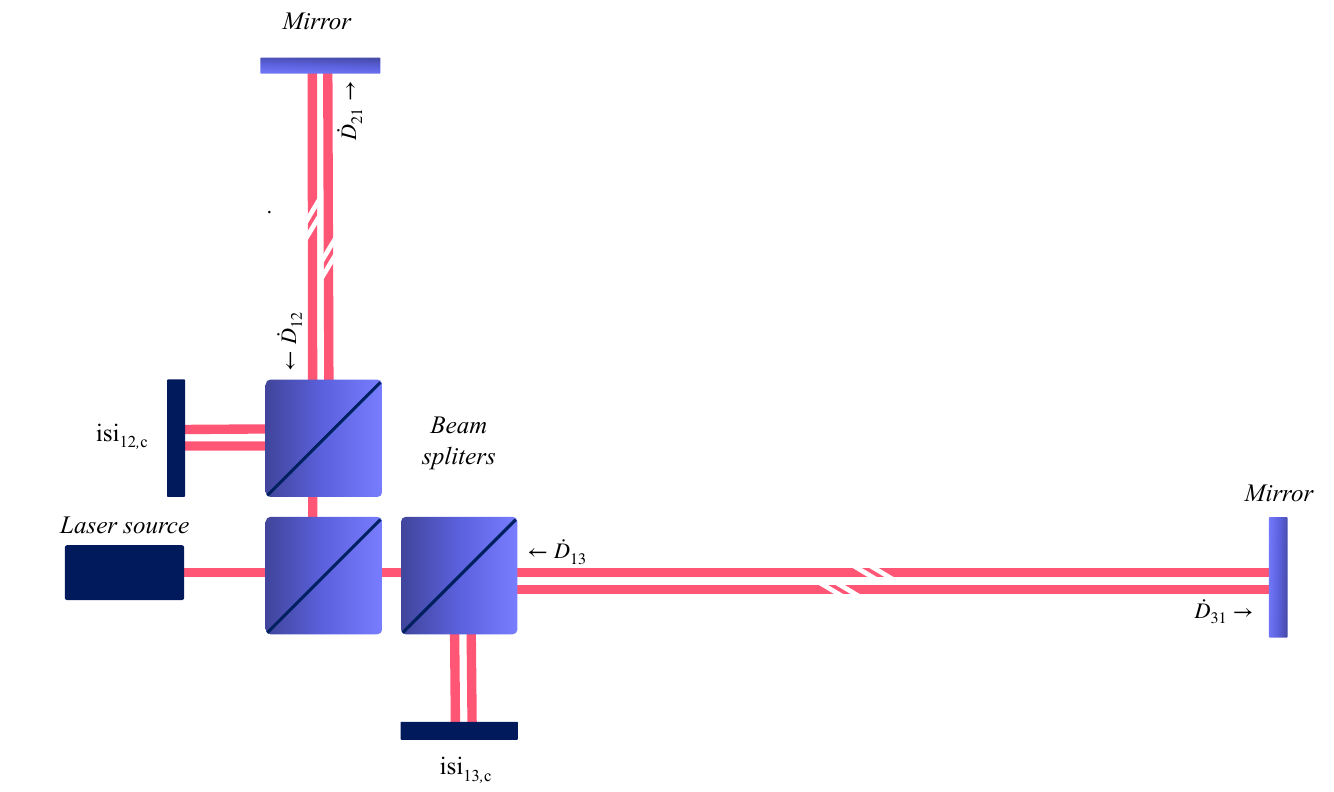}
	\caption{A simplified unequal-arm interferometer toy model used to illustrate the concept behind classical TDI and later to introduce TDI-$\infty$.}
	\label{im:ToyModel}
\end{figure}

We will consider a simplified toy model to illustrate the concept of first-generation TDI using basic equations. This toy model, shown in Fig.\ \ref{im:ToyModel}, involves a single laser source with beams propagating into two arms before reflecting back to the origin, experiencing round-trip delays denoted by \( \dot{\mathcal{D}}_{12} \dot{\mathcal{D}}_{21} \) and \( \dot{\mathcal{D}}_{13} \dot{\mathcal{D}}_{31} \), with  \( \dot{\mathcal{D}}_{12} \dot{\mathcal{D}}_{21} \neq \dot{\mathcal{D}}_{13} \dot{\mathcal{D}}_{31} \). The phases of the two returning beams are measured as \( \text{isi}_{12,\text{c}} \) and \( \text{isi}_{13,\text{c}} \), both influenced by the common laser noise \( p_1 \). In this example, for simplicity, we neglect measurement noise, other noise sources, and gravitational-wave contributions. Assuming a static configuration, the two measurements \( \text{isi}_{12,\text{c}} \) and \( \text{isi}_{13,\text{c}} \) can be described using equation  \eqref{eq:isi_12} as
\begin{align}
    \text{isi}_{12,\text{c}}  &= \mathbf{F} \dot{\mathbf{T}}_1 \left\{\dot{\mathcal{D}}_{12} \dot{\mathcal{D}}_{21} p_{1} - p_{1}     \right\},
    \label{eq:isi_12_toymodel} \\
    \text{isi}_{13,\text{c}}  &= \mathbf{F} \dot{\mathbf{T}}_1 \left\{\dot{\mathcal{D}}_{13} \dot{\mathcal{D}}_{31} p_{1} - p_{1}     \right\}.
\label{eq:isi_13_toymodel} 
\end{align}
By subtracting equation  \eqref{eq:isi_13_toymodel} from equation  \eqref{eq:isi_12_toymodel}, it becomes evident that the laser noise does not cancel automatically due to the mismatch in arm lengths. To achieve effective noise cancellation, we need to time-shift $\text{isi}_{12,\text{c}}$ by $\dot{\mathcal{D}}_{13} \dot{\mathcal{D}}_{31}$ and $ \text{isi}_{13,\text{c}}$ by $\dot{\mathcal{D}}_{12} \dot{\mathcal{D}}_{21}$:
\begin{align}
    \dot{\mathcal{D}}_{13} \dot{\mathcal{D}}_{31}\text{isi}_{12,\text{c}}  &= \mathbf{F} \dot{\mathbf{T}}_1 \left\{\dot{\mathcal{D}}_{13} \dot{\mathcal{D}}_{31}\dot{\mathcal{D}}_{12} \dot{\mathcal{D}}_{21} p_{1} - \dot{\mathcal{D}}_{13} \dot{\mathcal{D}}_{31}p_{1}     \right\},
    \label{eq:D_isi_12_toymodel} \\
    \dot{\mathcal{D}}_{12} \dot{\mathcal{D}}_{21}\text{isi}_{13,\text{c}}  &= \mathbf{F} \dot{\mathbf{T}}_1 \left\{\dot{\mathcal{D}}_{12} \dot{\mathcal{D}}_{21}\dot{\mathcal{D}}_{13} \dot{\mathcal{D}}_{31} p_{1} - \dot{\mathcal{D}}_{12} \dot{\mathcal{D}}_{21}p_{1}     \right\},
\label{eq:D_isi_13_toymodel} 
\end{align}
and build the  {TDI-1} Michelson channel $X_{1}$ via
\begin{align}
X_{1} = (\text{isi}_{12,\text{c}} + \dot{\mathcal{D}}_{12} \dot{\mathcal{D}}_{21}\text{isi}_{13,\text{c}} ) - (\text{isi}_{13,\text{c}} + \dot{\mathcal{D}}_{13} \dot{\mathcal{D}}_{31}\text{isi}_{12,\text{c}} ).
\label{eq:TDIX1}
\end{align}
In the static case, the concatenated delay operators commute, i.e., \( \dot{\mathcal{D}}_{12} \dot{\mathcal{D}}_{21} \dot{\mathcal{D}}_{13} \dot{\mathcal{D}}_{31} = \dot{\mathcal{D}}_{13} \dot{\mathcal{D}}_{31} \dot{\mathcal{D}}_{12} \dot{\mathcal{D}}_{21} \), allowing the laser noise terms in equation  \eqref{eq:TDIX1} to cancel pairwise. This results in the linear combination $X_{1}$ that effectively eliminates laser noise.
While  {TDI-1} provides adequate noise reduction in static scenarios, it proved insufficient for the realistic LISA case. LISA's arm-length variability necessitated further refinement to maintain effective noise suppression under realistic mission conditions, leading to second-generation TDI ( {TDI-2}\footnote{ {Throughout this text, the terms TDI-2 and classical TDI are used interchangeably.}}), as outlined in \cite{2003TDI2ndGenMotiv}. TDI-2 improves on TDI-1 by suppressing laser frequency noise to the required level for LISA, making it the current standard in LISA data analysis. The linear combinations of TDI-2 are more complex than the one in equation  \eqref{eq:TDIX1}, but they follow the same underlying principle of pairwise noise cancellation. Interested readers are encouraged to consult the relevant literature for further derivations.

While classical TDI effectively suppresses laser frequency noise by combining time-delayed phase measurements, data gaps pose a major challenge for downstream data analysis and parameter inference. In classical TDI, a single data gap impacts not only the immediate TDI output but affects also future TDI samples, as evident from equation  \eqref{eq:TDIX1} for  {TDI-1}. In TDI-2, which relies on multiple nested delays for effective noise suppression (of the order of 8 times the LISA light travel time, roughly 100 seconds) and on high-precision fractional delay filters with long interpolation kernels (often of order 31 at a 4 Hz sampling rate, corresponding to about 10 seconds), this cascading effect can render large portions of the data unusable depending on the pattern of the gaps, significantly degrading the overall quality and sensitivity of LISA's observations.
In the following, we introduce TDI-$\infty$, adapt it to the realistic LISA mission scenario, and demonstrate its advantages in handling data gaps compared to classical TDI throughout the paper. 

\subsection{TDI-$\infty$ for the LISA toy model}
In TDI-$\infty$, the raw LISA measurements are analyzed directly for gravitational-wave detection and parameter estimation without transforming them into TDI observables using predefined analytical expressions, as in classical TDI. The TDI-$\infty$ observables are computed numerically from the LISA arm lengths and can be folded into the likelihood calculation. In TDI-$\infty$, as introduced in \cite{PhysRevD.103.082001} for the toy model shown in Fig.\ \ref{im:ToyModel}, the measurement time series $\text{isi}_{12,\text{c}}$ and $\text{isi}_{13,\text{c}}$ are represented in vector form as
\begin{align}
\boldsymbol{y} = \left[\text{isi}_{12,\text{c}}(t_0), \text{isi}_{13,\text{c}}(t_0), \text{isi}_{12,\text{c}}(t_1), \dots, \text{isi}_{13,\text{c}}(t_{n-1})\right]^\top
\in \, \mathbb{R}^{2n\times1},\label{eq:y_toymodel}\end{align}
where $n$ denotes the number of valid measurement samples. The relations described by equations \eqref{eq:isi_12_toymodel} and \eqref{eq:isi_13_toymodel} can then be expressed in matrix-vector form as
\begin{align}
\boldsymbol{y} = M \boldsymbol{p},\label{eq:matrixvector_toymodel}
\end{align}
where $\boldsymbol{p} \in \, \mathbb{R}^{n\times1}$ represents the (unmeasured) discretized laser noise time series, and $M\in \, \mathbb{R}^{2n\times n}$ is a design matrix modeling the delayed finite differences from equations \eqref{eq:isi_12_toymodel} and \eqref{eq:isi_13_toymodel} using fractional-delay finite-impulse-response filters. When the delays correspond to integer multiples of the sampling time, the design matrix contains only $+1$, $-1$, and zeros, as no interpolation between noise samples is necessary. For fractional delays, the $+1$s are broken across the appropriate filter masks. In this work, we employ delay filters based on Lagrange interpolation, as described in \cite{PhysRevD.103.082001}.
Note that here we are using fractional-delay filters to model a physical delay induced by light travel rather than to delay time series in post-processing to build TDI combinations. The accuracy requirements that must be posed on the filters are the same in both uses.

The TDI-$\infty$ observable, which by construction suppresses laser noise fully, is defined as
\begin{align}
    \boldsymbol{o} = T\boldsymbol{y}\label{eq:o_TDIInf},
\end{align}
where $T\in \, \mathbb{R}^{n\times 2n}$ denotes the null space of $M^\top$ such that
\begin{align}
    TM = 0.\label{eq:TMcondition}
\end{align}
Note that TDI-$\infty$ generalizes to any time dependence of the arm lengths (not just linear variations as in TDI-2, or quadratic variations in a hypothetical TDI-3) -- hence the name TDI-$\infty$.

\subsection{TDI-$\infty$ for LISA}
\begin{figure}[]
	\centering
  \includegraphics[trim=180 220 30 80, clip,width=1.0\textwidth]{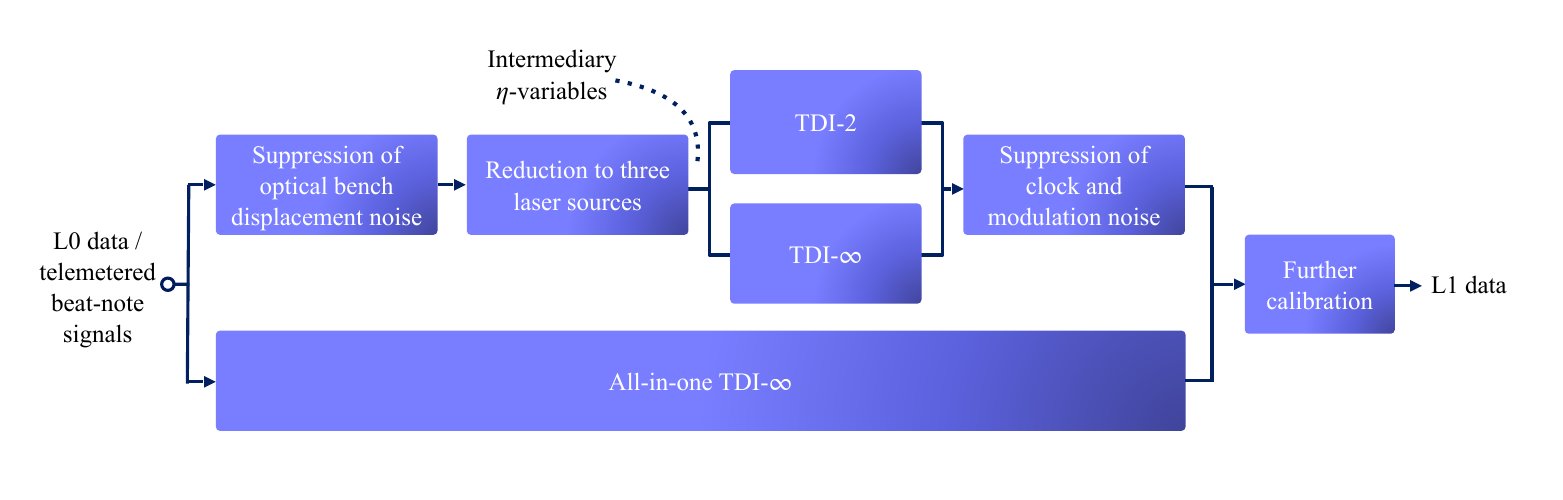}
	\caption{Schematic illustration of three different possible architectures for LISA's L0-L1 data-processing strategy.  The first two pipelines illustrate data-processing approaches that are well established in the literature, whereas the all-in-one TDI-$\infty$ pipeline is introduced in this paper.}
	\label{im:L0L1Architectures}
\end{figure}

When adapting TDI-$\infty$ to LISA, the core definition of the TDI-$\infty$ observable of equation  \eqref{eq:o_TDIInf} remains unchanged, along with the condition in equation  \eqref{eq:TMcondition} that determines the null-space matrix $T$. However, what differs is the setup of the measurement vector $\boldsymbol{y}$ in equation  \eqref{eq:y_toymodel}, as well as the noise vector $\boldsymbol{p}$ and the design matrix $M$. The specific formulation of these components depends on the objectives of the TDI-$\infty$ approach.  To explore this further, we refer to Fig.\ \ref{im:L0L1Architectures}, which illustrates three potential architectures for LISA's L0-L1 data-processing strategy. L0 data corresponds to the raw telemetered beat-note signals, while L1 data represents the calibrated TDI channels prepared for astrophysical analysis.
 {Figure \ref{im:L0L1Architectures} shows that L0-L1 data processing extends beyond laser-noise cancellation to encompass the suppression of optical-bench displacement noise, clock noise, and modulation noise. These steps are part of constellation-wide calibration (as opposed to local calibrations, such as converting differential wavefront sensing voltages to angular measurements or calibrating the test-mass stiffness).} The noise sources addressed during constellation-wide calibration can be suppressed since their effects are measured independently across different detectors at various times. 

 {The pipeline at the top of Fig.\ \ref{im:L0L1Architectures} incorporates TDI-2 and represents the current standard for LISA data processing \cite{https://doi.org/10.15488/8545,PhysRevD.105.122008, bayle:tel-03120731}. The process begins by suppressing optical-bench displacement noise through linear combinations of time-shifted interferometric measurements. Next, the pipeline reduces the number of free-running lasers to three by generating intermediate $\eta$-variables, derived from these time-shifted measurements and the output of the previous step. Then, the $\eta$-variables are inputs to TDI-2, which mitigates the remaining laser noise. In the final step, the impact of clock and modulation noise from the right-hand side MOSAs on the TDI-2 combinations is subtracted. Details on the algorithms can be found in  \cite{PhysRevD.103.123027}.}

One possible adaptation of TDI-$\infty$ for LISA is to replace the TDI-2 block in Fig.\ \ref{im:L0L1Architectures} with the TDI-$\infty$ framework while continuing to use the $\eta$-variables as inputs. This pipeline was previously introduced in \cite{Houba2023}. The advantage of this approach is that TDI-$\infty$ only needs to handle the remaining laser noise sources, thereby simplifying the null-space computation. However, a notable drawback is that the input consists of linear combinations rather than raw measurements, which reduces TDI-$\infty$'s effectiveness in dealing with data gaps. Since these linear combinations already incorporate time shifts, a gap in a raw measurement results in multiple invalid samples in the intermediate variables.  Due to this limitation, the focus of the paper is on the ``all-in-one TDI-$\infty$'' approach, as shown in the lower part of Fig.\ \ref{im:L0L1Architectures}.
In this case, TDI-$\infty$ receives all the raw interferometric measurements directly and is responsible for canceling all suppressible noise sources simultaneously. For now, the all-in-one approach does not include post-TDI calibration steps from the classical L0-L1 pipeline, such as tilt-to-length (TTL) noise mitigation. Since TTL exhibits different coupling behaviors on each spacecraft, they cannot be suppressed through simple linear combinations of detector measurements and must instead be explicitly estimated \cite{Houba2023, PhysRevD.106.042005}. The all-in-one framework is organized as follows. 

\paragraph{Measurement vector $\boldsymbol{y}\in\mathbb{R}^{24n\times1}$.} The measurement vector \( \boldsymbol{y} \) is constructed by organizing the individual measurements from different interferometers into a single vector. Specifically, \( \boldsymbol{y} \) contains the following components:
\begin{itemize}[label={-}]
    \item 6 inter-spacecraft interferometer measurements at the carrier frequency $\textrm{isi}_{ij,\text{c}}$,
    \item 6 inter-spacecraft interferometer measurements at the sideband frequency $\textrm{isi}_{ij,\text{sb}}$,
    \item 6 test-mass interferometer  measurements $\textrm{tmi}_{ij}$,
    \item 3 reference interferometer  measurements at the carrier frequency $\textrm{rfi}_{ij,\text{c}}$,
    \item 3 reference interferometer  measurements at the sideband frequency $\textrm{rfi}_{ij,\text{sb}}$.
\end{itemize}
\noindent
The inter-spacecraft interferometer measurements at the carrier frequency $\textrm{isi}_{ij,\text{c}}$ are derived from equation  \eqref{eq:isi_12} and summarized as
\begin{align}
\begin{aligned}
\boldsymbol{y}_{\text{isi},\text{c}} = [&\textrm{isi}_{12,\text{c}}(t_0),\dots , \textrm{isi}_{12,\text{c}}(t_{n-1}),\,\textrm{isi}_{23,\text{c}}(t_0),\dots , \textrm{isi}_{23,\text{c}}(t_{n-1}), \\ &\textrm{isi}_{31,\text{c}}(t_0),\dots , \textrm{isi}_{31,\text{c}}(t_{n-1}),\,\textrm{isi}_{13,\text{c}}(t_0),\dots , \textrm{isi}_{13,\text{c}}(t_{n-1}), \\ &\textrm{isi}_{32,\text{c}}(t_0),\dots , \textrm{isi}_{32,\text{c}}(t_{n-1}),\,\textrm{isi}_{21,\text{c}}(t_0),\dots , \textrm{isi}_{21,\text{c}}(t_{n-1}) ]^\top 
\in \, \mathbb{R}^{6n\times1}.\label{eq:y_isi_c_LISA}
\end{aligned}
\end{align}
 {Although clock noise and relativistic time-frame differences prevent the data from different spacecraft from being synchronized to the same absolute time, this mismatch is inherently accounted for by incorporating the measured pseudo-ranges as delays in the design matrix, similar to the approach used for TDI-2 in \cite{PhysRevD.105.122008}. Therefore, data-stream synchronization is not required to construct \( \boldsymbol{y} \). The measurements here are expressed in the onboard clock frames and are sampled within each clock frame on a regular time grid $t_0, t_1, \dots, t_{n-1}$.}

Based on equation  \eqref{eq:isi_12_sb}, we obtain $\boldsymbol{y}_{\text{isi},\text{sb}}$, summarizing the inter-spacecraft interferometer measurements at the sideband frequency $\textrm{isi}_{ij,\text{sb}}$:
\begin{align}
\begin{aligned}
\boldsymbol{y}_{\text{isi},\text{sb}} = [&\textrm{isi}_{12,\text{sb}}(t_0),\dots , \textrm{isi}_{12,\text{sb}}(t_{n-1}),\,\textrm{isi}_{23,\text{sb}}(t_0),\dots , \textrm{isi}_{23,\text{sb}}(t_{n-1}), \\ &\textrm{isi}_{31,\text{sb}}(t_0),\dots , \textrm{isi}_{31,\text{sb}}(t_{n-1}),\,\textrm{isi}_{13,\text{sb}}(t_0),\dots , \textrm{isi}_{13,\text{sb}}(t_{n-1}), \\ &\textrm{isi}_{32,\text{sb}}(t_0),\dots , \textrm{isi}_{32,\text{sb}}(t_{n-1}),\,\textrm{isi}_{21,\text{sb}}(t_0),\dots , \textrm{isi}_{21,\text{sb}}(t_{n-1}) ]^\top 
\in \, \mathbb{R}^{6n\times1}.\label{eq:y_isi_sb_LISA}
\end{aligned}
\end{align}
The test-mass interferometer measurements $\textrm{tmi}_{ij}$ with the model given in equation  \eqref{eq:tmi_12} are written as
\begin{align}
\begin{aligned}
\boldsymbol{y}_{\text{tmi}} = [&\textrm{tmi}_{12}(t_0),\dots , \textrm{tmi}_{12}(t_{n-1}),\,\textrm{tmi}_{23}(t_0),\dots , \textrm{tmi}_{23}(t_{n-1}), \\ &\textrm{tmi}_{31}(t_0),\dots , \textrm{tmi}_{31}(t_{n-1}),\,\textrm{tmi}_{13}(t_0),\dots , \textrm{tmi}_{13}(t_{n-1}), \\ &\textrm{tmi}_{32}(t_0),\dots , \textrm{tmi}_{32}(t_{n-1}),\,\textrm{tmi}_{21}(t_0),\dots , \textrm{tmi}_{21}(t_{n-1}) ]^\top 
\in \, \mathbb{R}^{6n\times1}.\label{eq:y_tmi_LISA}
\end{aligned}
\end{align}
For the reference interferometer measurements at the carrier and sideband frequencies $\textrm{rfi}_{ij,\text{c}}$ and $\textrm{rfi}_{ij,\text{sb}}$ from equations \eqref{eq:rfi_12} and \eqref{eq:rfi_12_sb}, only three measurements provide independent information. We select those corresponding to the left-sided MOSAs:
\begin{align}
\begin{aligned}
\boldsymbol{y}_{\text{rfi},\text{c}} = [&\textrm{rfi}_{12,\text{c}}(t_0),\dots , \textrm{rfi}_{12,\text{c}}(t_{n-1}), \\ &\textrm{rfi}_{23,\text{c}}(t_0),\dots , \textrm{rfi}_{23,\text{c}}(t_{n-1}), \\ &\textrm{rfi}_{31,\text{c}}(t_0),\dots , \textrm{rfi}_{31,\text{c}}(t_{n-1}) ]^\top 
\in \, \mathbb{R}^{3n\times1},\label{eq:y_rfi_c_LISA}
\end{aligned}
\end{align}
and
\begin{align}
\begin{aligned}
\boldsymbol{y}_{\text{rfi},\text{sb}} = [&\textrm{rfi}_{12,\text{sb}}(t_0),\dots , \textrm{rfi}_{12,\text{sb}}(t_{n-1}), \\ &\textrm{rfi}_{23,\text{sb}}(t_0),\dots , \textrm{rfi}_{23,\text{sb}}(t_{n-1}), \\ &\textrm{rfi}_{31,\text{sb}}(t_0),\dots , \textrm{rfi}_{31,\text{sb}}(t_{n-1}) ]^\top 
\in \, \mathbb{R}^{3n\times1}.\label{eq:y_rfi_sb_LISA}
\end{aligned}
\end{align}
The all-in-one measurement vector $\boldsymbol{y}$ is then constructed as:
\begin{equation}
\boldsymbol{y} = [\boldsymbol{y}_{\text{isi},\text{c}}, \,\boldsymbol{y}_{\text{isi},\text{sb}}, \,\boldsymbol{y}_{\text{tmi}}, \,\boldsymbol{y}_{\text{rfi},\text{c}}, \,\boldsymbol{y}_{\text{rfi},\text{sb}}]^\top\in\mathbb{R}^{24n\times1}\label{eq:y_composed_LISA}.
\end{equation}

\paragraph{Noise vector $\boldsymbol{p} \in\mathbb{R}^{18n\times1}$.}
The noise vector \( \boldsymbol{p} \) contains contributions from all noise sources that can be suppressed  {during constellation-wide calibration, except for TTL noise}. The noise components include:
\begin{itemize}[label={-}]
    \item 6 laser noise sources $\dot{p}_{ij}$,
    \item 6 jitter noise terms from LISA's optical benches $\dot{N}_{ij}^{\Delta}$,
    \item 3 clock noise terms $\dot{q}^{\epsilon}_{i}$,
    \item 3 modulation noise terms $\dot{M}_{ij}$  {(as explained in \cite{PhysRevD.103.123027}, only three of the six modulation noise sources can be eliminated. The remaining three contribute to unsuppressible secondary noise in TDI. This happens regardless of the TDI generation that is used).}
\end{itemize}
For the laser noise $\dot{p}_{ij}$, we write
\begin{align}
\begin{aligned}
\boldsymbol{p}_{\text{laser}} = [&\dot{p}_{12}(t_0),\dots , \dot{p}_{12}(t_{n-1}),\,\dot{p}_{23}(t_0),\dots , \dot{p}_{23}(t_{n-1}), \\ &\dot{p}_{31}(t_0),\dots , \dot{p}_{31}(t_{n-1}),\,\dot{p}_{13}(t_0),\dots , \dot{p}_{13}(t_{n-1}), \\ &\dot{p}_{32}(t_0),\dots , \dot{p}_{32}(t_{n-1}),\,\dot{p}_{21}(t_0),\dots , \dot{p}_{21}(t_{n-1}) ]^\top 
\in \, \mathbb{R}^{6n\times1}.\label{eq:p_laser}
\end{aligned}
\end{align}
Similarly, we write the translational jitter noise from the optical benches $\dot{N}^{\Delta}_{ij}$:
\begin{align}
\begin{aligned}
\boldsymbol{p}_{\text{jitter}} = [&\dot{N}^{\Delta}_{12}(t_0),\dots , \dot{N}^{\Delta}_{12}(t_{n-1}),\,\dot{N}^{\Delta}_{23}(t_0),\dots , \dot{N}^{\Delta}_{23}(t_{n-1}), \\ &\dot{N}^{\Delta}_{31}(t_0),\dots , \dot{N}^{\Delta}_{31}(t_{n-1}),\,\dot{N}^{\Delta}_{13}(t_0),\dots , \dot{N}^{\Delta}_{13}(t_{n-1}), \\ &\dot{N}^{\Delta}_{32}(t_0),\dots , \dot{N}^{\Delta}_{32}(t_{n-1}),\,\dot{N}^{\Delta}_{21}(t_0),\dots , \dot{N}^{\Delta}_{21}(t_{n-1}) ]^\top 
\in \, \mathbb{R}^{6n\times1}.\label{eq:p_jitter}
\end{aligned}
\end{align}
For the clock noise, we set
\begin{align}
\begin{aligned}
\boldsymbol{p}_{\text{clock}} = [&\dot{q}^{\epsilon}_{1}(t_0),\dots , \dot{q}^{\epsilon}_{1}(t_{n-1}), \\ &\dot{q}^{\epsilon}_{2}(t_0),\dots , \dot{q}^{\epsilon}_{2}(t_{n-1}), \\ &\dot{q}^{\epsilon}_{3}(t_0),\dots , \dot{q}^{\epsilon}_{3}(t_{n-1}) ]^\top 
\in \, \mathbb{R}^{3n\times1}.\label{eq:p_clock}
\end{aligned}
\end{align}
The three modulation noises that are chosen to be suppressed are:
\begin{align}
\begin{aligned}
\boldsymbol{p}_{\text{modulation}} = [&\dot{M}_{13}(t_0),\dots , \dot{M}_{13}(t_{n-1}), \\ &\dot{M}_{32}(t_0),\dots , \dot{M}_{32}(t_{n-1}), \\ &\dot{M}_{21}(t_0),\dots , \dot{M}_{21}(t_{n-1}) ]^\top 
\in \, \mathbb{R}^{3n\times1}.\label{eq:p_modulation}
\end{aligned}
\end{align}
The all-in-one noise vector $\boldsymbol{p}$ then is:
\begin{equation}
\boldsymbol{p} = [\boldsymbol{p}_{\text{laser}}, \,\boldsymbol{p}_{\text{jitter}}, \,\boldsymbol{p}_{\text{clock}}, \,\boldsymbol{p}_{\text{modulation}}]^\top\in\mathbb{R}^{18n\times1}.\label{eq:p_composed_LISA}
\end{equation}

\paragraph{Design matrix \( M \in\mathbb{R}^{24n\times18n} \).} The design matrix \( M \) defines the relationship between the measurement vector \( \boldsymbol{y} \) and the noise vector \( \boldsymbol{p} \). It is constructed by combining the noise contributions for each measurement at each time step.
The entries in \( M \) are calculated using equations  \eqref{eq:isi_o_12} to \eqref{eq:rfi_12_sb}. The delayed terms are considered in $M$ with the appropriate Lagrange interpolation filter masks. 
The inter-spacecraft interferometer carrier measurements are captured by the vector \( \boldsymbol{y}_{\text{isi,}\text{c}} \). The measurement equation is given by
\begin{equation}
\boldsymbol{y}_{\text{isi,}\text{c}} = M_{\text{isi,}\text{c}} \boldsymbol{p},
\end{equation}
where \( M_{\text{isi},\text{c}}\in\mathbb{R}^{6n\times 18n}\) is the submatrix that maps the noise sources in \( \boldsymbol{p} \) to the inter-spacecraft interferometer carrier measurements following the measurement model of equation  \eqref{eq:isi_12}. For the sideband measurements of the inter-spacecraft interferometer, represented by \( \boldsymbol{y}_{\text{isi}, sb} \), the measurement equation is
\begin{equation}
\boldsymbol{y}_{\text{isi,}\text{sb}} = M_{\text{isi,}\text{sb}} \boldsymbol{p},
\end{equation}
where \( M_{\text{isi,}\text{sb}} \in\mathbb{R}^{6n\times 18n}\) maps the noise sources in \( \boldsymbol{p}\) to \( \boldsymbol{y}_{\text{isi,}\text{sb}} \) following equation  \eqref{eq:isi_12_sb}.
The test-mass interferometer measurements \( \boldsymbol{y}_{\text{tmi}} \) can be stated in matrix-vector form as
\begin{equation}
\boldsymbol{y}_{\text{tmi}} = M_{\text{tmi}} \boldsymbol{p},
\end{equation}
where \( M_{\text{tmi}}\in\mathbb{R}^{6n\times 18n} \) represents the mapping of noise sources to the test-mass interferometer measurements following equation  \eqref{eq:tmi_12}. The reference interferometer carrier measurements \( \boldsymbol{y}_{\text{rfi,}\text{c}} \) are related to the all-in-one noise vector $\boldsymbol{p}$ by
\begin{equation}
\boldsymbol{y}_{\text{rfi,}\text{c}} = M_{\text{rfi,}\text{c}} \boldsymbol{p},
\end{equation}
with $M_{\text{rfi,}\text{c}}\in\mathbb{R}^{3n\times 18n}$. Analogously, for the reference interferometer sideband meas\-ure\-ments \( \boldsymbol{y}_{\text{rfi,}\text{sb}} \) we set
\begin{equation}
\boldsymbol{y}_{\text{rfi,}\text{sb}} = M_{\text{rfi,}\text{sb}} \boldsymbol{p},
\end{equation}
with $M_{\text{rfi,}\text{sb}}\in\mathbb{R}^{3n\times 18n}$. By stacking the individual measurement equations for each type of interferometer, the full measurement system can be expressed as
\begin{align}
\boldsymbol{y} &= \begin{pmatrix}
    \boldsymbol{y}_{\text{isi,}\text{c}\phantom{,}} \\
    \boldsymbol{y}_{\text{isi,}\text{sb}} \\
    \boldsymbol{y}_{\text{tmi}\phantom{,,}} \\
    \boldsymbol{y}_{\text{rfi,}\text{c}\phantom{,}} \\
    \boldsymbol{y}_{\text{rfi,}\text{sb}}
\end{pmatrix} =
\begin{pmatrix}
    M_{\text{isi,}\text{c}\phantom{b}} \\
    M_{\text{isi,}\text{sb}} \\
    M_{\text{tmi\phantom{,,}}} \\
    M_{\text{rfi,}\text{c}\phantom{b}} \\
    M_{\text{rfi,}\text{sb}}
\end{pmatrix}
\begin{pmatrix}
    \boldsymbol{p}_{\text{laser}} \\
    \boldsymbol{p}_{\text{jitter}} \\
    \boldsymbol{p}_{\text{clock}} \\
    \boldsymbol{p}_{\text{modulation}} 
\end{pmatrix} = M \boldsymbol{p}.\label{eq:TDIinfinity_fullsystem}
\end{align}

\paragraph{Null-space matrix \( T \in\mathbb{R}^{6n\times24n} \).} 
To construct the TDI-$\infty$ observable \( \boldsymbol{o} \), we need to find the matrix \( T \) whose rows span the null space of the transposed design matrix, \( M^T \). The matrix \( T \) then satisfies
\begin{align}
    T M &= T 
\begin{pmatrix}
    M_{\text{isi,}\text{c}\phantom{b}} \\
    M_{\text{isi,}\text{sb}} \\
    M_{\text{tmi\phantom{,,}}} \\
    M_{\text{rfi,}\text{c}\phantom{b}} \\
    M_{\text{rfi,}\text{sb}}
\end{pmatrix} = 0,
\label{eq:TMcondition_allinone}
\end{align}
so \( T \) defines the transformation needed to project the measurement vector \( \boldsymbol{y} \) into the null space of \( M^\top \), canceling all noise sources contained in \( \boldsymbol{p} \), analogously to the toy model. The TDI-$\infty$ observable \( \boldsymbol{o} \) is then constructed according to equation  \eqref{eq:o_TDIInf}. 

If $M$ is full rank, then $\boldsymbol{o} \in \mathbb{R}^{6n\times1}$, and the space of TDI-$\infty$ observables is spanned by six independent TDI generators. Remember, however, that the generators were only three for TDI-2 \cite{Prince2002}. The reason for this difference becomes clear when we construct the matrix \( T \) using algorithms that produce sparse banded matrices. One such algorithm is the turnback algorithm, efficiently implemented and kindly provided to us by Kai Pfeiffer
\cite{pfeiffer_turnbackLU}.
Since the full design matrix \( M \) becomes sparse and banded when its rows and columns are reordered to group together measurement and noise categories belonging to the same time step, this algorithm can, under certain conditions, yield a sparse and banded structure for \( T \) as well.
Then \( T \) exhibits a repeating pattern of six elements, three of which encode observables that are sensitive to gravitational-wave signals, while the other three are also noise-suppressing but are insensitive to gravitational waves because they consist of local differences of interferometric measurements  {(i.e., differences formed from interferometric measurements on the same spacecraft).}
A more detailed analysis of the composition of the TDI-$\infty$ observable \( \boldsymbol{o} \) is given by Fig.\ \ref{im:PatternsNullSpaceTurnback} in the Appendix.

\subsection{Validation of noise suppression with all-in-one TDI-$\infty$}
\begin{figure}[]
	\centering
  \includegraphics[trim=0 0 0 0, clip,width=0.7\textwidth]{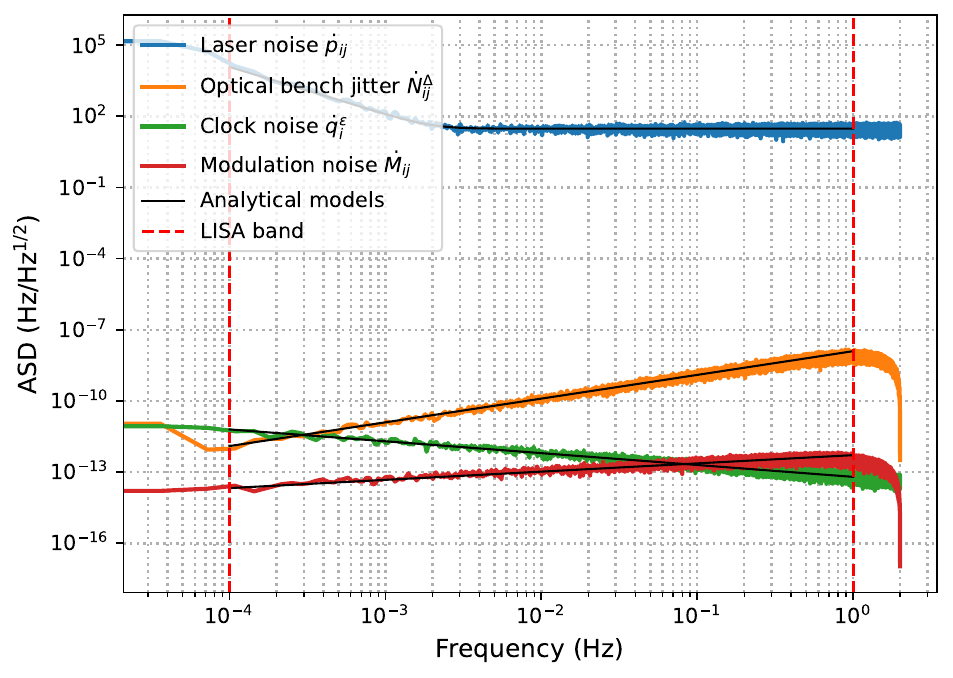}
    \caption{Amplitude spectral density (ASD) of the four primary noise sources that are suppressed by the all-in-one TDI-$\infty$ pipeline: laser noise, optical-bench jitter noise, clock noise, and modulation noise. The plot compares the analytical noise models with the simulated noise time series used as input for the TDI-$\infty$ framework.}
	\label{im:NoiseSpecifications_plot}
\end{figure}

\begin{figure}[htp]
    \centering
    \begin{subfigure}[b]{1\textwidth}
        \centering
        \includegraphics[trim=0 24 0 24, clip,width=0.93\textwidth]{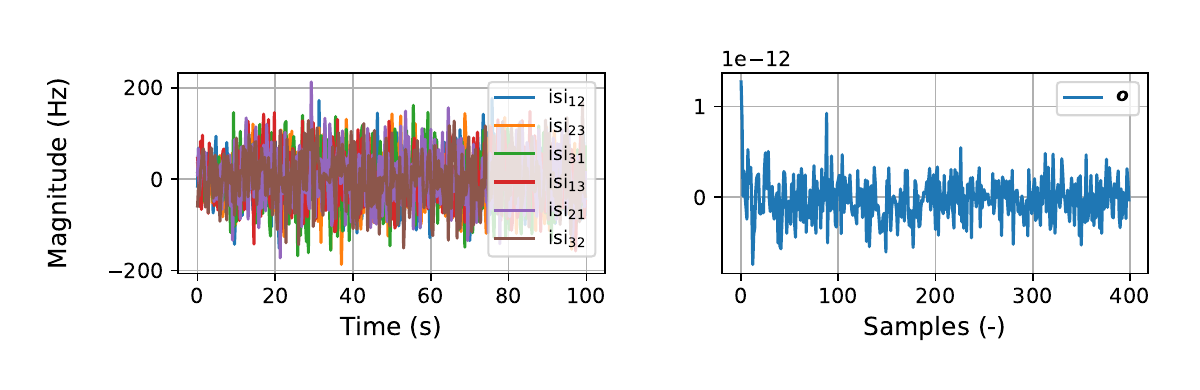}
        \caption*{(a)}
    \end{subfigure}
    \begin{subfigure}[b]{1\textwidth}
        \centering
        \includegraphics[trim=0 24 0 24, clip,width=0.93\textwidth]{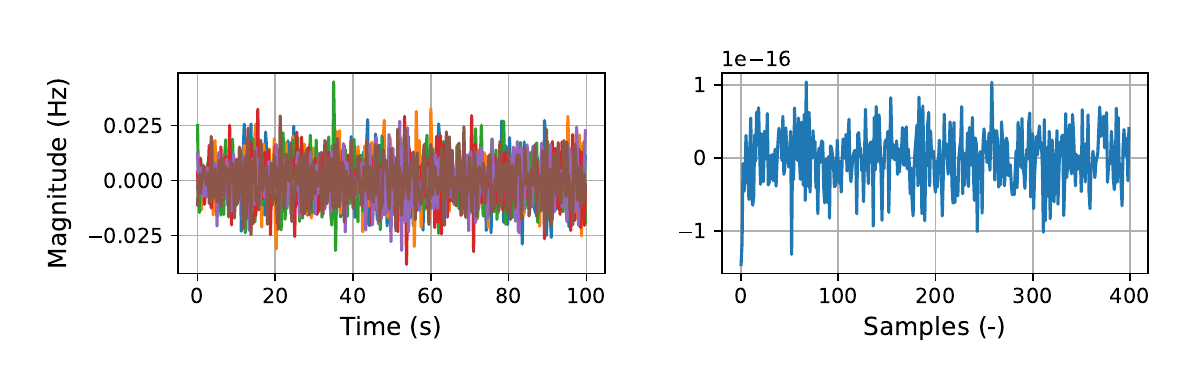}
        \caption*{(b)}
    \end{subfigure}
    \begin{subfigure}[b]{1\textwidth}
        \centering
        \includegraphics[trim=0 24 0 24, clip,width=0.93\textwidth]{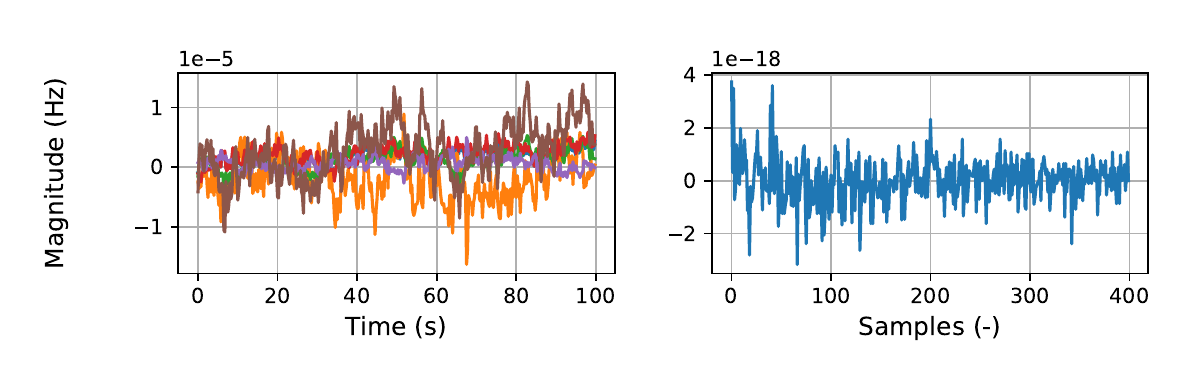}
        \caption*{(c)}
    \end{subfigure}
    \begin{subfigure}[b]{1\textwidth}
        \centering
        \includegraphics[trim=0 24 0 24, clip,width=0.93\textwidth]{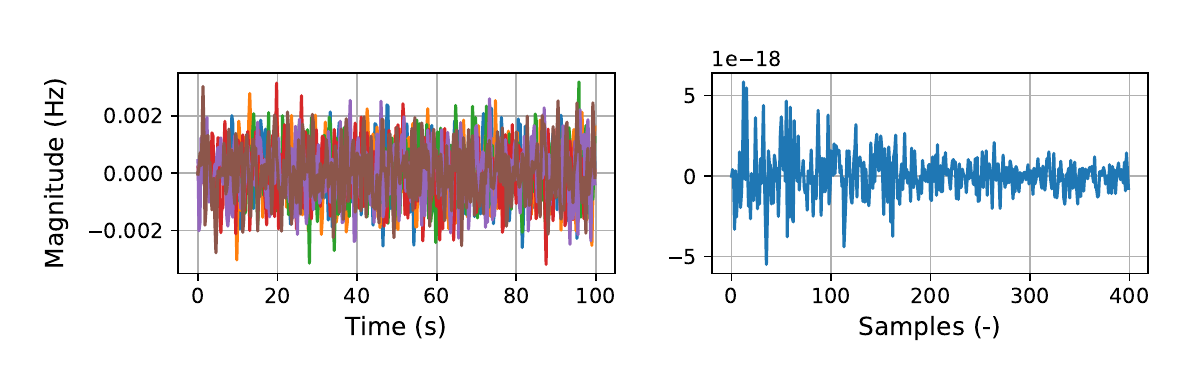}
        \caption*{(d)}
    \end{subfigure}
    \begin{subfigure}[b]{1\textwidth}
        \centering
        \includegraphics[trim=0 24 0 24, clip,width=0.93\textwidth]{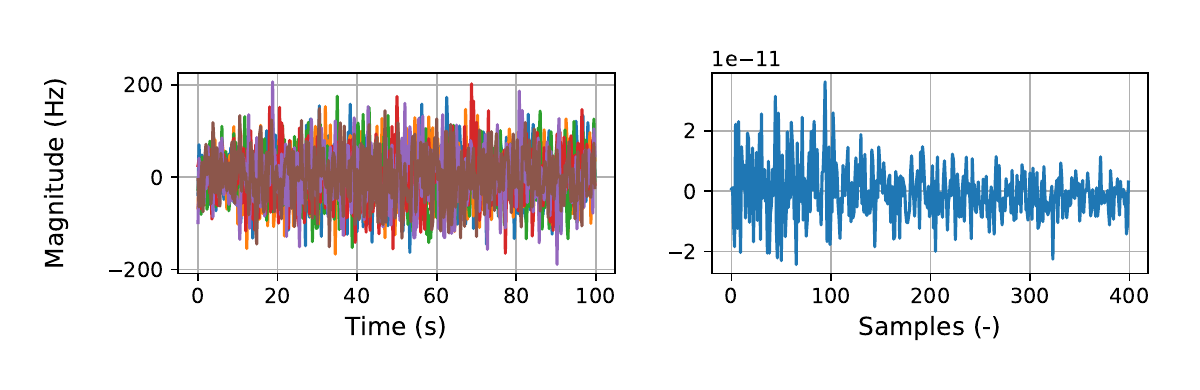}
        \caption*{(e)}
    \end{subfigure}
    \caption{Illustration of the noise-affected inter-spacecraft interferometer measurements (left) and the corresponding TDI-$\infty$ observable $\boldsymbol{o}$ (right). Five scenarios are presented: (a) Laser noise only, (b) Optical-bench noise only, (c) Clock noise only, (d) Modulation noise only, and (e) All noise sources combined.}
    \label{im:SuppressionPerformances}
\end{figure}

We validate the all-in-one TDI-$\infty$ framework by demonstrating its noise suppression capabilities through numerical simulations. Figure \ref{im:NoiseSpecifications_plot} highlights the four primary noise sources considered within the framework (laser noise, optical-bench noise, clock noise, and modulation noise): the effective suppression of these noises is crucial for the mission’s performance and the accuracy of gravitational-wave detection.
Figure \ref{im:SuppressionPerformances} illustrates the impact of the different noise sources on the interferometric measurements in the time domain and the effectiveness of the TDI-$\infty$ pipeline in suppressing them. The inter-spacecraft interferometer measurements displayed here represent a subset of the data processed within the TDI-$\infty$ framework. 

Our validation covers five distinct noise scenarios: laser noise only, optical-bench noise only, clock noise only, modulation noise only, and the combination of all noise sources. The sampling rate for the raw interferometric measurements is set to 4 Hz, matching LISA’s expected downlink rate. The arm lengths are unequal and static in this first example. Realistic LISA orbits will be considered in Section \ref{sec:Simulation}. Note that specific time steps cannot be directly assigned to the TDI-$\infty$ observable.  {Consequently, obtaining a Fourier domain spectrum becomes non-trivial and is omitted.}
\subsection{Computational complexity of all-in-one TDI-$\infty$}
The all-in-one TDI-$\infty$ framework requires that we handle large matrix systems. Specifically, we account for 18 different noise sources, including laser noise, optical-bench jitter, clock noise, and modulation noise from all three LISA spacecraft, as well as 24 interferometric measurements. The full design matrix \( M \in \mathbb{R}^{24n \times 18n} \) grows significantly with the number of time samples, making it computationally challenging to compute its null space for a large number of samples $n$.\enlargethispage{\baselineskip}

\begin{figure}[]
	\centering
  \includegraphics[trim=0 0 0 0, clip,width=0.64\textwidth]{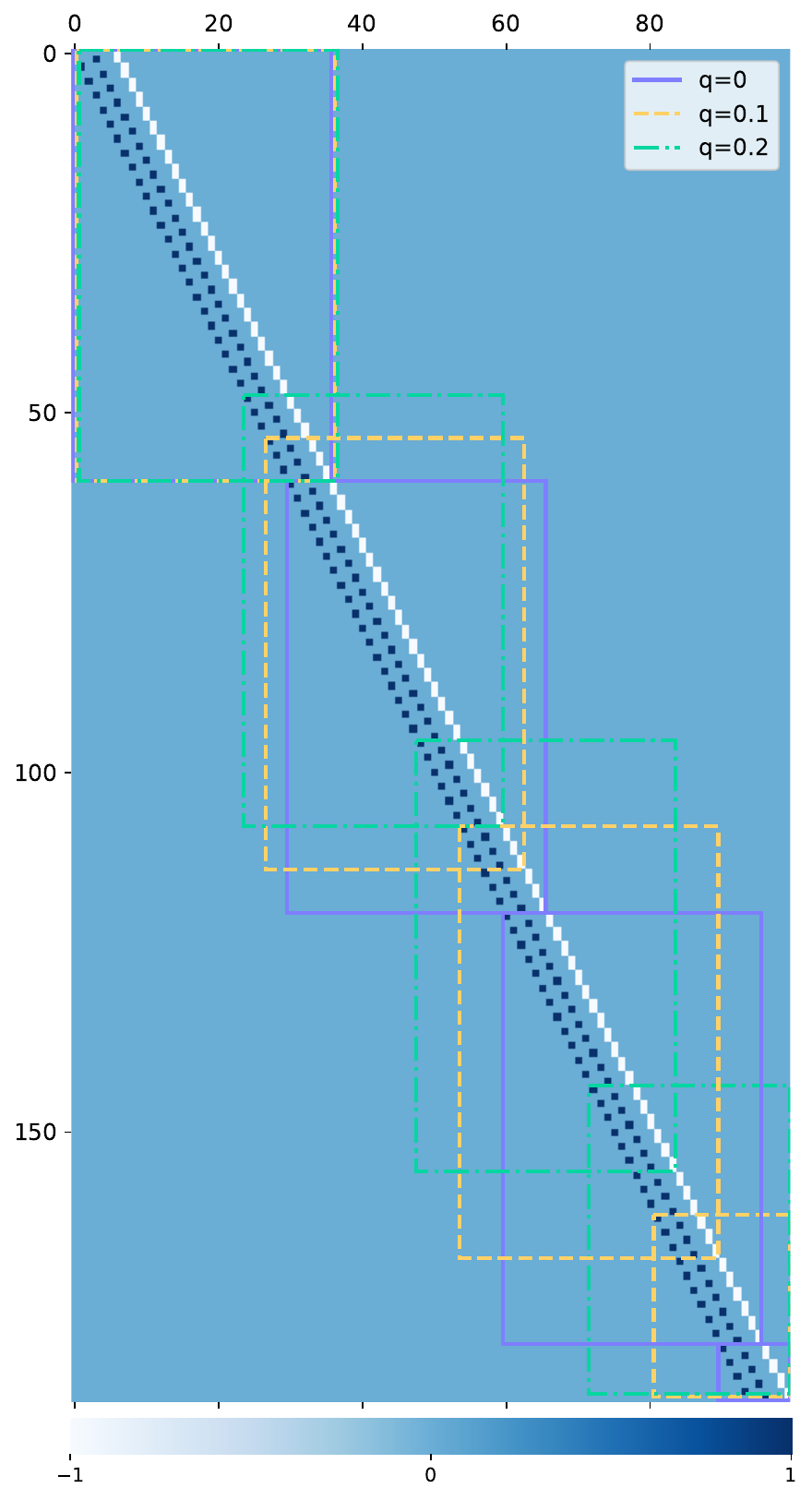}
  \caption{Illustration of the chunking process for the design matrix $M$ in the LISA toy model, considering the measurements $\textrm{isi}_{12,\text{c}}$ and $\textrm{isi}_{13,\text{c}}$ with a common laser noise source $p_1$. The round-trip delays are static, with $\dot{\mathcal{D}}_{12}\dot{\mathcal{D}}_{21} = 6\Delta t$ and $\dot{\mathcal{D}}_{13}\dot{\mathcal{D}}_{31} = 3\Delta t$, where $\Delta t$ is the sampling interval. The chunking is illustrated for different values of the row overlap parameter $0\leq q < 1$ and $l=40$. When  $q=0$, the rows do not overlap, meaning each measurement in $\boldsymbol{y}$ is used exactly once. In this example, we opted for $n = 100 -\dot{\mathcal{D}}_{\textrm{max}}/\Delta t$ valid samples per measurement, where the maximum round-trip delay is given by $\dot{\mathcal{D}}_{\textrm{max}} = \max(\dot{\mathcal{D}}_{12} \dot{\mathcal{D}}_{21}, \, \dot{\mathcal{D}}_{13} \dot{\mathcal{D}}_{31}) = 6\Delta t$.}
	\label{im:MMatrixChunking}
\end{figure}

\newpage\noindent
To address this, we divide the design matrix into smaller submatrices and compute \( T \) locally for each submatrix. This \emph{chunking} technique reduces the computational burden. In chunking, the design matrix \( M \) is split into several submatrices \( M_{\text{sub}}^i \), each handling a portion of the overall measurements and noise sources while ensuring that boundary conditions are maintained. The $i$-th submatrix is
\begin{equation}
\vspace{5pt}
    M_{\text{sub}}^i = M[mli \cdot (1 - q) : ml \cdot (i + 1 - qi), rli \cdot (1 - q) : rl \cdot (i + 1 - qi) + r\dot{\mathcal{D}}_{\textrm{max}}/\Delta t], 
    \label{eq:chunking-def}
\end{equation}
where \( m \) represents the number of measurements, which is 24 for the all-in-one TDI-$\infty$ approach, and \( r \) represents the 18 noise sources to be suppressed. The parameter \( l \) is a free variable that controls the size of each submatrix and can be adjusted based on computational requirements. The matrix index $i$ ranges from 0 to $N_{\text{max}}-1$, where $N_{\text{max}}-1$ is the index of the last submatrix that fits within $M$, even if only partially. The overlap between consecutive submatrices is governed by \( q \), an adjustable factor that impacts estimation accuracy during Bayesian inference  {by reducing boundary effects in the associated chunked noise covariance matrices. Note that when $q=0$, the assembled set of submatrices already retains the full information from the global design matrix. However, chunking the covariance matrix this way leads to a partial loss of cross-correlation information and can be attenuated with $q>0$. This topic is discussed in greater detail in Section \ref{ss:loglike_tdiinf}, which introduces the likelihood of the TDI-$\infty$ concept.} Finally, $\dot{\mathcal{D}}_{\textrm{max}}$ represents the maximum inter-spacecraft delay in the system, including the interpolation order in case of fractional delays, and \( \Delta t \) is the sampling time. Figure \ref{im:MMatrixChunking} illustrates the chunking process, showing how the design matrix \( M \) is split into smaller submatrices for the LISA toy model in Fig.\ \ref{im:ToyModel} with \( m = 2 \), \( r = 1 \), and \( l = 40 \). The same process is applied to the all-in-one framework, but due to its complexity, the visualization is omitted. In practice, the full design matrix is not constructed first; instead, the submatrices \( M_{\text{sub}}^i \) are generated directly to reduce computational overhead.

Chunking, as defined by equation  \eqref{eq:chunking-def}, ensures that the boundary conditions are correctly handled in each submatrix. This process assumes that only valid measurement data is used to construct $\boldsymbol{y}$ in equation  \eqref{eq:y_composed_LISA}. Specifically, ``valid'' refers to measurements taken from $t = \dot{\mathcal{D}}_{\textrm{max}}/\Delta t$ to $t_{n-1}$, where both remote and local noise contributions are present  for successful noise suppression.  Excluding invalid samples prevents a situation analogous to the initialization phase in TDI-2, where a waiting period is needed for light to propagate across the constellation after activating the laser sources. By omitting these initial invalid samples after switch-on, we avoid repeating initialization phases when processing each new submatrix in TDI-$\infty$.
 {Consequently, only valid samples are shown in Fig.\ \ref{im:MMatrixChunking}.}

Note that the rows of $M$ must be arranged so that measurements from the same time step are grouped together. If the rows were instead organized by measurement type, for example, placing all inter-spacecraft interferometer measurements first, followed by all test-mass interferometer measurements, the resulting submatrices would not contain the required information to compute a valid null space. Therefore, reordering the rows correctly is crucial for the noise suppression process to work while chunking.

\newpage\noindent
For each chunked submatrix \( M_{\text{sub}}^i \), we compute a corresponding null-space matrix \( T^i \). Similarly, the measurement vector \( \boldsymbol{y} \) is divided into sub-vectors \( \boldsymbol{y}_{\text{sub}}^i \), allowing to compute the TDI-$\infty$ observable for the current submatrix:
\begin{equation}
\boldsymbol{o}^i = T^i \boldsymbol{y}_{\text{sub}}^i.  \label{eq:TDIInf_sub}
\end{equation}
After calculating \( \boldsymbol{o}^i \) for each submatrix, we concatenate all contributions to obtain the final TDI-$\infty$ observable \( \boldsymbol{o} \):
\begin{equation}
\boldsymbol{o} = \left[ \boldsymbol{o}^0, \boldsymbol{o}^1, \dots, \boldsymbol{o}^{N_{\text{max}}-1} \right]. \label{eq:TDIInf_composed}
\end{equation}
 {The result obtained with chunking generally differs from that without chunking, depending heavily on the characteristics of the null-space algorithm used. When the algorithm produces sparse or banded null spaces, the null space of each chunk remains a geometric invariant. In this case, increasing $q$ leads to a repetition of TDI observables. However, for algorithms that produce dense null spaces -- which mix samples across a wide range of time steps -- the situation becomes more complex. The impact of this on Bayesian inference will be studied in the future in more detail.}

Since the TDI-$\infty$ observable is not interpreted as a continuous time-series signal, there are no concerns about discontinuities when concatenating the \( \boldsymbol{o}^i \)s. However, applying the same chunking method to the templates during parameter inference is paramount to ensure consistency between the data and the model processing.

\section{Bayesian parameter inference with TDI-$\infty$}\label{sec:Bayesian}
Bayesian inference is the standard approach to extract physical parameters from high-dimensional, noisy data in gravitational-wave astronomy \cite{Moulin_Veeravalli_2018, Bailer-Jones2017-iw, Tang_2022, Keskin2023}. The goal is to determine the posterior distribution of parameters that describe the signal of interest in the recorded data. In this section, we introduce the theoretical foundations of Bayesian inference and explain how it is applied to recover astrophysical information from LISA's measurements with the TDI-$\infty$ framework.

\subsection{Principles of Bayesian inference}
In the context of gravitational-wave astronomy, the detection and characterization of signals are framed as parameter-estimation problems. Given a set of noisy data, we want to infer the most probable values of parameters that describe the physical processes that generated the observed data. This process lends itself naturally to the framework of Bayesian inference. At its heart is Bayes' theorem \cite{10.1093/biomet/45.3-4.296}
\begin{equation}
p(\theta \mid d) = \frac{p(d \mid \theta) p(\theta)}{p(d)},
\end{equation}
where \( p(\theta \mid d) \) is the {posterior distribution}, representing the probability of the parameters \( \theta \) given the data \( d \); \( p(d \mid \theta) \) is the {likelihood}, the probability of observing the data \( d \) given the model parameters \( \theta \);  \( p(\theta) \) is the prior probability distribution, representing our knowledge about the parameters before we consider the data; and \( p(d) \) is the {evidence}, a normalizing constant that ensures the posterior is a valid probability distribution.
In parameter estimation, the goal is to compute the posterior distribution \( p(\theta \mid d) \) given the observed data and the prior distribution.

\paragraph{Likelihood.} The likelihood \( p(d \mid \theta) \) represents the probability of observing the data \( d \) given a specific set of parameters \( \theta \). For most applications, the likelihood depends on the noise properties of the measurement and on the model of the signal of interest. In the case of Gaussian-distributed additive noise, the likelihood is typically expressed in terms of a {noise-weighted inner product}:
\begin{equation}
\log p(d \mid \theta) \propto -\frac{1}{2} \langle d - h(\theta), d - h(\theta) \rangle_{\text{noise}},\label{eq:log_likelihood}
\end{equation}
where \( h(\theta) \) is the predicted signal for a given parameter set \( \theta \), and \( \langle \cdot ,\hspace{-2pt}\cdot \rangle_{\text{noise}} \) is the noise-weighted inner product defined as
\begin{equation}
\langle x, y \rangle_{\text{noise}} = 4 \, \Re \int_0^\infty \frac{\tilde{x}(f) \tilde{y}^*(f)}{S_\textrm{noise}(f)} \, \mathrm{d}f.\label{eq:inner_product}
\end{equation}
Here, \( \tilde{x}(f) \) and \( \tilde{y}(f) \) are the Fourier transforms of the signals \( x(t) \) and \( y(t) \), and \( S_\textrm{noise}(f) \) is the noise power spectral density of the detector.

\paragraph{Priors.} The prior \( p(\theta) \) encapsulates our knowledge about the parameters before we take the data into account. In some cases, a well-motivated prior based on theoretical expectations may be available. In the absence of prior knowledge, uninformative or flat priors can be used to allow the data to dominate the inference process. Priors can also be more complex, incorporating constraints or physically motivated limits on the parameters.

\paragraph{Evidence.} The evidence \( p(d) \) is the normalization factor that ensures the posterior integrates to 1:
\begin{equation}
p(d) = \int p(d \mid \theta) p(\theta) d\theta.
\end{equation}
While the evidence plays a crucial role in model comparison, it cancels out when comparing the relative probability of different parameter values within a single model. 

The main goal of Bayesian inference is to compute the {posterior distribution} \( p(\theta \mid d) \), which contains all the information about the parameters to be estimated. From the posterior, summary statistics such as the {mean}, {median}, or {mode} of the distribution can be computed, along with {credible intervals}, which provide a range of parameter values that contain a certain probability mass.
Computing the posterior distribution analytically is often infeasible, especially when the parameter space is high-dimensional or when the likelihood function is complex. To address this, {sampling methods} such as {Markov chain Monte Carlo (MCMC)} are employed, which allow the posterior distribution to be approximated through random sampling \cite{noauthor_2023-fe, vanOijen2024, Spiegelhalter-95, Hanada2022, Hancock2004}.

\subsection{Markov chain Monte Carlo sampling}
MCMC methods are a family of algorithms used to sample from complex probability distributions, especially when direct computation of the posterior is not possible.  MCMC generates a sequence of samples that approximate the target posterior distribution after a sufficient number of iterations. One of the most commonly used MCMC algorithms, and the one that is used in this paper, is the {Metropolis--Hastings algorithm} \cite{Robert2015-pq, Berg2004-fr, Minh2015-ts}. At each iteration, the algorithm proposes a new set of parameters \( \theta_{\text{new}} \) based on the current parameter values \( \theta_{\text{old}} \), and then decides whether to accept or reject the new parameters based on a probabilistic criterion. The algorithm proceeds as follows:
\begin{enumerate}
    \item \textit{Propose a new state:} given the current state \( \theta_{\text{old}} \), propose a new state \( \theta_{\text{new}} \) using a proposal distribution \( q(\theta_{\text{new}} \mid \theta_{\text{old}}) \).
    \item \textit{Calculate acceptance probability:} compute the acceptance ratio $\alpha$, which compares the likelihood of the proposed state to the current state. This ratio helps determine whether to accept the proposed move, ensuring the algorithm explores the parameter space effectively. The ratio is defined as
    \begin{equation}
    \alpha = \min \left( 1, \,\frac{p(\theta_{\text{new}} \mid d) q(\theta_{\text{old}} \mid \theta_{\text{new}})}{p(\theta_{\text{old}} \mid d) q(\theta_{\text{new}} \mid \theta_{\text{old}})} \right).
    \end{equation}
    \item \textit{Accept or reject:} Draw a random number \( u \) from a uniform distribution \( u \sim \mathcal{U}(0,1) \). If \( u < \alpha \), accept the new state \( \theta_{\text{new}} \); otherwise, keep the old state \( \theta_{\text{old}} \).
    \item \textit{Repeat:} Iterate the process to build up a chain of samples.
\end{enumerate}
In practice, the early part of the MCMC chain,  the {burn-in} period, may not represent the true posterior distribution, as the chain is still exploring the parameter space. This initial set of samples is discarded to ensure that the remaining samples are drawn from the equilibrium distribution.
Convergence diagnostics, such as the {Gelman-Rubin statistic} \cite{Gelman2011, https://doi.org/10.48550/arxiv.1812.09384}, can be used to assess whether the MCMC chain has converged to the target posterior distribution. Once convergence is achieved, the remaining samples are considered representative of the posterior distribution.

\subsection{Log-likelihood calculation with classic TDI combinations}
The key objective of this study is to compare the posterior distributions for MBHB parameters obtained via classical TDI and TDI-$\infty$ in the presence of data gaps and noise, so we need to set up the likelihood computation in both frameworks.
For TDI-2, the process is straightforward  in the absence of data gaps and is well established in the literature. Assuming the observed data \( d \) and the model prediction \( h(\theta) \), the log-likelihood follows the standard form of equation  \eqref{eq:log_likelihood}. The total log-likelihood for TDI-2 is modeled as the sum of the individual log-likelihoods for the three quasi-orthogonal TDI channels  {\( A \), \( E \), and \( T \)},  {which are constructed as linear combinations of the TDI Michelson channels \( X \), \( Y \), and \( Z \) as described in \cite{Prince2002}. These channels are designed to yield a diagonal noise-correlation matrix under certain conditions\footnote{When considering unequal arm lengths or realistic orbits of the LISA spacecraft, the \( A \), \( E \), and \( T \) channels are not strictly orthogonal, making the summation an approximation.}.
 The orthogonality property enables us to compute the overall log-likelihood of TDI-2 as
\begin{equation}
\log p(d \mid \theta) = -\frac{1}{2} \sum_{ } \langle d_i - h_i(\theta), d_i - h_i(\theta) \rangle_{\text{noise}},
\label{eq:log_likelihood_TDI2.0_sum}
\end{equation}
where \( d_A \), \( d_E \), and \( d_T \) represent the observed data, and \( h_A(\theta) \), \( h_E(\theta) \), and \( h_T(\theta) \) are the corresponding templates for these channels based on the parameter set \( \theta \). Here \( \langle \cdot , \cdot \rangle_{\text{noise}} \) denotes the noise-weighted inner product, as defined in equation  \eqref{eq:inner_product}, using power spectral densities specific to each channel. The formulation uses the power spectral density \( S_\text{noise}(f) \) of the noise sources propagated through TDI, ensuring that the frequency-dependent noise properties of the LISA detector are appropriately accounted for in the likelihood. 

In the presence of data gaps, computing the log-likelihood for classical TDI becomes more complicated. Handling gaps might require specialized techniques, such as gap filling or interpolation, to reconstruct the missing data in a way that preserves the statistical properties of the noise and the signal. Comprehensive methods for addressing data gaps in classical TDI are an ongoing area of research and are beyond the scope of this paper. Instead, here we adopt a straightforward approach, primarily for comparison with the data gap handling method in TDI-$\infty$, which will be presented later: gaps in the data naturally divide the continuous time series of each TDI channel into distinct segments. The log-likelihood calculation for classical TDI is then performed independently on each uninterrupted data segment. Specifically, for each gap-free segment ${(s)}$ of the time series \(d_i^{(s)}\) of TDI variables $i = $ \(A\), \(E\), and \(T\), the log-likelihood is computed for that segment, and the results for all segments are summed together. This segmented approach allows to avoid dealing with gap interpolation methods and instead leverages the data points that are available without making assumptions about the missing data -- assumptions that TDI-$\infty$ will avoid naturally. The total log-likelihood can then be formulated as the sum of the log-likelihoods for each segment across all TDI channels.  {This segment-wise summation approach represents the simplest method for handling the likelihood calculation, although it introduces some loss of information: the time series of noise across neighboring segments are not statistically independent.} For TDI-2 in the presence of data gaps, we express the log-likelihood as
\begin{equation}
\log p(d^{\text{gaps}} \mid \theta)
 = -\frac{1}{2} \sum_{i=A,\,E,\,T} \sum_{s=0}^{N_{\text{seg},i}-1} \langle d_i^{(s)} - h_i^{(s)}(\theta), d_i^{(s)} - h_i^{(s)}(\theta) \rangle_{\text{noise}},
\label{eq:log_likelihood_TDI2.0_with_gaps}
\end{equation}
where \( d_i^{(s)} \) represents the \(s\)-th gap-free segment of the data for TDI variable \(i\),  \( h_i^{(s)}(\theta) \) represents the model prediction for the corresponding segment, and \( N_{\text{seg},i} \) is the number of segments for the TDI variable \(i\).

\subsection{Log-likelihood calculation with TDI-$\infty$} \label{ss:loglike_tdiinf}
TDI-$\infty$ offers a natural solution to handle data gaps without requiring additional gap-filling techniques. This is achieved by systematically removing the corresponding invalid measurements from the measurement vector \( \boldsymbol{y} \), defined in equation  \eqref{eq:y_composed_LISA}, as well as eliminating the corresponding rows in the design matrix \( M \) from equation  \eqref{eq:TDIinfinity_fullsystem}. Consequently, gaps are handled elegantly in the TDI-$\infty$ framework, ensuring that no special treatment is needed when constructing the likelihood function.

To derive the likelihood function for TDI-$\infty$, it is assumed that the noise in the measurements is Gaussian and characterized by a zero mean and covariance matrix \( N \). The log-likelihood in this case can be written as
\begin{equation}
    \log p(d \mid \theta) = 
    -\frac{1}{2} \Delta \boldsymbol{o}^\dagger \left( T N T^\dagger \right)^{-1} \Delta \boldsymbol{o}
    - \frac{1}{2} \log \left( |2 \pi T N T^\dagger| \right),
    \label{eq:TDIinfinity_loglikelihood}
\end{equation}
where \( \Delta \boldsymbol{o} = T (\boldsymbol{y} - \boldsymbol{y}_{\text{GW}}(\theta)) \) is the residual between the projected measurement vector and the model, mapped to the TDI-$\infty$ observable using the null-space matrix $T$ \cite{PhysRevD.103.082001}. The covariance matrix \( N \in\mathbb{R}^{24n\times24n} \) describes the noise properties of the measurement system, and \( T N T^\dagger \) is the effective noise covariance in the TDI-$\infty$ space. Note that \( T \) and \( TNT^\dagger \) do not depend on \(\theta\) and only need to be computed once.

An important distinction between the likelihood calculation in TDI-2 and TDI-$\infty$ is the domain in which they operate. While the TDI-2 likelihood is computed in the frequency domain, TDI-$\infty$ works in the time domain. Consequently, the noise covariance matrix $N$ of equation  \eqref{eq:TDIinfinity_loglikelihood} must also be expressed in the time domain. To construct it we first define the noise covariance matrix \( \tilde{N} \) in the frequency domain. The diagonal elements of \( \tilde{N} \) represent the power spectral density \( S_\textrm{noise}(f) \) of the secondary noise sources, as observed in the various interferometer measurements, evaluated at each frequency. We then define the discrete Fourier transform (DFT) matrix \( F \) for the times represented in a TDI-$\infty$ chunk and for the appropriate set of frequencies, and transform the covariance matrix into the time domain via
\begin{equation}
N = F^\dagger \tilde{N} F,\label{eq:DFTConversion}
\end{equation}
where \( F^\dagger \) denotes the conjugate transpose of the DFT matrix.

When chunking is applied to handle long measurement time series, the total likelihood is computed by summing the likelihood contributions from each individual chunk:
\begin{equation}
    \log p(d^\text{chunked} \mid \theta) = \sum_{i=0}^{N_\textrm{max}-1} \left( 
    -\frac{1}{2} \Delta \boldsymbol{o}^i{}^\dagger \left( T^i N_{\text{sub}}^i T^i{}^\dagger \right)^{-1} \Delta \boldsymbol{o}^i
    - \frac{1}{2} \log \left( |2 \pi T^i N_{\text{sub}}^i T^i{}^\dagger| \right)
    \right).
    \label{eq:TDIinfinity_loglikelihood_chunked}
\end{equation}
Chunking of \( N \) follows the approach similar to equation  \eqref{eq:chunking-def} and illustrated in Fig.\ \ref{im:MMatrixChunking} for the LISA toy model. The submatrices obtained from $N$ are defined by
\begin{equation}
    N_{\text{sub}}^i = N[mli \cdot (1 - q) : ml \cdot (i + 1 - qi), mli \cdot (1 - q) : ml \cdot (i + 1 - qi)],
    \label{eq:chunking-def-covariancematrix}
\end{equation}
with the parameter definitions equal to equation  \eqref{eq:chunking-def}. Unlike the design matrix \( M \), where entries outside the submatrices are zero, the noise covariance matrix \( N \) inevitably contains cross-correlations between measurement samples that extend beyond the submatrices obtained from equation  \eqref{eq:chunking-def-covariancematrix}. As a result, chunking \( N \) leads to a loss of covariance information. To partially mitigate this issue, the row overlap factor \( q \) can chosen to be greater than zero. This parameter allows for some overlap between submatrices, helping to reduce the loss of information and retain cross-correlation information between adjacent chunks.  {The selection of $q$ and the chunk length $l$ involves a trade-off between computational complexity, signal-to-noise ratio loss, and the double-counting of data in overlapping matrix regions.}

Note that handling data gaps under chunking remains straightforward, with some appropriate adjustments. The design matrix \( M \) needs to be first partitioned into submatrices \( M_{\text{sub}}^i \) according to equation  \eqref{eq:chunking-def}. Once chunking has been applied, the gap indices -- originally defined with respect to the dimensions of \( M \) -- must be translated into the local submatrix frame, which is determined by the size of each submatrix and its specific identifier. After this transformation, the relevant rows in both \( M_{\text{sub}}^i \) and \( \boldsymbol{y}_{\text{sub}}^i \) can be removed. The same applies to \( N_{\text{sub}}^i \). Performing chunking prior to gap removal ensures that the matrix boundary conditions necessary for effective noise suppression are preserved.

\section{Simulation results}\label{sec:Simulation}
This section presents the simulation results of our study. To provide context, we briefly introduce the two main data-processing pipelines that will be developed for LISA: the global-fit and the low-latency pipelines.
\begin{enumerate}
\item  The global-fit pipeline is designed to analyze extensive LISA datasets, aiming to identify and model multiple overlapping signals from a diverse range of astrophysical sources. Its main challenge is disentangling these signals from a persistent global background created by millions of individually unresolved sources. The global fit typically requires months or years of data for thorough analysis. The global-fit pipeline also needs to account for the instrumental noise (and potential stochastic signals), which adds to its complexity.

\item In contrast, LISA's low-latency pipeline focuses on the rapid detection and analysis of transient gravitational-wave events such as the mergers of MBHBs. The primary objective of this pipeline is to provide early warnings of such events, which is particularly important when there could be electromagnetic counterparts: for example, jets or accretion disks formed during the merger. Early alerts enable ground- and space-based observatories to prepare for follow-up observations. Due to the urgency of these detections, the low-latency pipeline prioritizes speed over the comprehensive analysis performed by the global fit, working with limited data to quickly extract key information, such as the time of coalescence of MBHBs.  For the alert pipeline, we can assume the noise to be known, which renders TDI-\(\infty\) feasible.
\end{enumerate}
 {Within this context, our simulation scenario focuses on parameter estimation for MBHBs under the constraints of limited data, aligning with the objectives of a low-latency pipeline. In such low-latency applications, data often arrive as a quasi-live stream where some packets may be missing. When data are scarce, the impact of gaps becomes especially significant. Requesting retransmission of lost data could introduce delays of several hours, which is unfeasible for real-time or near-real-time analysis. The goal of the simulations described here is to assess the performance of TDI-\(\infty\) and compare its accuracy in parameter inference to classical TDI, particularly in the presence of data gaps.}

\subsection{Astrophysical parameters of MBHBs}
In the context of low-latency analysis for LISA, MBHBs are among the most important sources of transient gravitational-wave signals. These systems, which consist of two supermassive black holes orbiting each other, are expected to produce strong gravitational-wave signals as they spiral inward and eventually merge in the LISA band. 
The key astrophysical parameters of interest for MBHBs include the following:
\begin{itemize}
    \item {Coalescence time} \( t_c \): the time until the two black holes merge.\vspace{1pt}
    \item {Sky localization} (ecliptic latitude \( \beta \) and longitude \( \lambda \)): the position of the binary on the sky.\vspace{2pt}
    \item {Masses} \( m_1 \) and \( m_2 \): the masses of the individual black holes.\vspace{1pt}
    \item {Spins} \( \chi_1 \) and \( \chi_2 \): the dimensionless spins of the black holes, affecting both the waveform and the final spin of the remnant black hole after the merger.\vspace{1pt}
    \item {Redshift} \( z \): the redshift of the system, which indicates its distance and helps place the merger in a cosmological context.\vspace{1pt}
    \item {Luminosity distance} \( D_L \): the inferred distance to the binary system.
\end{itemize}
The ability to quickly extract these parameters from LISA data, even with limited observational windows, underscores the importance of efficient algorithms such as TDI-$\infty$. The rapid and precise identification of MBHB mergers ensures that critical opportunities for multi-messenger astronomy are not missed.

\subsection{Simulation scenario}
In this simulation, we analyze one hour of data sampled at 4 Hz, containing a single MBHB injection, and compare the performance of TDI-2 and TDI-$\infty$, both with and without data gaps.
In the scenario with gaps, six one-sample gaps are introduced at intervals of 300 samples within the inter-spacecraft interferometer measurements from optical benches 12, 23, and 31, located near the merger event.
A second scenario, discussed later, will explore the impact of the same gap pattern occurring during the inspiral phase.  {Gaps are represented by ``NaN'' values in the time series of raw interferometer measurements, which are then propagated through the L0--L1 processing pipelines of Fig.\ \ref{im:L0L1Architectures}.} The parameters for the injected MBHB system are summarized in Table \ref{tab:injected_params}.
\begin{table}[]
    \centering
    \caption{Injected parameters for the MBHB system.}
    \small 
    \begin{tabular}{ll}
        \toprule
        \textbf{Parameter} & \textbf{Value} \\
        \midrule
        Ecliptic latitude & -0.1895 \\
        Ecliptic longitude & 0.2230 \\
        Polar angle of spin 1 & 1.2075 \\
        Polar angle of spin 2 & 1.2279 \\
        Spin 1 & 0.8240 \\
        Spin 2 & 0.6162 \\
        Mass 1 & 100,565.05 \( M_{\odot} \) \\
        Mass 2 & 193,254.58 \( M_{\odot} \) \\
        Coalescence time & 1 hour* \\
        Phase at coalescence & 2.0290 \\
        Initial polar angle  & 0.6304 \\
        Initial azimuthal angle  & 1.0909 \\
        Redshift & 1.013 \\
        Luminosity distance & 31,489.75 Mpc \\
        Signal-to-noise ratio & ca. 300 \\
        \bottomrule
    \end{tabular}
    \label{tab:injected_params}
        \vspace{0.2cm}
    \begin{minipage}{0.55\textwidth}
        \scriptsize * The coalescence time is measured from the initial sample in the data stream.
    \end{minipage}
\end{table}
Figures \ref{im:TDI20_GapsIllustration_TimeDomain} and \ref{im:TDIInf_GapsIllustration_TimeDomain} demonstrate the impact of data gaps on both TDI-2 and TDI-$\infty$ for the first simulation scenario. Figure \ref{im:TDI20_GapsIllustration_TimeDomain} presents the TDI-2 Michelson channels $X$, $Y$, and $Z$ showing both the noisy dataset and the TDI-processed injection alone (both centered around the gaps region for clarity). For TDI-$\infty$, Fig.\ \ref{im:TDIInf_GapsIllustration_TimeDomain} shows the full dataset considered in this study.
\begin{figure}[]
    \centering
    \includegraphics[width=0.9\textwidth]{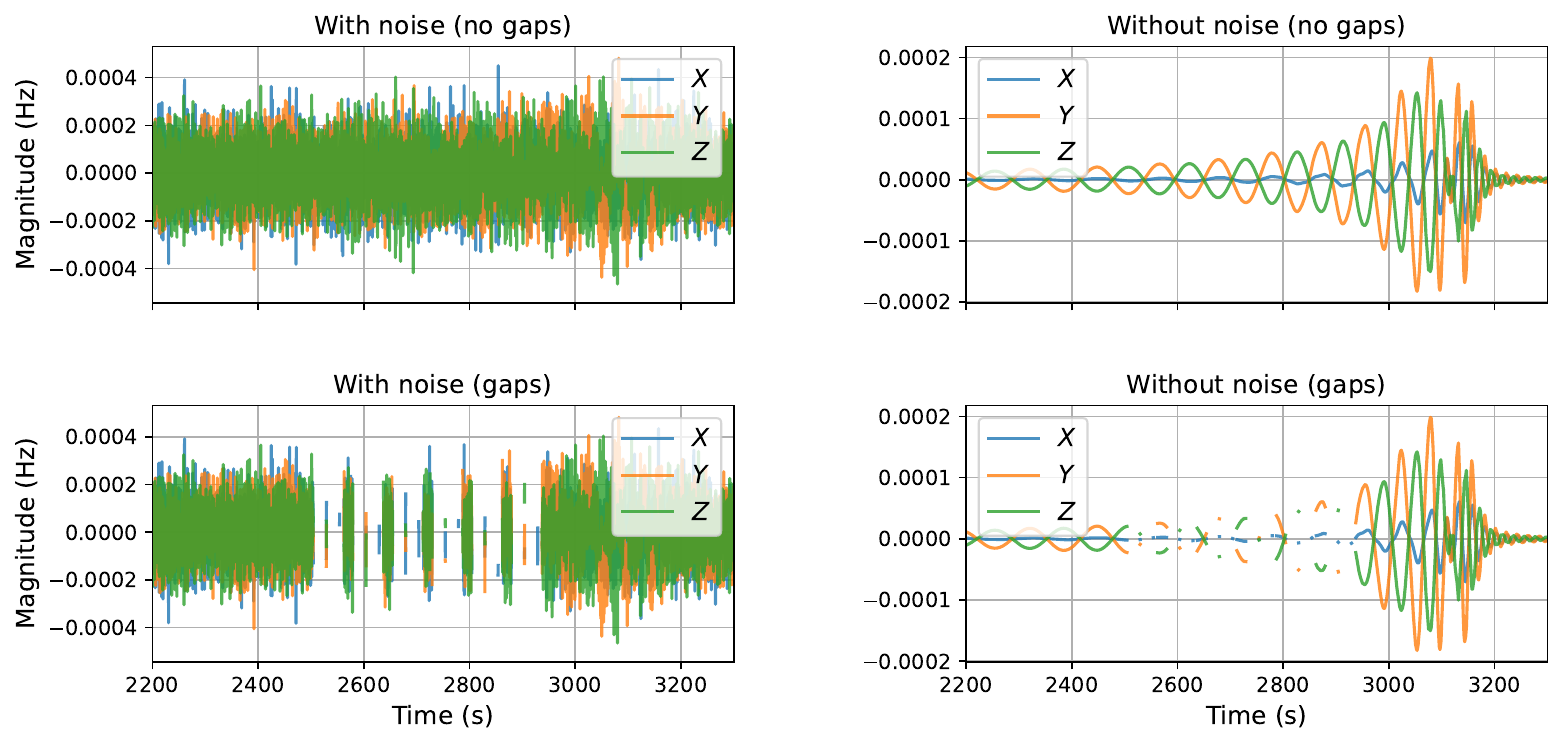}
    \caption{TDI-2 Michelson channels $X$, $Y$, and $Z$ for the gap-free (top) and gap-affected (bottom) cases considered in this study. 
     {For clarity, the time series are shown only around the regions affected by gaps.}
    The data gaps, injected near the merger event as described in the main text, cause substantial data corruption in TDI-2, affecting approximately 400 seconds of data. The left panels illustrate the scenario with secondary noise from the optical metrology system, while the right panels show the noiseless case.}
    \label{im:TDI20_GapsIllustration_TimeDomain}
    \vspace{45pt}
    \centering
    \includegraphics[trim=0 0 0 0, clip,width=0.9\textwidth]{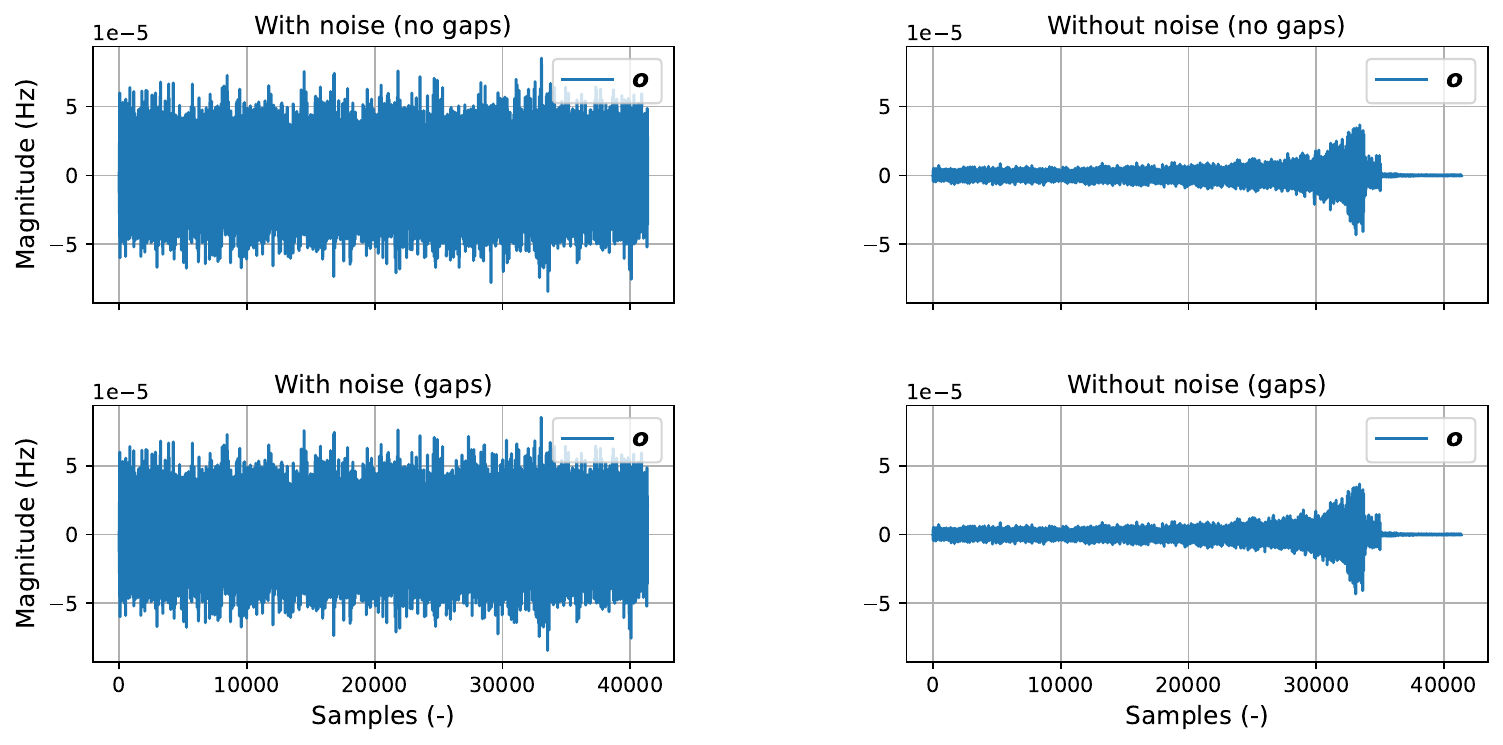}
    \caption{TDI-$\infty$ observable 
    $\boldsymbol{o}$ in the gap-unaffected (top) and gap-affected (bottom) cases. The left panels illustrate the scenario with secondary noise from the optical metrology system, while the right panels show the noiseless case. Unlike classical TDI, gaps do not appear in the TDI-$\infty$ output because they are eliminated from the design matrix prior to null-space computation, as explained in Section \ref{sec:Bayesian}.  {Since the gap-free and gap-included cases have different numbers of samples, plotting their absolute differences to visualize discrepancies is not feasible and omitted here.}}
    \label{im:TDIInf_GapsIllustration_TimeDomain}
\end{figure}
In the presence of data gaps, TDI-2 is significantly impacted, with approximately 400 seconds of data being corrupted. In TDI-$\infty$, the number of lost measurements corresponds roughly to the number of introduced gaps. Differences between the gap-free and gap-included cases are observed  {mainly} in the chunks where gaps are present.  {The discrepancy in sample counts makes direct comparison challenging, as aligning data points between the two cases is non-trivial in regions affected by gaps.}

\subsection{Classical TDI framework}
The analytical power spectral density models for secondary noise in the inter-spacecraft and reference interferometers, collectively summarized as optical metrology system (OMS) noise, are provided in equations \eqref{eq:isi_OMS} and \eqref{eq:rfi_OMS} in units $\textrm{Hz}^2/\textrm{Hz}$. These models are applied in Figs. \ref{im:TDI20_GapsIllustration_TimeDomain} and \ref{im:TDIInf_GapsIllustration_TimeDomain} and will also be used in the MCMC simulations of Section \ref{subsec:Posteriors} to construct the TDI-$\infty$ noise covariance matrix $N$, as defined in equation  \eqref{eq:DFTConversion}:
\begin{align}
    \widetilde{\text{isi}}_{\text{OMS}}(f) &= \Big( 2 \pi \cdot 6.35\,\frac{\textrm{pm}}{\sqrt{\textrm{Hz}}} \Big)^2 \cdot \left( f^2 + \frac{(2\,\textrm{mHz})^4}{f^2} \right) \cdot \bigg( \frac{\nu_0}{c} \bigg)^2\label{eq:isi_OMS} \\
    \widetilde{\text{rfi}}_{\text{OMS}}(f)&= \Big( 2 \pi \cdot 3.32\,\frac{\textrm{pm}}{\sqrt{\textrm{Hz}}} \Big)^2 \cdot \left( f^2 + \frac{(2\,\textrm{mHz})^4}{f^2} \right) \cdot \bigg( \frac{\nu_0}{c} \bigg)^2 \label{eq:rfi_OMS}
\end{align}
In the equations above, $c$ denotes the speed of light.
To construct $S_\textrm{noise}(f)$ from equation  \eqref{eq:inner_product} and compute the log-likelihood for classical TDI of equation  \eqref{eq:log_likelihood_TDI2.0_sum}, the OMS noise contributions must be propagated through the algebraic TDI algorithm. The power spectral density model for the OMS noise in the TDI-2 Michelson variable \( X \), based on the nominal LISA arm length of \( L_0 = 2.5 \) Gm, is given by
\begin{align}
\widetilde{\text{X}}_{\text{OMS}}(f) = 4 &\cdot \left| e^{\frac{4 \pi j f L_0}{c}} - 1 \right|^4 \cdot \left| e^{\frac{4 \pi j f L_0}{c}} + 1 \right|^2 \cdot \Big( \widetilde{\text{isi}}_{\text{OMS}}(f) + \widetilde{\text{rfi}}_{\text{OMS}}(f) \Big).\label{eq:tdi_OMS}
\end{align}
The analytical models are depicted in Fig.\ \ref{im:ScondaryNoiseSpecifications} as amplitude spectral densities, alongside numerical simulations generated using \texttt{lisainstrument} and \texttt{pyTDI} \cite{lisainstrument22, https://doi.org/10.5281/zenodo.6351736}.

\begin{figure}[]
	\centering
  \includegraphics[trim=0 0 0 0, clip,width=0.68\textwidth]{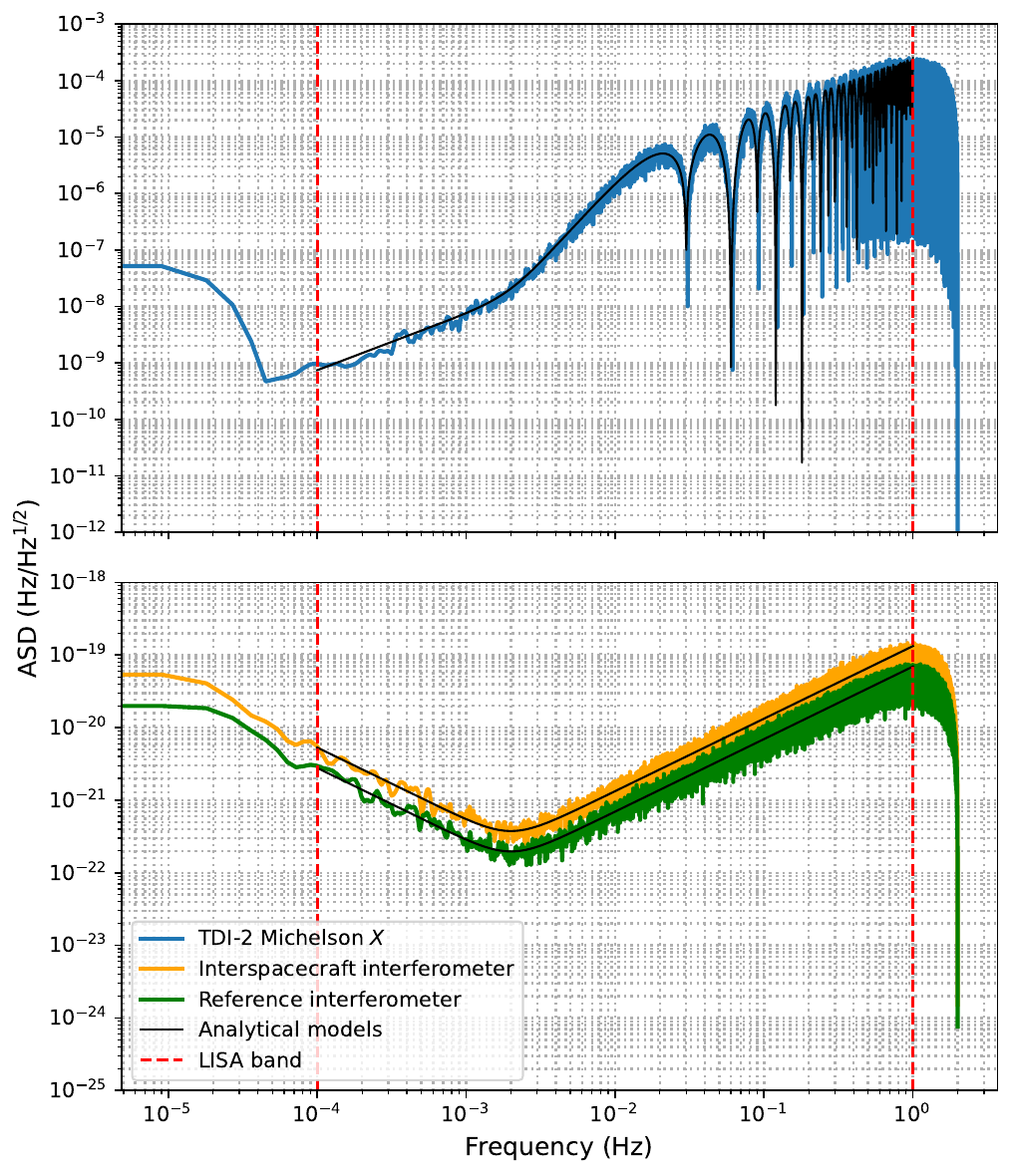}
    \caption{Amplitude spectral density of the optical metrology system noise within the inter-spacecraft interferometer and the reference interferometer, along with their propagation into TDI-2 Michelson $X$. Unlike the noise sources shown in Fig.\ \ref{im:NoiseSpecifications_plot}, OMS noise cannot be suppressed and must be endured during the parameter inference process. The plot shows a comparison between the analytical noise models and the simulated noise time series. The analytical models for the raw noise components are used to construct the TDI-$\infty$ noise covariance matrix $N$ in equation  \eqref{eq:DFTConversion}, while the analytical model for classical TDI is  used to define \( S_\textrm{noise}(f) \) in equation  \eqref{eq:inner_product}.}
	\label{im:ScondaryNoiseSpecifications}
\end{figure}

The $A$, $E$, and $T$ equivalents of equation \eqref{eq:tdi_OMS} are used to derive \( S_{\textrm{noise}}(f) \), which is required for the noise-weighted inner product in the classical TDI log-likelihood of equation \eqref{eq:log_likelihood_TDI2.0_with_gaps}. The complete derivation is omitted here.

An important complication in classical TDI arises from the assumptions made when modeling the contribution of OMS noise to $S_{\textrm{noise}}(f)$. The modeling assumes a linear time-invariant system, implying a stationary LISA configuration, often with equal arm \newline\newline lengths as assumed in equation  \eqref{eq:tdi_OMS}. However, in reality, the LISA arms are constantly changing due to the orbital dynamics. While necessary to define the TDI propagation behavior analytically, these assumptions do not fully reflect the non-stationary nature of the actual data.  The resulting mismatch is already apparent in Fig.\ \ref{im:ScondaryNoiseSpecifications}, where the zeros of the analytical model -- positioned at multiples of $c/4L_0$ -- exhibit a slight deviation when compared to the zeros of the numerical data generated with a dynamic orbit.

In contrast, TDI-$\infty$ avoids this issue by constructing its noise covariance matrix $N$ from the noise model of the raw measurements; that is, operating locally rather than at the constellation level.


\subsection{Reduced TDI-$\infty$ framework}
The key distinction between the toy model for TDI-$\infty$ and the all-in-one framework of equations \eqref{eq:TDIinfinity_fullsystem} and \eqref{eq:TMcondition_allinone} lies in the structure of the design matrix. In the toy model, the design matrix primarily contains entries of $-1$ and $0$ as well as Lagrange-interpolation weights to account for fractional delays of the delayed laser. In the all-in-one framework, we now also incorporate optical-bench displacement noise, clock noise, and modulation noise. These additional noise sources introduce a wider dynamic range in the $M$ matrix entries, as their contributions to the measurement model involve prefactors that depend on parameters such as the speed of light, the nominal laser frequency, and offset frequencies based on a predetermined frequency plan.  {This wider range of values makes it numerically challenging to compute accurate null spaces for the design matrix\footnote{ {The limitation is not fundamental to the all-in-one TDI-$\infty$ approach but stems from the current implementation of the design matrix $M$. In future work, we will apply a re-normalization process to reduce the dynamic range in $M$, ensuring the null-space calculation remains robust and accurate. An initial version of this has already proven successful.}}.}

Figure \ref{im:NullspaceComparison} illustrates a comparison of noise-suppression performance using the turnback null-space algorithm from \cite{pfeiffer_turnbackLU} versus SciPy's null-space algorithm within the all-in-one TDI-$\infty$ framework. All suppressible noise sources are active. The subfigures present results for (a) a static orbit and (b) a Keplerian LISA orbit. Significant differences regarding the noise residual in $\boldsymbol{o}$ can be observed depending on the null-space algorithm, even with identical input signals.
\begin{figure}[]
    \centering
    \begin{subfigure}{0.88\textwidth}
        \centering
        \includegraphics[width=\textwidth]{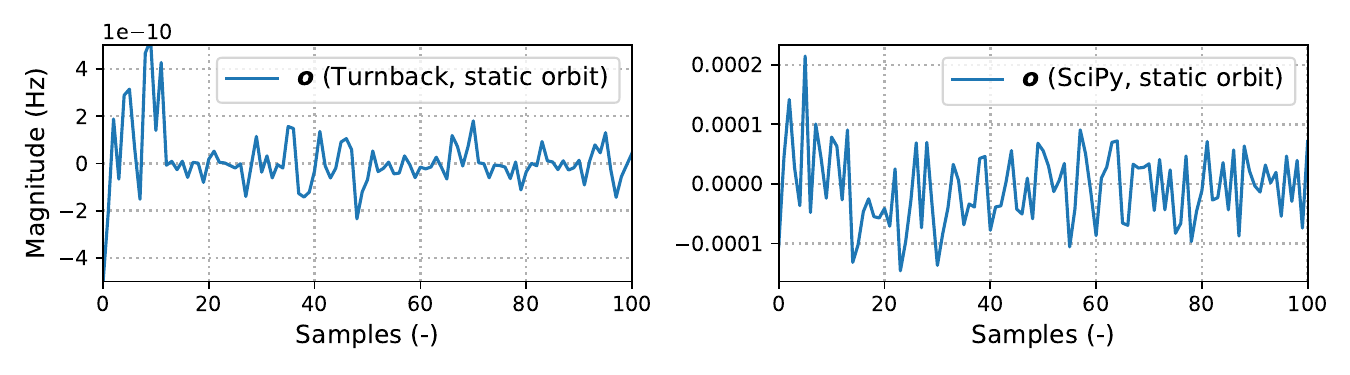}\vspace{-10pt}
        \caption{}
        \label{im:NullspaceComparison_StaticOrbit}
    \end{subfigure}
    \begin{subfigure}{0.88\textwidth}
        \centering
        \includegraphics[width=\textwidth]{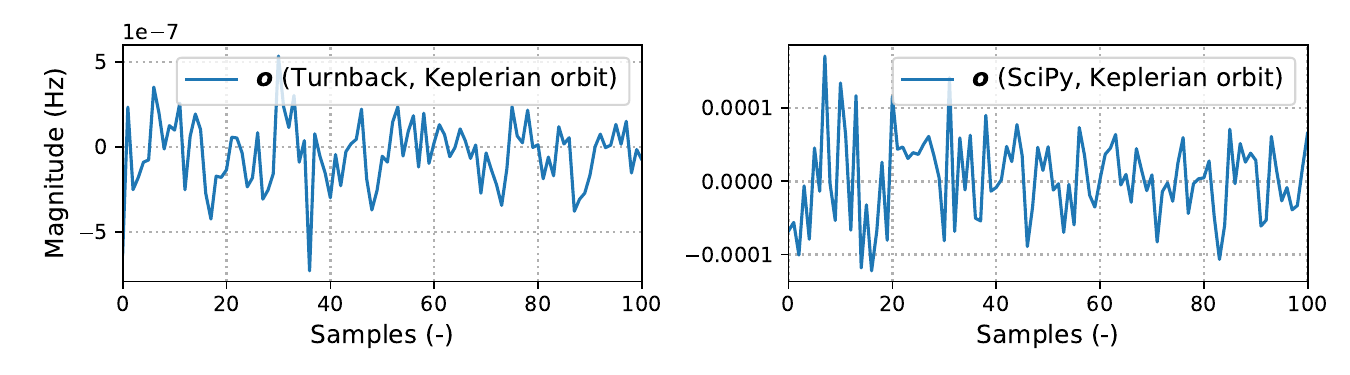}\vspace{-10pt}
        \caption{}
        \label{im:NullspaceComparison_KeplerianOrbit.pdf}
    \end{subfigure}
    \caption{Comparison of the noise suppression performance using the turnback null-space algorithm from \cite{pfeiffer_turnbackLU} versus SciPy's null-space algorithm within the all-in-one TDI-$\infty$ framework, with all suppressible noise sources active. The subfigures show results for (a) the static orbit and (b) the Keplerian LISA orbit. Both algorithms are tested with identical input data.}
    \label{im:NullspaceComparison}
\end{figure}

In this paper, we focus on applying the TDI-$\infty$ approach for Bayesian parameter inference, marking the first practical use of this method. While the impact of different null-space algorithms on parameter recovery performance is an important topic, it falls outside the scope of this initial study. Instead, our goal is to highlight the advantage of the TDI-$\infty$ methodology in performing accurate parameter inference in the presence of data gaps, compared to TDI-2, without confounding effects from null-space algorithm performance.  For this reason, we consider only six laser noise sources for both TDI-2 and TDI-$\infty$, deactivating optical-bench, clock, and modulation noise to reduce the dynamic range in $M$. The non-suppressible (secondary) noise sources are activated depending on the simulation scenario. To achieve laser-noise suppression with TDI-$\infty$, we use a reduced version of the all-in-one framework that focuses on the six inter-spacecraft interferometer measurements and the three reference interferometer measurements at the carrier frequency, treating all other measurements in the setup as zero.

\subsection{Comparison of posterior distributions for classical TDI and TDI-$\infty$}\label{subsec:Posteriors}
Our Bayesian analysis employs uniform priors with limits of $\pm 1\%$ of the values listed in Table \ref{tab:injected_params}. This study focuses on estimating three parameters: the coalescence time and the component masses. We leave the estimation of all the parameters of Table \ref{tab:injected_params} to future work, which will require more advanced MCMC samplers, such as \texttt{eryn} \cite{Karnesis_Nikolaos_2023}. For the current analysis, we use a basic implementation of the \texttt{emcee} sampler \cite{Foreman-Mackey_Daniel_2013}. This affine-invariant \emph{ensemble} sampler explores parameter space with a cloud of ``walkers,'' which provide a Dirac-delta representation of the target distribution.
We do not apply special techniques such as parallel tempering \cite{geyer1991markov}, simulated annealing \cite{10.5555/2073946.2073948}, or heterodyning \cite{PhysRevD.104.104054}.  The MCMC is run with 60 walkers over 500 steps, with the first 100 steps discarded as a burn-in phase.
We obtain the final samples for parameter estimation after thinning the chains by a factor of 15. Despite the limited number of iterations and walkers, the run achieves sufficient convergence for estimating the three parameters considered in this study. To assess convergence the criterion mentioned in the previous section is used. We plot posterior distributions obtained using TDI-2 and TDI-$\infty$ likelihoods for three scenarios:
\begin{enumerate}
    \item without noise, without gaps (Fig.\ \ref{im:MCMC_nonoise_nogaps_1hour});
    \item with noise, without gaps (Fig.\ \ref{im:MCMC_noise_nogaps_1hour});
    \item with noise and gaps during merger (Fig.\ \ref{im:MCMC_noise_gaps_1hour}) or during inspiral (Fig.\ \ref{im:MCMC_noise_gaps_1hour_inspiral}).
\end{enumerate}
In the first scenario with no noise and no data gaps, shown in Fig.\ \ref{im:MCMC_nonoise_nogaps_1hour}, both TDI-2 and TDI-$\infty$ recover the parameters of interest effectively. Note that axis limits are kept the same across all figures in this section, and contours are shown for 68\% and 95\% iso-posterior credible regions. In the absence of noise and gaps, the posterior distributions for classical TDI and TDI-$\infty$ are almost identical. Any discrepancies arise from factors such as the inherent randomness of the MCMC algorithm, inaccuracies in the discrete Fourier transform and the chunking process. The impact of chunking on the TDI-$\infty$ posteriors will be investigated more extensively in future studies.
\begin{figure}[b!]
    \centering
    \begin{subfigure}[b]{0.48\textwidth}
        \centering
        \includegraphics[trim=0 0 0 0, clip, width=\textwidth]{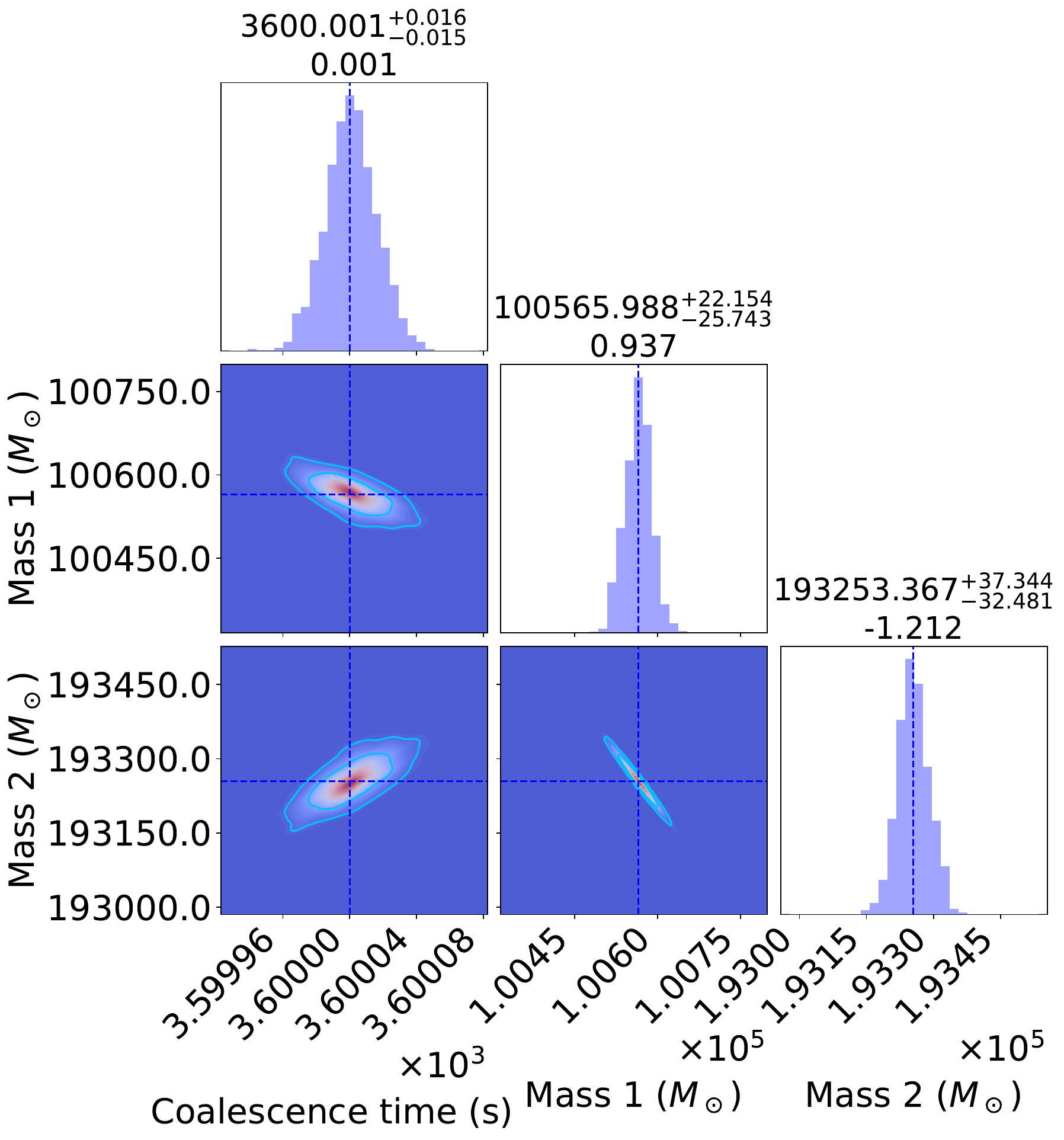}
        \caption{Classical TDI}
        \label{im:MCMC_TDI20_nonoise_nogaps_1hour}
    \end{subfigure}
    \hfill
    \begin{subfigure}[b]{0.48\textwidth}
        \centering
        \includegraphics[trim=0 0 0 0, clip, width=\textwidth]{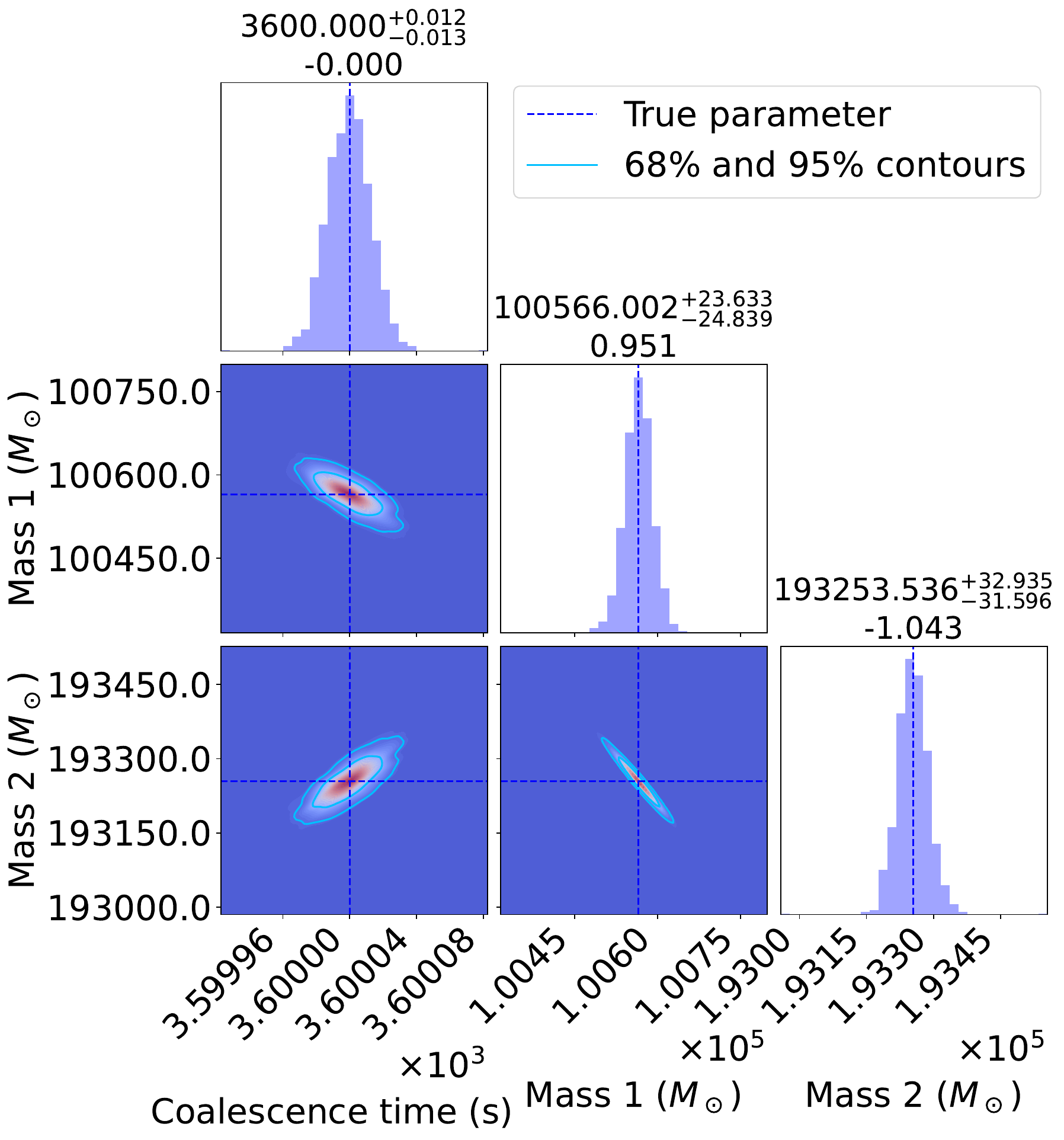}
        \caption{TDI-$\infty$}
        \label{im:MCMC_TDIInf_nonoise_nogaps_1hour}
    \end{subfigure}
    
        \caption{Posterior distributions of the coalescence time and mass parameters for classical TDI (left) and TDI-$\infty$ (right; segment length $l=500$ and row overlap factor $q=0.1$ applied for this and all subsequent cases), without secondary noises and gaps (scenario i). This idealized scenario provides a baseline for evaluating parameter recovery in the absence of disturbances.
         {The numbers above each histogram show the mean value and their uncertainties (upper row), and the deviation of the mean value from the true parameter (lower row) in the respective units. The uncertainties shown are the 1-sigma credible intervals derived from the 16th and 84th percentiles of the posterior distribution, representing the range around the median that contains 68\% of the probability mass.}}\label{im:MCMC_nonoise_nogaps_1hour}
        \end{figure}
        \begin{figure}
    \centering
    \begin{subfigure}[b]{0.48\textwidth}
        \centering
        \includegraphics[trim=0 0 0 0, clip, width=\textwidth]{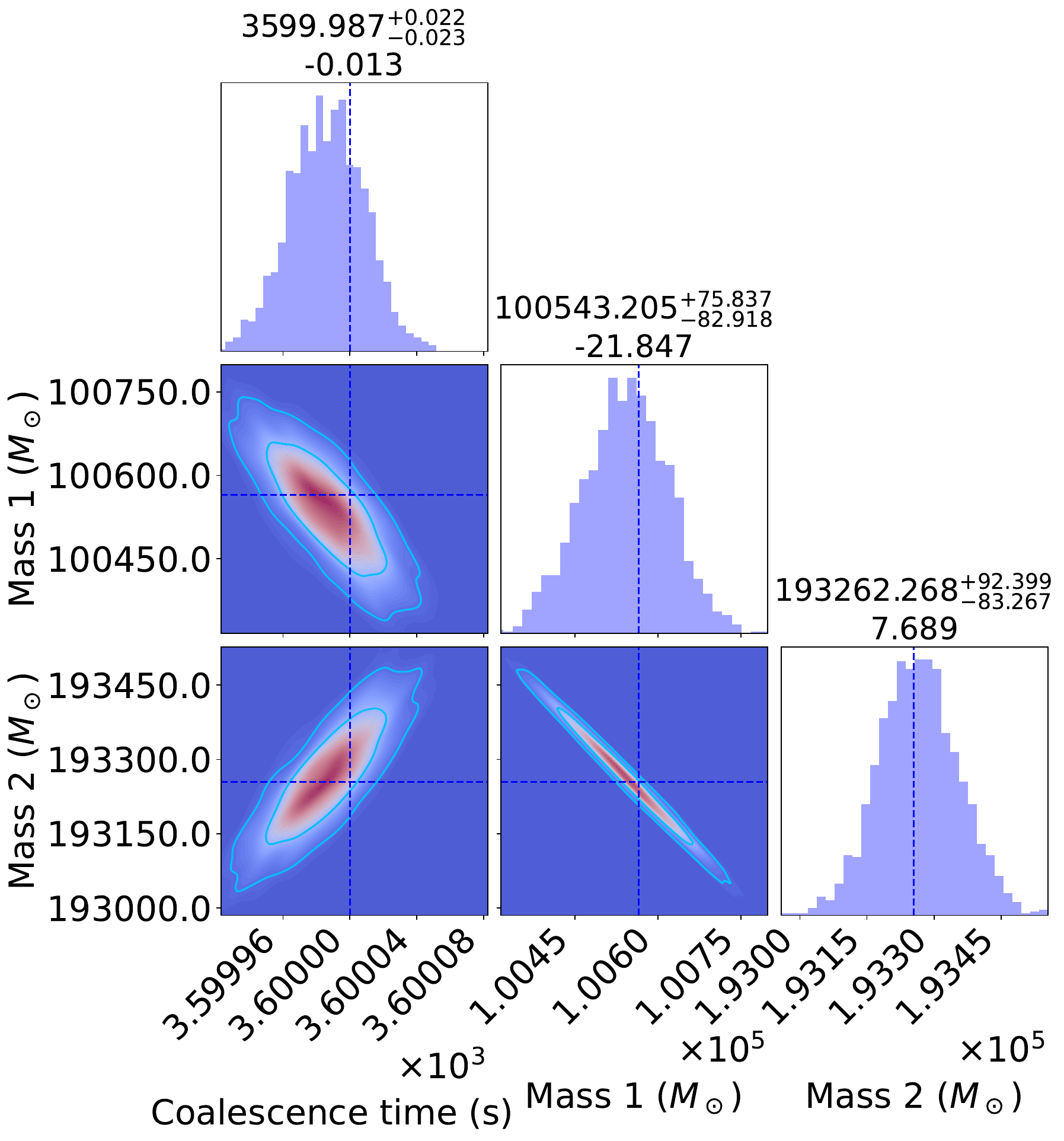}
        \caption{Classical TDI (frequency domain)}
\label{im:MCMC_TDI20_noise_nogaps_1hour}
    \end{subfigure}
    \hfill
        \begin{subfigure}[b]{0.48\textwidth}
        \centering
        \includegraphics[trim=0 0 0 0, clip, width=\textwidth]{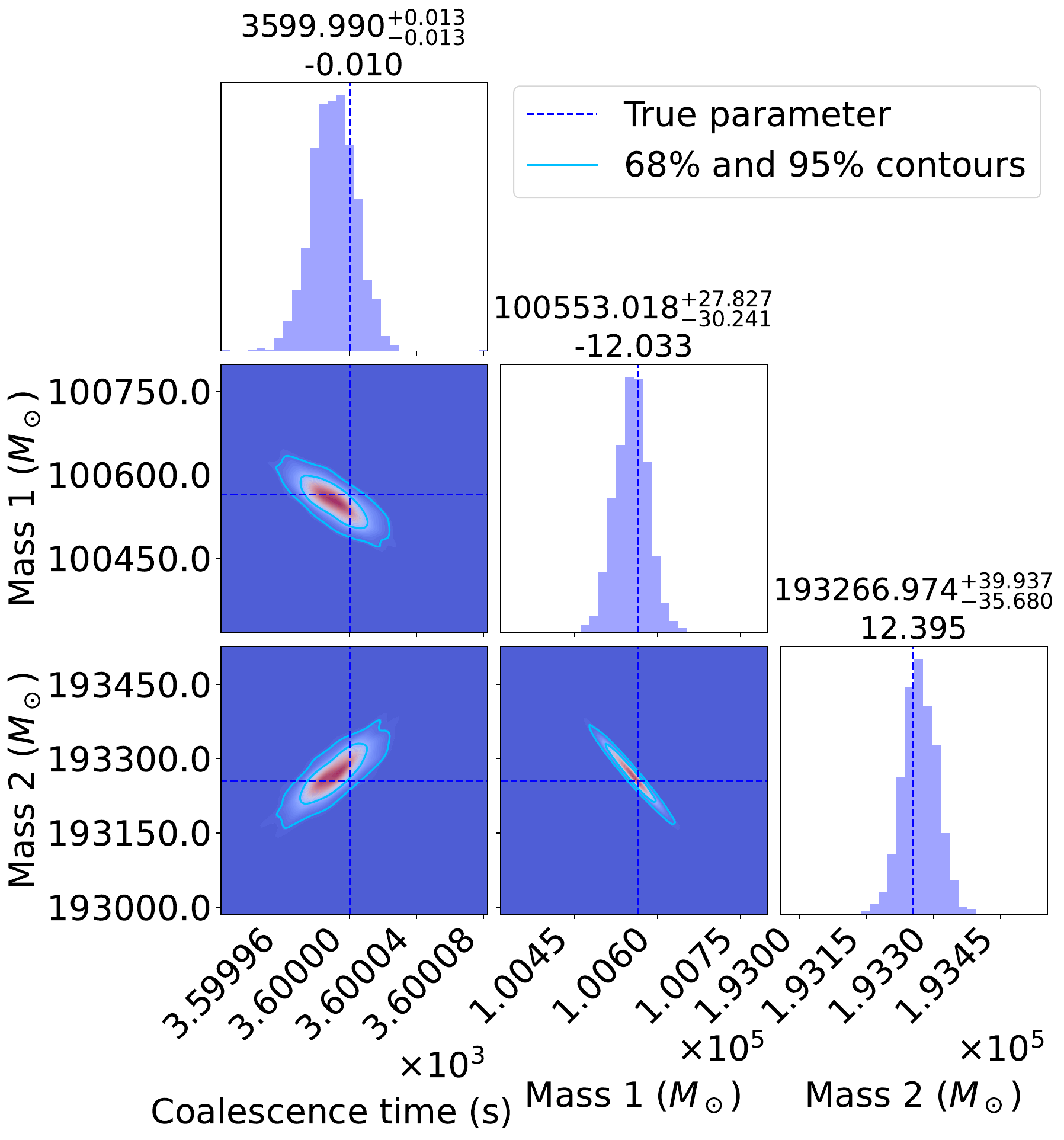}
        \caption{Classical TDI (time domain)}
\label{im:MCMC_TDI20Time_noise_nogaps_1hour}
    \end{subfigure}
    \hfill \vspace{10pt}
    \begin{subfigure}[b]{0.48\textwidth}
        \centering
        \includegraphics[trim=0 0 0 0, clip, width=\textwidth]{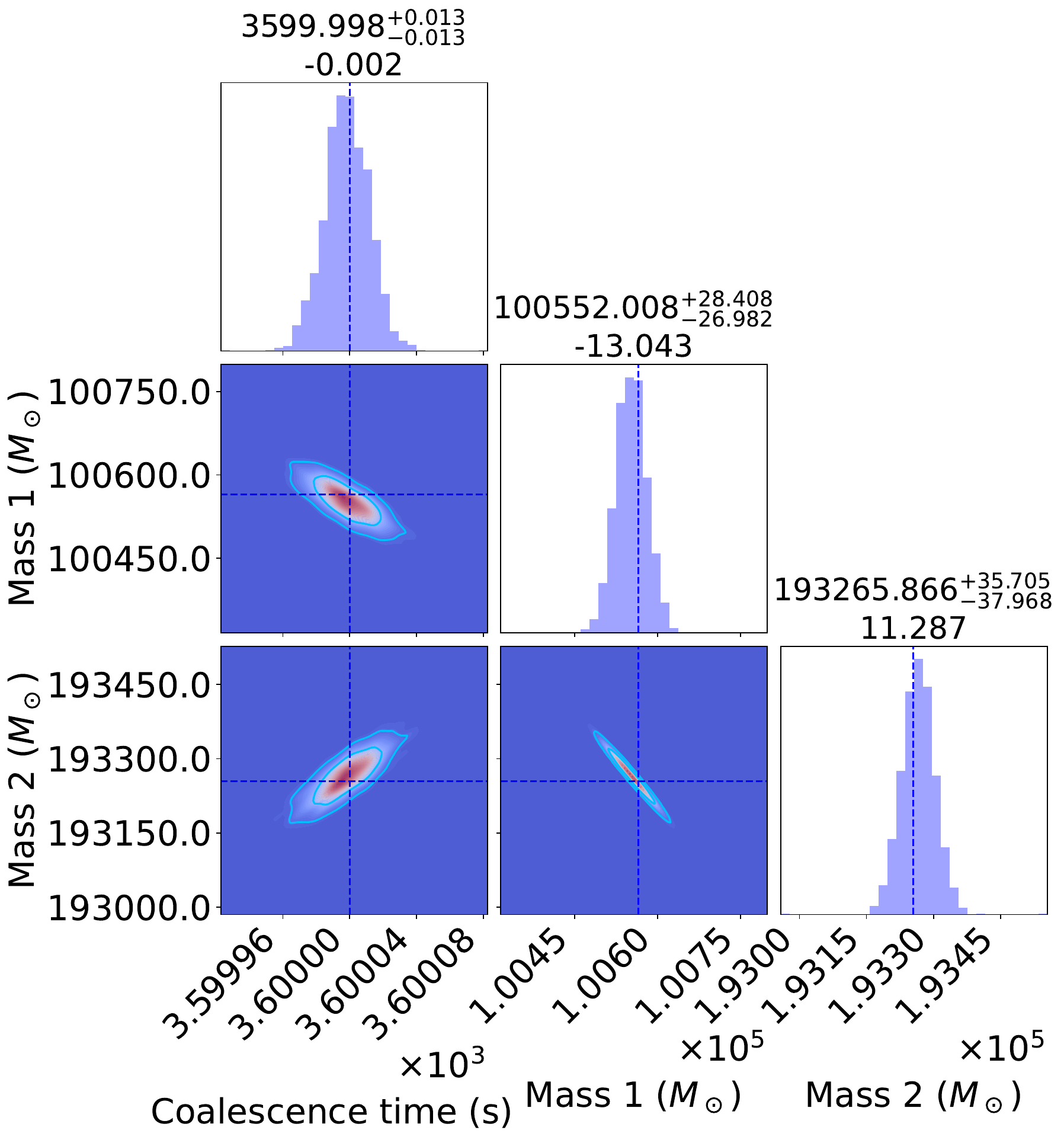}
        \caption{TDI-$\infty$}
        \label{im:MCMC_TDIInf_noise_nogaps_1hour}
    \end{subfigure}
    \caption{ {Posterior distributions with secondary noises and no gaps (scenario ii). The same noise realization is applied to classical TDI and TDI-$\infty$. While  biases induced by noise are present in both methods, the posterior distributions for classical TDI in the frequency domain are more dispersed compared to TDI-$\infty$. This discrepancy arises because  the first few frequency bins are dominated by noise, skewing the likelihood calculation, and must be discarded, which results in the loss of low-frequency signal content, particularly for the 1-hour time series considered here. In contrast, the time-domain approach for TDI-2, shown in panel (b), avoids this issue and closely matches the results from TDI-$\infty$.}}\label{im:MCMC_noise_nogaps_1hour}
\end{figure}

 {In the second scenario, shown in Fig.\ \ref{im:MCMC_noise_nogaps_1hour}, noise is introduced while no data gaps are present. Under these conditions, both classical TDI and TDI-$\infty$ successfully recover the source parameters. However, while the noise-free scenario (Fig.\ \ref{im:MCMC_nonoise_nogaps_1hour}) shows comparable posterior performance between classical TDI applied in the frequency domain and TDI-$\infty$, this equivalence breaks down when secondary noises are present. The posteriors obtained with classical TDI applied in the frequency domain are more dispersed than those from TDI-$\infty$, see Fig.\ \ref{im:MCMC_noise_nogaps_1hour} (a) and (c), despite using identical noise realizations for both methods to ensure a fair comparison. \nh{When the Fourier-transformed TDI-2 time series, contaminated by secondary noise, is used for parameter estimation, the lowest frequency bins are  dominated by noise. In theory, incorporating a noise model with an equally large power spectral density in the inner product computation should suppress the influence of low-frequency secondary noise. However, in practice, spectral leakage and bias in power spectral density estimation cause these low-frequency contributions to persist, distorting the likelihood function. When the power spectral density is estimated from finite-length data, the limited number of samples introduces bias, as the estimate cannot perfectly capture the true noise characteristics. This bias is particularly severe at low frequencies, where fewer cycles fit within the observation window, resulting in poor spectral resolution. Consequently, during MCMC analysis, the likelihood calculation can become skewed as noise power at low frequencies overwhelms the signal power. This leads to inaccurate parameter estimation and convergence issues.}
To mitigate these problems, discarding  the first few noise-dominated frequency bins becomes necessary\footnote{\nh{Of the 7195 frequency bins in the 1-hour signal sampled at 4 Hz, we discard the first 20.}}. While this approach supports chain convergence, it comes at the cost of losing some low-frequency signal information, which degrades the posterior distribution obtained using classical TDI in the frequency domain. The relationship between the increase in posterior spread and the removal of low-frequency samples can be confirmed by performing the same operation on the noise-free TDI-2 dataset shown in Fig.\ \ref{im:MCMC_nonoise_nogaps_1hour} (a). When applying cuts analogous to those in Fig.\ \ref{im:MCMC_noise_nogaps_1hour} (a), an equivalent degradation in performance is observed. This plot is not shown here.

The discrepancy in the frequency-domain analysis can be avoided if the time series is sufficiently long to ensure that discarded bins lie outside the signal band. However, this is not always possible when data gaps limit the measurement window, as in the scenario considered here.  In such cases, time-domain MCMC approaches are preferable since they do not require discarding samples, whether using TDI-$\infty$ or TDI-2. The result of the TDI-2-based MCMC analysis performed in the time domain is shown in Fig.\ \ref{im:MCMC_noise_nogaps_1hour} (b). It closely matches the one obtained with TDI-$\infty$. In this time-domain analysis, we use the same log-likelihood definition as for TDI-$\infty$, see equation \eqref{eq:TDIinfinity_loglikelihood}. Here, \( \Delta \boldsymbol{o} \) is replaced with the time-domain TDI-2 samples of \( A \), \( E \), and \( T \), and the null-space matrix is replaced by the identity matrix since the suppressible noises have already been removed in \( A \), \( E \), and \( T \). The time-domain noise covariance matrix incorporates the OMS noise contributions from the inter-spacecraft and reference interferometers. These noise contributions are propagated to  {the power spectral densities for the \( A \), \( E \), and \( T \) observables in analogy with} equation  \eqref{eq:tdi_OMS}. From this point onward, we will present classical TDI results using both the frequency-domain and time-domain approaches. }
\begin{figure}[]
    \centering
    \begin{subfigure}[b]{0.48\textwidth}
        \centering
        \includegraphics[trim=0 0 0 0, clip, width=\textwidth]{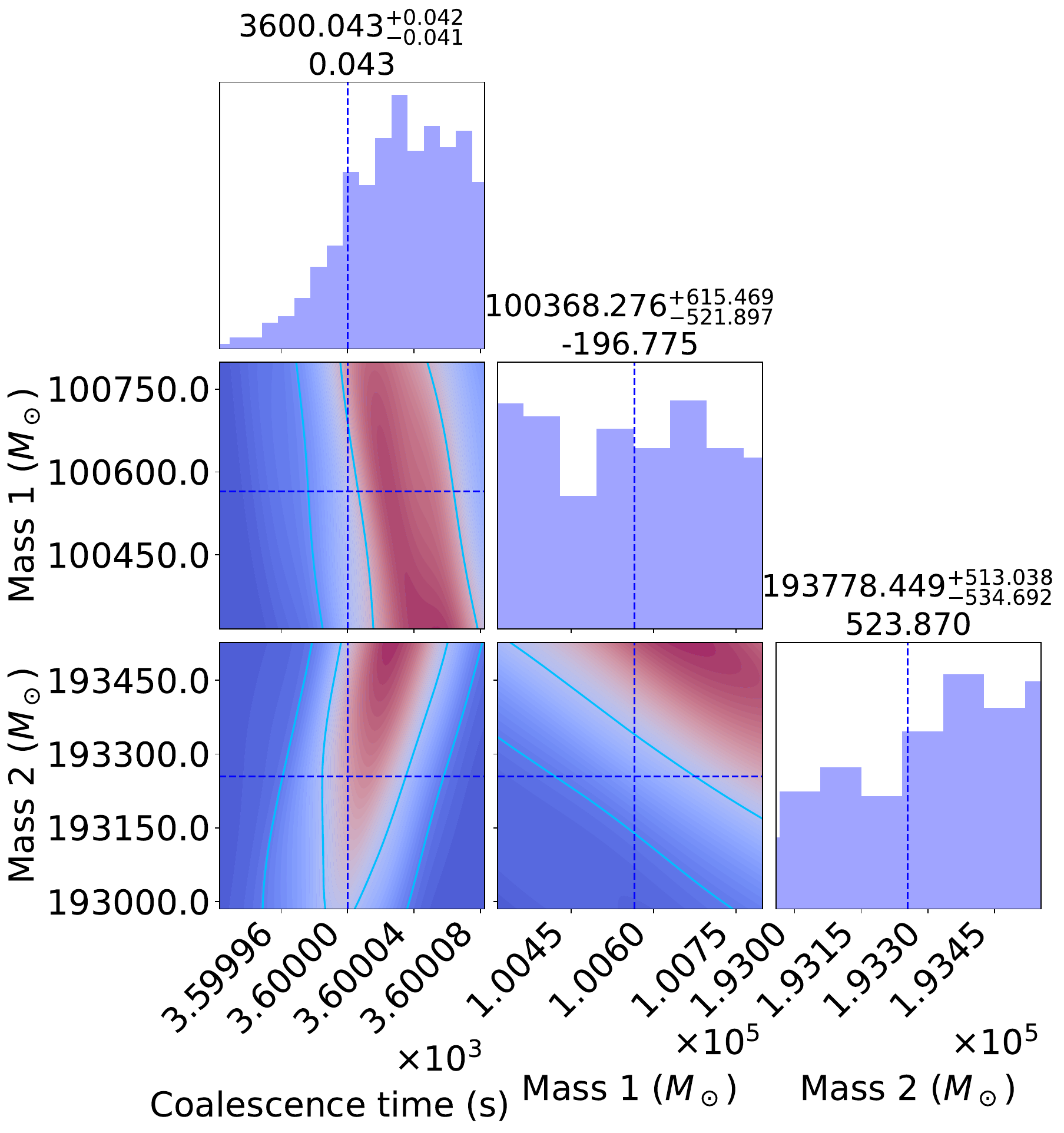}
        \caption{Classical TDI (frequency domain)}
        \label{im:MCMC_TDI20_noise_gaps_1hour}
    \end{subfigure}
    \hfill    
    \begin{subfigure}[b]{0.48\textwidth}
        \centering
        \includegraphics[trim=0 0 0 0, clip, width=\textwidth]{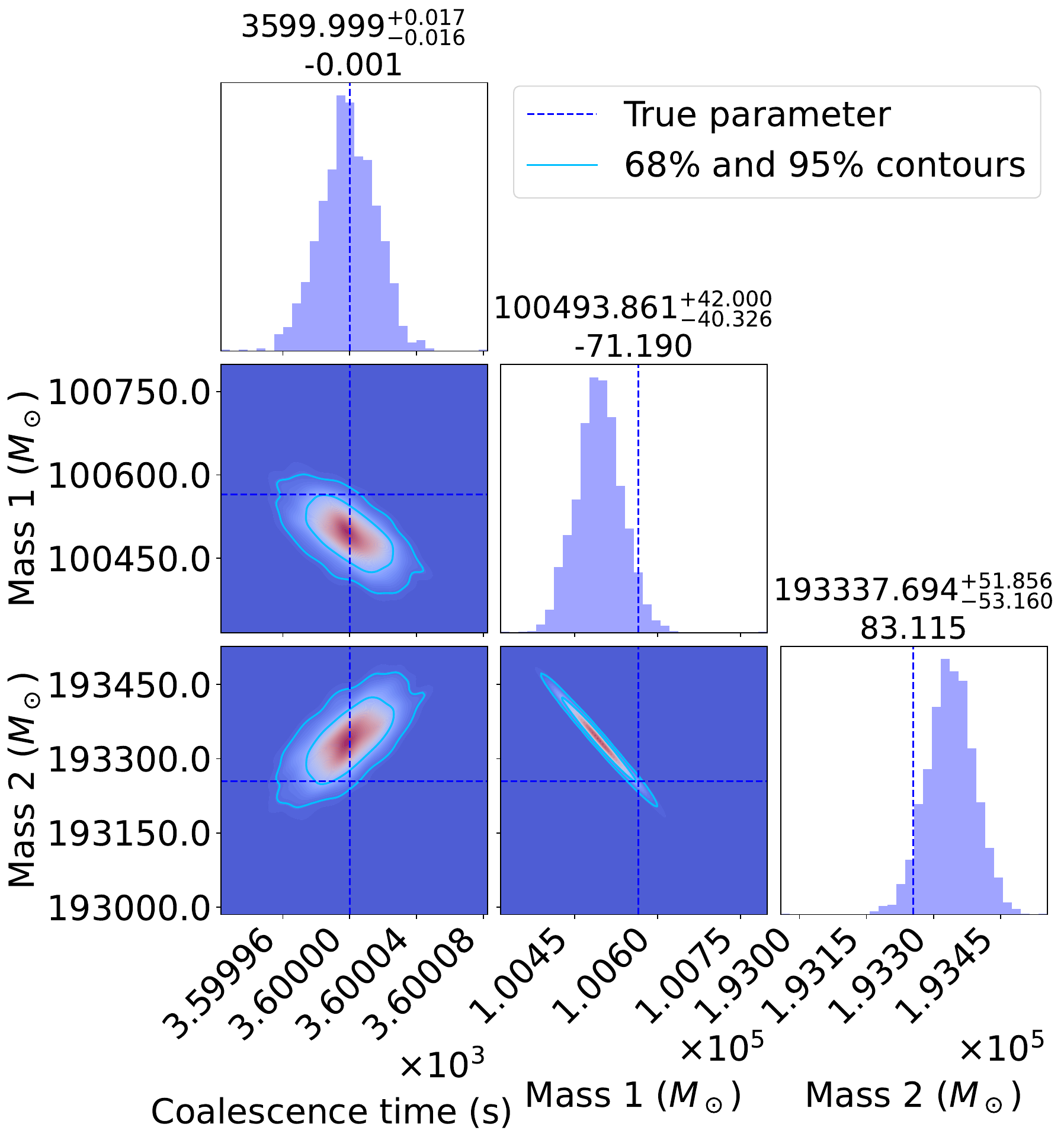}
        \caption{Classical TDI (time domain)}
        \label{im:MCMC_TDI20Time_noise_gaps_1hour}
    \end{subfigure}
    \hfill\vspace{10pt}
    \begin{subfigure}[b]{0.48\textwidth}
        \centering
        \includegraphics[trim=0 0 0 0, clip, width=\textwidth]{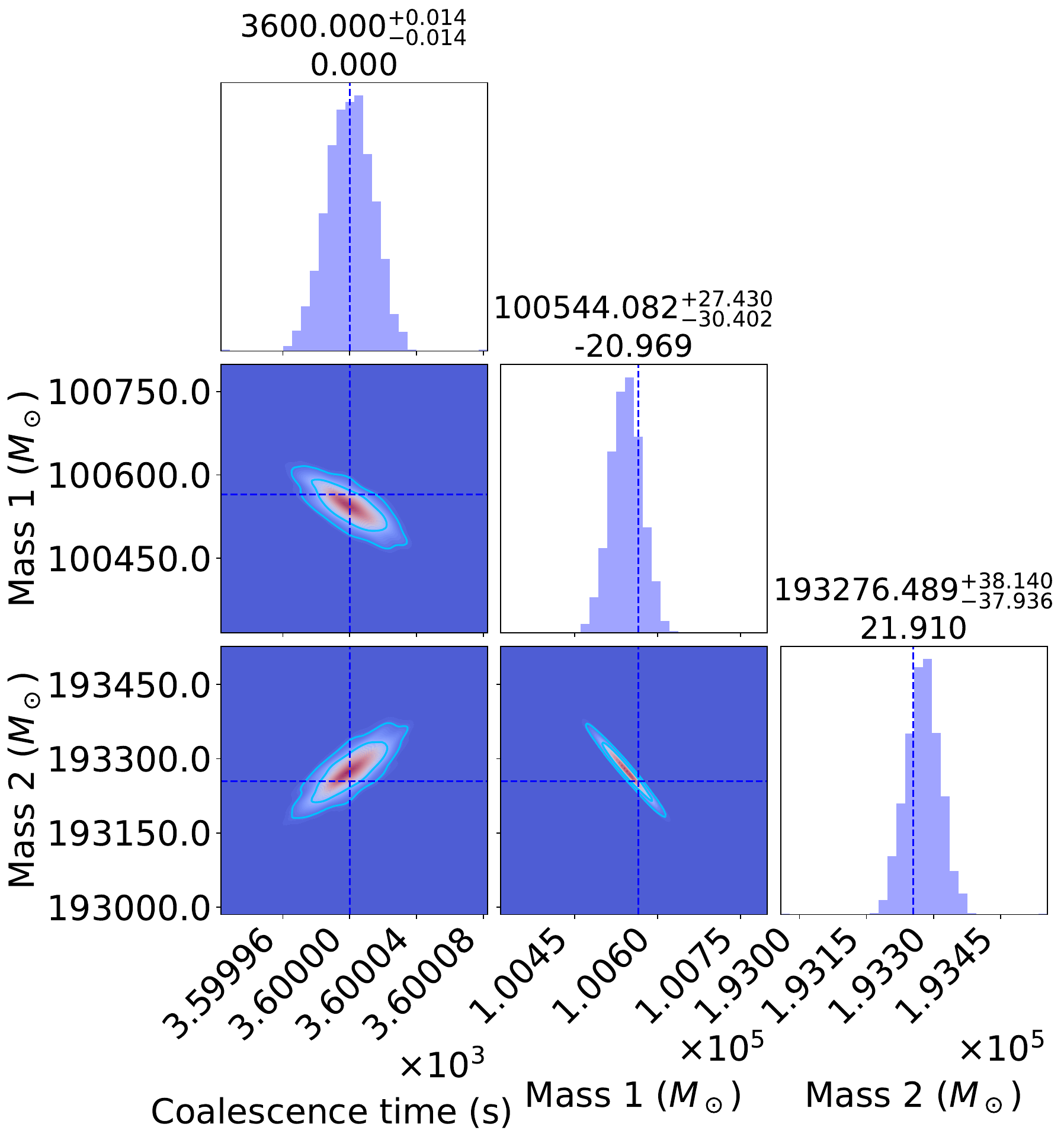}
        \caption{TDI-$\infty$}
        \label{im:MCMC_TDIInf_noise_gaps_1hour}
    \end{subfigure}
    \caption{ {Posterior distributions with secondary noise and gaps close to the merger, as illustrated in Figs.\ \ref{im:TDI20_GapsIllustration_TimeDomain} and \ref{im:TDIInf_GapsIllustration_TimeDomain} (scenario iii-a). The same noise realization is applied as in Fig.\ \ref{im:MCMC_noise_nogaps_1hour}. The posteriors obtained with TDI-$\infty$ remain largely unaffected by the gaps, showing only a slight increase in estimation biases. In contrast, classical TDI exhibits significant deterioration in parameter recovery, with posterior distributions showing substantial spreading and pronounced biases. This degradation is especially pronounced in the frequency-domain approach. These results highlight the advantage of TDI-$\infty$ in handling incomplete data compared to classical TDI for the gap scenario under consideration.}}\label{im:MCMC_noise_gaps_1hour}\end{figure}
\begin{figure}
    \centering
    \begin{subfigure}[b]{0.48\textwidth}
        \centering
        \includegraphics[trim=0 0 0 0, clip, width=\textwidth]{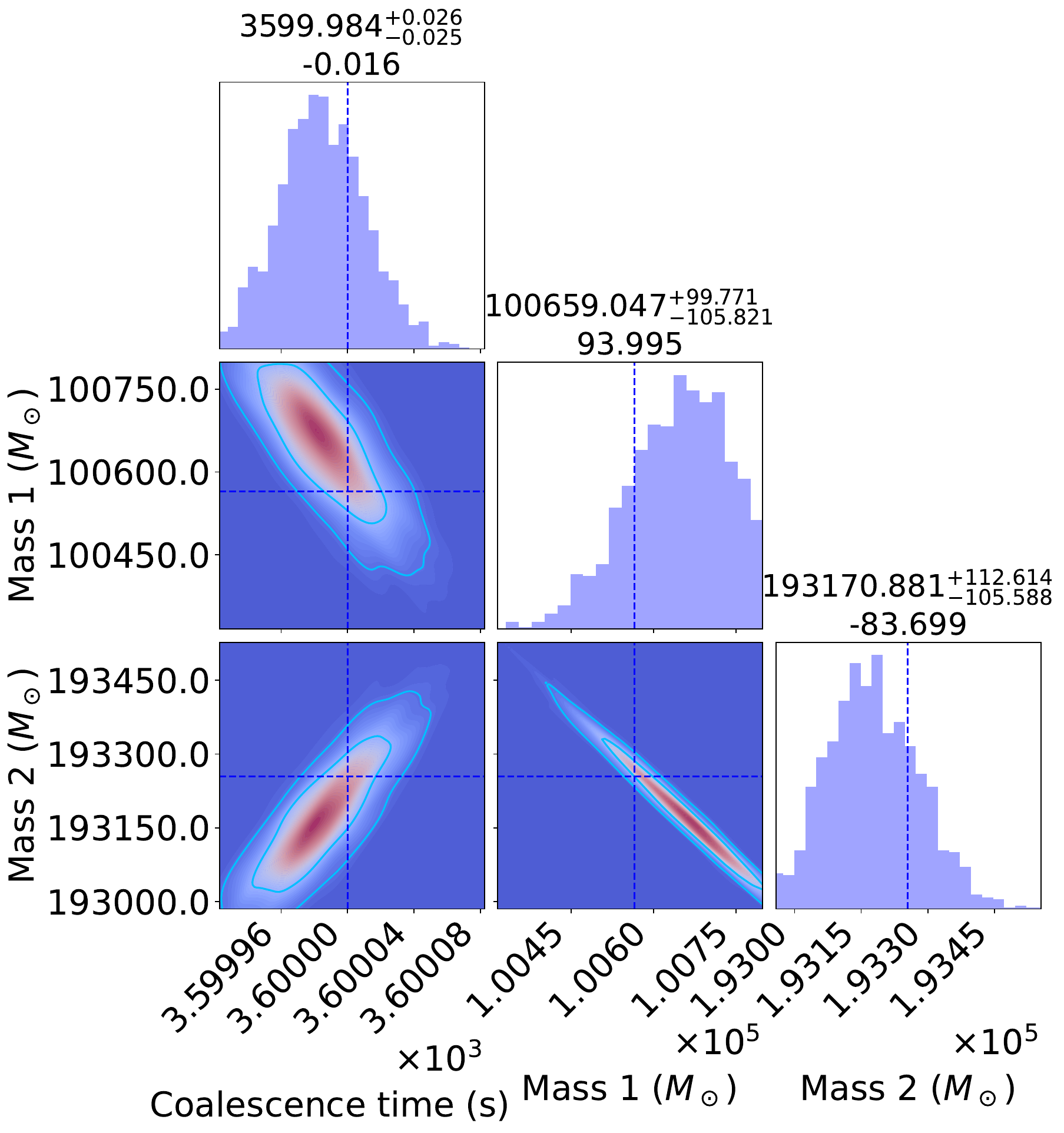}
        \caption{Classical TDI}
        \label{im:MCMC_TDI20_noise_gaps_1hour_inspiral}
    \end{subfigure}
    \hfill
        \begin{subfigure}[b]{0.48\textwidth}
        \centering
        \includegraphics[trim=0 0 0 0, clip, width=\textwidth]{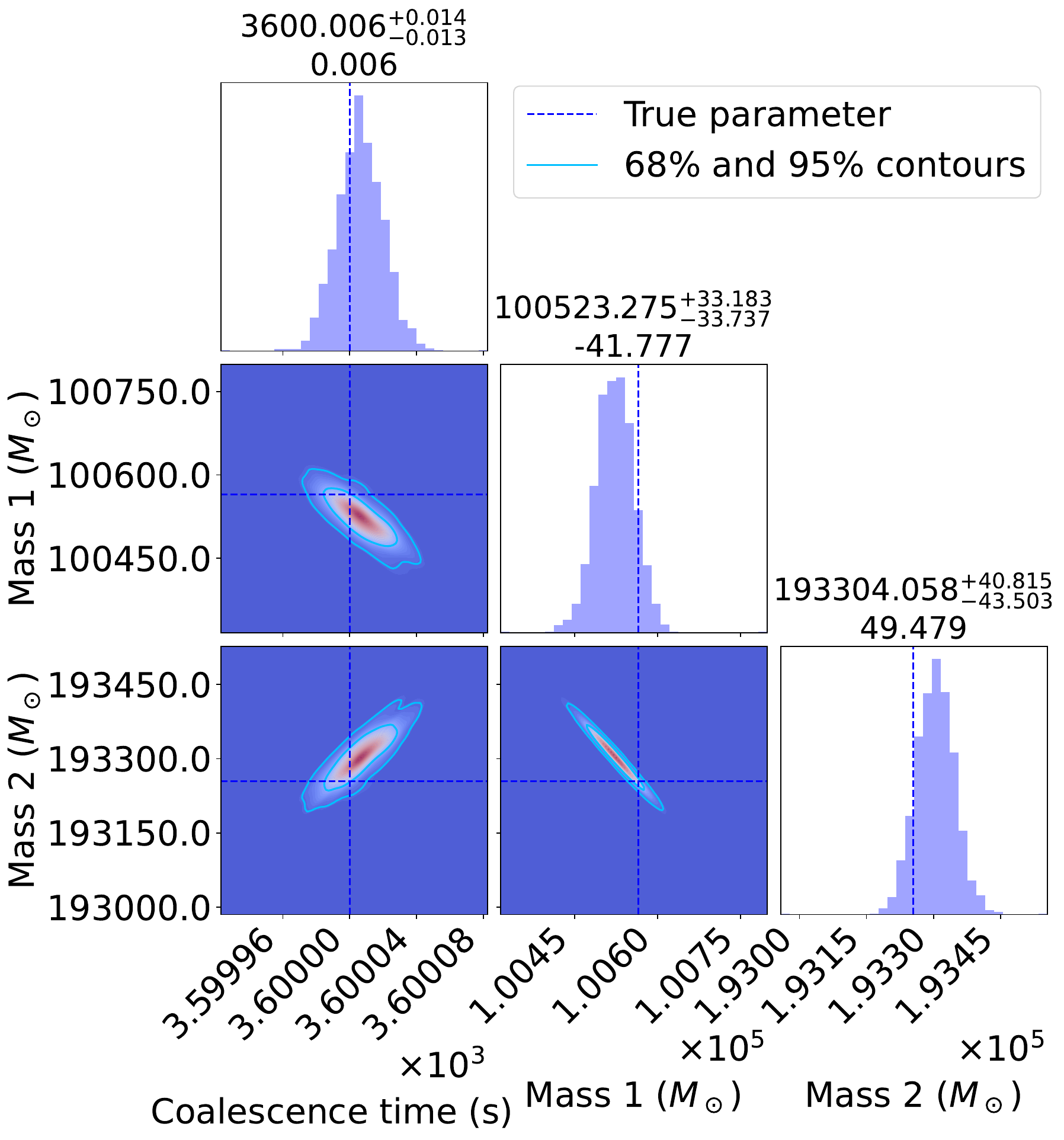}
        \caption{Classical TDI}
        \label{im:MCMC_TDI20Time_noise_gaps_1hour_inspiral}
    \end{subfigure}
    \hfill\vspace{10pt}
    \begin{subfigure}[b]{0.48\textwidth}
        \centering
        \includegraphics[trim=0 0 0 0, clip, width=\textwidth]{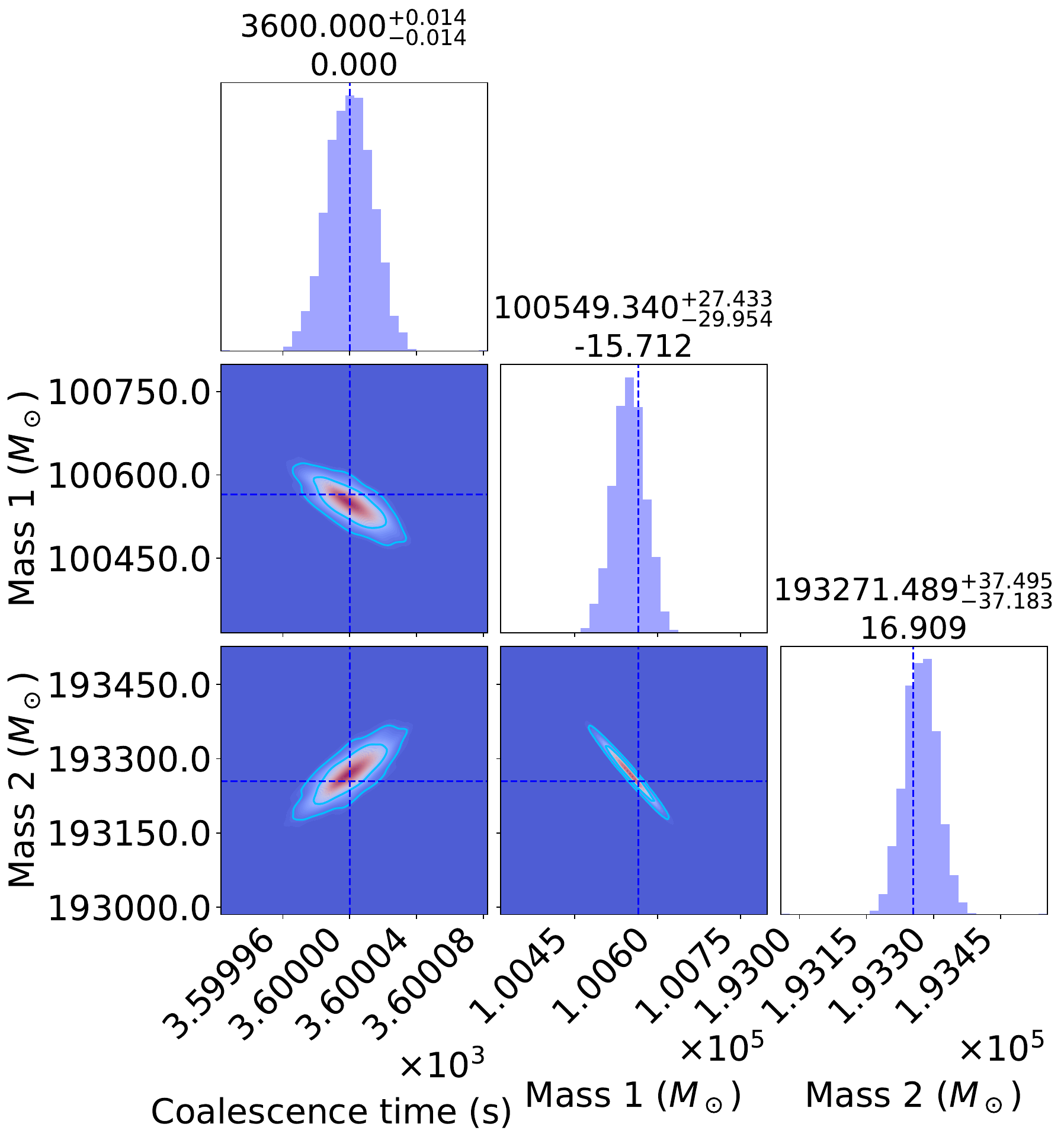}
        \caption{TDI-$\infty$}
        \label{im:MCMC_TDIInf_noise_gaps_1hour_inspiral}
    \end{subfigure}
    \caption{ {Posterior distributions with secondary noise and gaps during the inspiral phase (scenario iii-b). The same noise realization is applied as in Figs.\ \ref{im:MCMC_noise_nogaps_1hour} and \ref{im:MCMC_noise_gaps_1hour}. The posterior distributions improve compared to Fig.\ \ref{im:MCMC_noise_gaps_1hour} because the critical information from the merger phase remains intact. Since the merger signal typically has the highest signal-to-noise ratio, gaps during the inspiral phase have a less pronounced impact on parameter recovery than those occurring during the merger and ringdown phases.}}\label{im:MCMC_noise_gaps_1hour_inspiral}
\end{figure}

 {Fig.\ \ref{im:MCMC_noise_gaps_1hour} presents the third scenario, where both secondary noises and data gaps are included. In this case, TDI-$\infty$ demonstrates notable resilience to the presence of gaps. The posterior distributions for TDI-$\infty$ remain relatively stable, with only a slight increase in estimation biases compared to the results from the gap-free scenario. This suggests that TDI-$\infty$ can effectively handle incomplete data without substantial degradation in performance. In contrast, classical TDI shows a marked sensitivity to data gaps. We present results for classical TDI using both the frequency-domain approach, where initial frequency bins are discarded due to noise contamination, and the time-domain approach, where no data is discarded. The posterior distributions obtained from classical TDI exhibit considerable spreading and significant biases, particularly when analyzed in the frequency domain. Even with the time-domain approach, classical TDI struggles to maintain accuracy due to its vulnerability to gaps, as the algorithm's inherent delays invalidate multiple TDI-2 output samples, as discussed throughout the article. These results highlight the advantage of TDI-$\infty$ when dealing with incomplete data compared to classical TDI, at least in the scenario considered here. The ability of TDI-$\infty$ to produce reliable parameter estimates despite data gaps underscores its robustness in scenarios where continuous data acquisition cannot be guaranteed.}

 {Results for the third scenario with data gaps during the inspiral phase are shown in Fig.\ \ref{im:MCMC_noise_gaps_1hour_inspiral}. Compared to Fig.\ \ref{im:MCMC_noise_gaps_1hour}, all posterior distributions show improvement because the critical information from the merger phase remains unaffected. Since the merger signal generally has the highest signal-to-noise ratio, gaps during the inspiral phase have a less pronounced impact on the parameter recovery process than gaps occurring during the merger and ringdown phases, as expected. While it will probably be feasible in the future to achieve performances comparable to TDI-$\infty$ using classical TDI in the time domain combined with sophisticated gap-handling techniques, TDI-$\infty$ inherently manages data gaps without the need for such additional methods. This built-in capability simplifies the data analysis process, offering a considerable benefit over classical TDI approaches that require extra processing.
}

\section{Conclusion}\label{sec:Conclusion}
LISA will open a new window into the millihertz frequency band of gravitational waves, enabling observations of sources like MBHBs, EMRIs, and compact binaries.
Two key challenges for LISA data analysis are the cancellation of laser frequency noise and the handling of data gaps due to instrument downtime or communication interruptions. Classical TDI has been developed to address the first challenge, suppressing the otherwise overwhelming laser noise. While classical TDI is very successful in this task, its sensitivity to data gaps is problematic, especially in low-latency applications that require near real-time responses with limited data.

This study extends the TDI-$\infty$ framework, originally developed for a simplified toy model, to the full LISA mission scenario. We explore the application of TDI-$\infty$ to the processing of L0-L1 LISA data, both for the reduction of laser noise and in an all-in-one framework that targets all other suppressible noise sources (optical-bench noise, clock noise, and modulation noise). The TDI-$\infty$ framework operates by marginalizing numerically over these noises, avoiding the need for explicit algebraic combinations like those used in classical TDI. The formulation leads to significant improvements in handling measurement gaps, as TDI-$\infty$ accounts for them naturally by analyzing actual data only; by contrast, TDI-2 requires additional gap-filling or interpolation techniques.

We demonstrate the integration of TDI-$\infty$ into a Bayesian parameter-inference pipeline, setting up the likelihood function in the time domain (rather than in the frequency domain as typical for classical TDI). This shift in domain simplifies the treatment of data gaps; invalid measurements are easily removed from the design matrix before computing the null space, eliminating them from the likelihood calculation.
We performed a number of simulations to demonstrate that the accuracy of astrophysical parameter recovery using TDI-2 degrades significantly in the presence of gaps, while it remains robust with TDI-$\infty$, providing accurate posterior distributions even with incomplete data.

 {Our analysis indicates that the performance of classical TDI varies between frequency-domain and time-domain implementations, especially in the presence of data gaps and limited measurement data. In the frequency-domain approach, secondary noise can dominate  the first few frequency bins of the Fourier-transformed TDI-2 time series, leading to skewed likelihoods. This dominance necessitates discarding these noise-affected bins to achieve convergence in MCMC sampling, but doing so can result in the loss of low-frequency signal information and degrades parameter estimation. This issue occurs when data gaps limit the measurement window, preventing the discarded bins from lying outside the signal band. In contrast, time-domain implementations of classical TDI-based MCMC can handle data gaps more effectively, as no valid data needs to be cut. However, due to the inherent delays in classical TDI algorithms, data gaps result in a higher number of invalid samples, leading to degraded performance compared to TDI-$\infty$. Therefore, both time-domain and frequency-domain implementations of classical TDI necessitate additional gap-handling techniques to achieve performance levels comparable to TDI-$\infty$. TDI-$\infty$ inherently manages data gaps without the need for supplementary methods, streamlining the data analysis process and possibly reducing computational complexity. This built-in capability emphasizes the potential of TDI-$\infty$ as a compelling alternative to classical TDI, particularly for applications necessitating low-latency analysis. The framework's ability to handle measurement discontinuities presents a major advantage.}

This study serves as a foundation for future work to further refine TDI-$\infty$. This includes exploring optimal null-space algorithms for the all-in-one TDI-$\infty$ approach, enhancing its computational efficiency, and testing its performance with larger datasets and complex astrophysical scenarios of overlapping sources. Additionally, we intend to employ a more sophisticated MCMC sampler configuration and to utilize broader priors to fully assess the capabilities of the extended TDI-$\infty$ framework. The insights gained from this work pave the way for more resilient data-analysis techniques capable of extracting the most from LISA's unprecedented access to the low-frequency gravitational-wave universe.

\section*{Acknowledgments}
The authors thank the organizers of the 2024 LISA Sprint at Caltech, particularly Katerina Chatziioannou, for offering the opportunity to kick off this research. The authors sincerely appreciate the insightful discussions with Luigi Ferraioli, Martin Staab, and Olaf Hartwig. The authors thank the LISA Simulation Working Group and the LISA Simulation Expert Group for the lively discussions on all simulation-related activities. The authors further extend their gratitude to Kai Pfeiffer from Nanyang Technological University, Singapore, for providing the implementation of the turnback algorithm and for his continued dedication to addressing our questions. Niklas Houba acknowledges the support of GW-Learn, a project funded by a Sinergia grant from the Swiss National Science Foundation. Niklas is especially grateful to Domenico Giardini for funding his research visit to Caltech in Spring 2024, which allowed him to focus on this work without distraction. Jean-Baptiste Bayle acknowledges the support of the UK Space Agency, UKSA, through grant ST/Y004906/1.  Copyright 2024. All rights reserved.

\section*{Data availability statement}
The data that support the findings of this study are available upon reasonable request from the authors.

\section*{Appendix}\label{sec:appendix}
{Figure \ref{im:PatternsNullSpaceTurnback} illustrates the repeating patterns observed in the null-space matrix \( T \), plotted against the individual interferometer measurements (the components of $\boldsymbol{y}$ of equation  \eqref{eq:TDIinfinity_fullsystem}). To construct this matrix, we employ the turnback algorithm \cite{pfeiffer_turnbackLU}, designed to produce a banded sparse null-space matrix.

\begin{figure}[]
    \centering
    \includegraphics[width=1\textwidth]{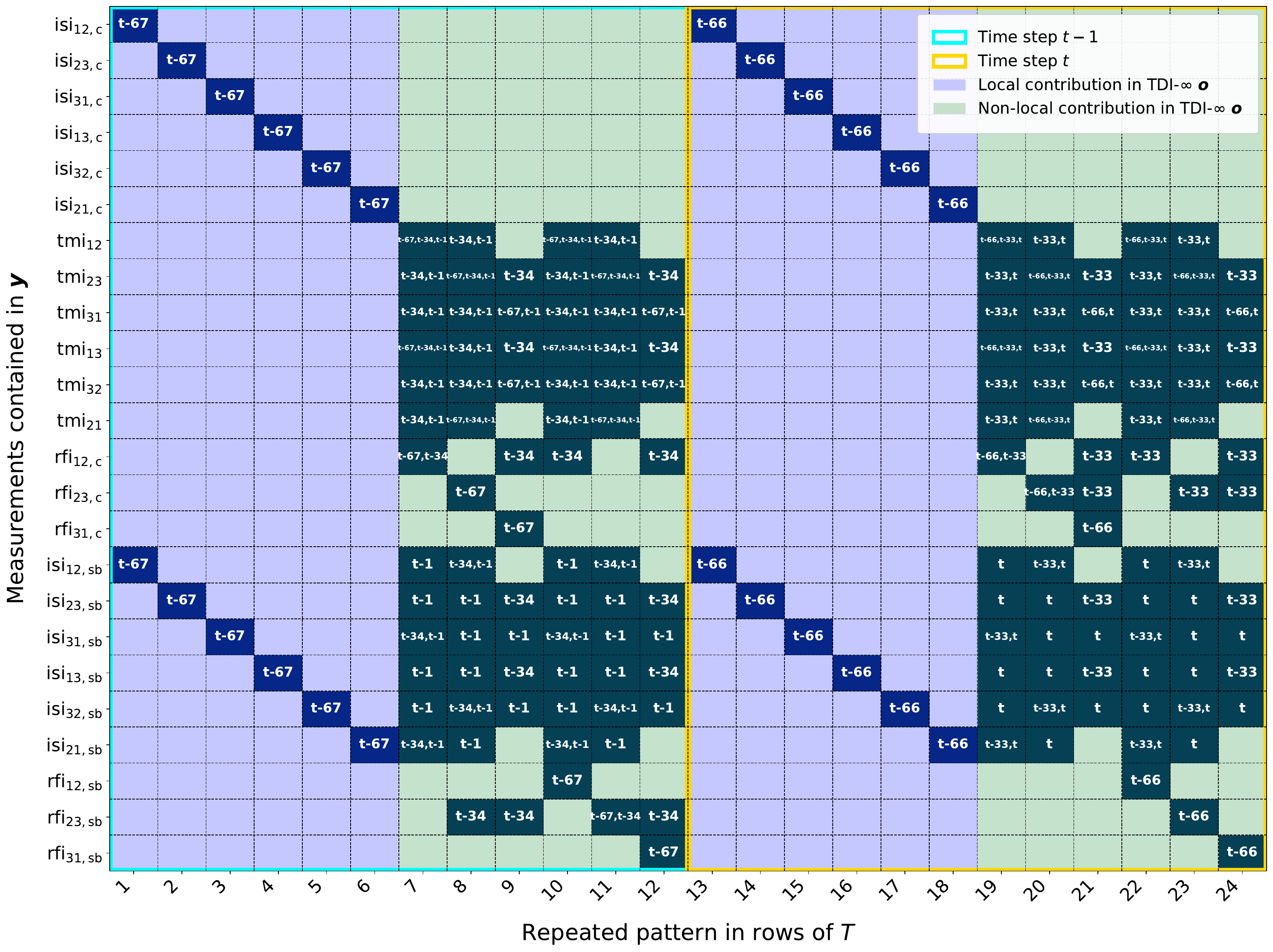}
    \caption{Illustration of repeating patterns in the null-space matrix \( T \) of the all-in-one TDI-$\infty$ framework, constructed using the turnback algorithm for a simplified case with laser noise only, static and equal arm lengths. Patterns repeat every 12 rows in \( T \), as highlighted by the {cyan box} (time step \( t-1 \)) and {yellow box} (time step \( t \)). The cell entries indicate the time step(s) of the respective measurement used to construct the corresponding output in the TDI-$\infty$ observable $\boldsymbol{o}$. Local contributions in $\boldsymbol{o}$ (highlighted in transparent blue) cancel laser noise but also suppress the gravitational-wave signal, making them unsuitable for astrophysical inference. Non-local contributions (highlighted in transparent green) suppress laser noise while preserving the gravitational-wave signal.}
    \label{im:PatternsNullSpaceTurnback}
\end{figure}

The null-space matrix exhibits a clear repeating pattern every 12 rows. This repetition reflects the structure of the TDI-$\infty$ observable $\boldsymbol{o}$. The thin colored boxes in Fig.\ \ref{im:PatternsNullSpaceTurnback} highlight specific elements of interest within the matrix: the {cyan box} encloses the 12 elements associated with the  time step \( t-1 \), while the {yellow box} encloses the 12 elements corresponding to the current time step \( t \). Within each box, the first six columns represent local contributions to the TDI-$\infty$ observable $\boldsymbol{o}$. These contributions are simple differences between the inter-spacecraft interferometer measurements at the carrier frequency and the sideband frequency for the same time step. When modulation noise and clock noise are not present, these differences effectively cancel noise. However, they also suppress the gravitational-wave signal. As such, these contributions in $\boldsymbol{o}$ are not useful for astrophysical parameter inference. The columns labeled 7 to 12 (for time step \( t-1 \)) and 19 to 24 (for time step \( t \)) reveal non-local contributions. These combinations suppress laser noise while preserving the gravitational-wave signal. A closer inspection reveals pairwise equivalences within these non-local combinations. For example, column 7 is equivalent to column 10, column 8 to column 11, and so on.
The only distinction between these pairs is whether some measurements are taken at the carrier or sideband frequency. 
The figure illustrates that there are three non-local combinations effectively spanning the TDI-$\infty$ space. This conceptually resembles the three generators of classical TDI described in \cite{TDITINTO2005}.

Finding a sparse banded null matrix is an NP-hard problem \cite{doi:10.1137/0608037}:
it becomes computationally prohibitive for large or complex instances, as the resources required to find an exact solution grow exponentially with the size of the problem. Consequently, exact methods become impractical as the problem scale increases. This intractability drives the need for heuristic approaches such as the turnback algorithm.

It is important to note that a banded sparse null space is not required for noise suppression in the proposed all-in-one TDI-$\infty$ setup. Any valid null-space matrix will suffice for this purpose. 
The desire for sparsity and bandedness is primarily motivated by the need for analytical clarity, because these structures facilitate the interpretation of how the individual measurements contribute to the composition of the TDI-$\infty$ output $\boldsymbol{o}$ of equation  \eqref{eq:TDIInf_composed}. Despite the utility of the turnback algorithm, its current implementation faces challenges in producing a sparse null space when both modulation and clock noise are active, likely due to the broken symmetry in the noise-suppression framework introduced by these noise sources. For this reason, Fig.\ \ref{im:PatternsNullSpaceTurnback} presents a simplified case where only laser noise is present and static and equal arm lengths are assumed.}

\section*{References}
\bibliographystyle{unsrt}
\bibliography{sn-bibliography} 

\begin{thebibliography}{10}

\bibitem{colpi2024lisa}
Monica Colpi, Karsten Danzmann, Martin Hewitson, Kelly Holley-Bockelmann, Philippe Jetzer, Gijs Nelemans, Antoine Petiteau, David Shoemaker, Carlos Sopuerta, Robin Stebbins, et~al.
\newblock {LISA Definition Study Report}, 2024.

\bibitem{PhysRevLett.116.061102}
Bruce~P. Abbott, Richard Abbott, Thomas~D. Abbott, Matthew~R. Abernathy, Francesco Acernese, Katherine Ackley, Carl Adams, Thomas Adams, Paolo Addesso, Rana~X. Adhikari, et~al.
\newblock {Observation of Gravitational Waves from a Binary Black Hole Merger}.
\newblock {\em Phys. Rev. Lett.}, 116:061102, Feb 2016.

\bibitem{PhysRevLett.116.241103}
Bruce~P. Abbott, Richard Abbott, Thomas~D. Abbott, Matthew~R. Abernathy, Francesco Acernese, Katherine Ackley, Carl Adams, Thomas Adams, Paolo Addesso, Rana~X. Adhikari, et~al.
\newblock {GW151226: Observation of Gravitational Waves from a 22-Solar-Mass Binary Black Hole Coalescence}.
\newblock {\em Phys. Rev. Lett.}, 116:241103, Jun 2016.

\bibitem{PhysRevLett.118.221101}
Bruce~P. Abbott, Richard Abbott, Thomas~D. Abbott, Francesco Acernese, Katherine Ackley, Carl Adams, Thomas Adams, Paolo Addesso, Rana~X. Adhikari, Varun~B. Adya, et~al.
\newblock {GW170104: Observation of a 50-Solar-Mass Binary Black Hole Coalescence at Redshift 0.2}.
\newblock {\em Phys. Rev. Lett.}, 118:221101, Jun 2017.

\bibitem{Abbott_2017}
Bruce~P. Abbott, Richard Abbott, Thomas~D. Abbott, Francesco Acernese, Katherine Ackley, Carl Adams, Thomas Adams, Paolo Addesso, Rana~X. Adhikari, and Varun~B. Adya.
\newblock {GW170608: Observation of a 19 Solar-mass Binary Black Hole Coalescence}.
\newblock {\em The Astrophysical Journal Letters}, 851(2):L35, dec 2017.

\bibitem{PhysRevLett.119.141101}
Bruce~P. Abbott, Richard Abbott, Thomas~D. Abbott, Francesco Acernese, Katherine Ackley, Carl Adams, Thomas Adams, Paolo Addesso, Rana~X. Adhikari, and Varun~B. Adya.
\newblock {GW170814: A Three-Detector Observation of Gravitational Waves from a Binary Black Hole Coalescence}.
\newblock {\em Phys. Rev. Lett.}, 119:141101, Oct 2017.

\bibitem{PhysRevLett.119.161101}
Bruce~P. Abbott, Richard Abbott, Thomas~D. Abbott, Francesco Acernese, Katherine Ackley, Carl Adams, Thomas Adams, Paolo Addesso, Rana~X. Adhikari, Varun~B. Adya, et~al.
\newblock {GW170817: Observation of Gravitational Waves from a Binary Neutron Star Inspiral}.
\newblock {\em Phys. Rev. Lett.}, 119:161101, Oct 2017.

\bibitem{PhysRevX.9.031040}
Bruce~P. Abbott, Richard Abbott, Thomas~D. Abbott, Suraj Abraham, Francesco Acernese, Katherine Ackley, Carl Adams, Rana~X. Adhikari, Varun~B. Adya, Carsten Affeldt, et~al.
\newblock {GWTC-1: A Gravitational-Wave Transient Catalog of Compact Binary Mergers Observed by LIGO and Virgo during the First and Second Observing Runs}.
\newblock {\em Phys. Rev. X}, 9:031040, Sep 2019.

\bibitem{Abbott_2020}
Bruce~P. Abbott, Richard Abbott, Thomas~D. Abbott, Suraj Abraham, Francesco Acernese, Katherine Ackley, Carl Adams, Rana~X. Adhikari, Varun~B. Adya, Carsten Affeldt, et~al.
\newblock {GW190425: Observation of a Compact Binary Coalescence with Total Mass $\sim$ 3.4 $M_{\odot}$ }.
\newblock {\em The Astrophysical Journal Letters}, 892(1):L3, mar 2020.

\bibitem{PhysRevD.102.043015}
Richard Abbott, Thomas~D. Abbott, Suraj Abraham, Francesco Acernese, Katherine Ackley, Carl Adams, Rana~X. Adhikari, Varun~B. Adya, Carsten Affeldt, Michalis Agathos, et~al.
\newblock {GW190412: Observation of a binary-black-hole coalescence with asymmetric masses}.
\newblock {\em Phys. Rev. D}, 102:043015, Aug 2020.

\bibitem{Abbott_2020_2}
Richard Abbott, Thomas~D. Abbott, Suraj Abraham, Francesco Acernese, Katherine Ackley, Carl Adams, Rana~X. Adhikari, Varun~B. Adya, Carsten Affeldt, Michalis Agathos, et~al.
\newblock {GW190814: Gravitational Waves from the Coalescence of a 23 Solar Mass Black Hole with a 2.6 Solar Mass Compact Object}.
\newblock {\em The Astrophysical Journal Letters}, 896(2):L44, jun 2020.

\bibitem{PhysRevLett.125.101102}
Richard Abbott, Thomas~D. Abbott, Suraj Abraham, Francesco Acernese, Katherine Ackley, Carl Adams, Rana~X. Adhikari, Varun~B. Adya, Carsten Affeldt, and Michalis Agathos.
\newblock {GW190521: A Binary Black Hole Merger with a Total Mass of $150\text{ }\text{ }{M}_{\ensuremath{\bigodot}}$}.
\newblock {\em Phys. Rev. Lett.}, 125:101102, Sep 2020.

\bibitem{SciReqDoc}
{LISA Science Study Team}.
\newblock {LISA Science Requirements document}.
\newblock Technical Report, European Space Agency, 2018.
\newblock Technical Report ESA-L3-EST-SCI-RS-001.

\bibitem{Amaro-Seoane_2012}
Pau Amaro-Seoane, Sofiane Aoudia, Stanislav Babak, Pierre Binétruy, Emanuele Berti, Alejandro Bohé, Chiara Caprini, Monica Colpi, Neil~J Cornish, Karsten Danzmann, et~al.
\newblock {Low-frequency gravitational-wave science with eLISA/NGO}.
\newblock {\em Classical and Quantum Gravity}, 29(12):124016, jun 2012.

\bibitem{PhysRevD.81.104014}
Jonathan~R. Gair, Christopher Tang, and Marta Volonteri.
\newblock Lisa extreme-mass-ratio inspiral events as probes of the black hole mass function.
\newblock {\em Phys. Rev. D}, 81:104014, May 2010.

\bibitem{Mangiagli2022}
Alberto Mangiagli, Chiara Caprini, Marta Volonteri, Sylvain Marsat, Susanna Vergani, Nicola Tamanini, and Lorenzo Speri.
\newblock {Cosmology with massive black hole binary mergers in the LISA era.}
\newblock In {\em Proceedings of 41st International Conference on High Energy physics — PoS(ICHEP2022)}, ICHEP2022. Sissa Medialab, October 2022.

\bibitem{amaroseoane2017laser}
Pau Amaro-Seoane, Harry Audley, Stanislav Babak, John Baker, Peter~Bender Enrico~Barausse, Emanuele Berti, Michael~Born Pierre~Binetruy, Daniele Bortoluzzi, et~al.
\newblock {Laser Interferometer Space Antenna}.
\newblock {\em arXiv e-prints}, page arXiv:1702.00786, February 02/2017.

\bibitem{Martens2021}
Waldemar Martens and Eric Joffre.
\newblock {Trajectory Design for the {ESA} {LISA} Mission}.
\newblock {\em The Journal of the Astronautical Sciences}, 68(2):402--443, June 2021.

\bibitem{Schuldt_2009}
Thilo Schuldt, Martin Gohlke, Dennis Weise, Ulrich Johann, Achim Peters, and Claus Braxmaier.
\newblock {Picometer and nanoradian optical heterodyne interferometry for translation and tilt metrology of the LISA gravitational reference sensor}.
\newblock {\em Classical and Quantum Gravity}, 26(8):085008, apr 2009.

\bibitem{PhysRevD.99.082001}
Michele Armano, Harry Audley, Jim Baird, Pierre Binétruy, Michael Born, Daniele Bortoluzzi, Enrico Castelli, Angelo Cavalleri, Andrea Cesarini, Andrew~M. Cruise, et~al.
\newblock {LISA Pathfinder platform stability and drag-free performance}.
\newblock {\em Phys. Rev. D}, 99:082001, Apr 2019.

\bibitem{Carbone2006-kz}
Luigi Carbone, Angelo Cavalleri, Giacomo Ciani, Riccardo Dolesi, Mauro Hueller, Davide Tombolato, Stefano Vitale, and William~J. Weber.
\newblock {Torsion pendulum facility for direct force measurements of {LISA} {GRS} related disturbances}.
\newblock In {\em {AIP} Conference Proceedings}. AIP, 2006.

\bibitem{BenderArt}
Peter~L. Bender, Alain Brillet, Ignazio Ciufolini, Andrew~M. Cruise, Curt Cutler, Karsten Danzmann, et~al.
\newblock {LISA. Laser Interferometer Space Antenna for the detection and observation of gravitational waves. An international project in the field of Fundamental Physics in Space}.
\newblock Max-Planck-Institut für Quantenoptik, München, Germany, 1998.
\newblock {Pre-Phase A Report}.

\bibitem{Dhurandhar2010}
Sanjeev Dhurandhar, K.~Rajesh Nayak, and Jacques Vinet.
\newblock {Time-delay interferometry for LISA with one arm dysfunctional}.
\newblock {\em Classical and Quantum Gravity}, 27:135013, 2010.

\bibitem{Bayle2018a}
Jean-Baptiste Bayle, Marc Lilley, Antoine Petiteau, and Henri Halloin.
\newblock {Analytic Model and Simulations of Residual Laser Noise after Time-Delay Interferometry in LISA}.
\newblock {\em arXiv: Instrumentation and Methods for Astrophysics}, 2018.

\bibitem{Bayle2018b}
Jean-Baptiste Bayle, Marc Lilley, Antoine Petiteau, and Henri Halloin.
\newblock {Effect of filters on the time-delay interferometry residual laser noise for LISA}.
\newblock {\em Physical Review D}, 99:084023, 2018.

\bibitem{Tinto2002TDI1stGen}
Massimo Tinto, Frank~B. Estabrook, and John~W. Armstrong.
\newblock {Time-delay interferometry for {{LISA}}}.
\newblock {\em {Physical Review D}}, {65}({8}), April {2002}.

\bibitem{TDITINTO2005}
Massimo Tinto and Sanjeev~V. Dhurandhar.
\newblock {Time-Delay Interferometry}.
\newblock {\em Living Reviews in Relativity}, 8(1), Jul 2005.

\bibitem{PhysRevD.103.042006}
Quentin Baghi, James~Ira Thorpe, Jacob Slutsky, and John Baker.
\newblock Statistical inference approach to time-delay interferometry for gravitational-wave detection.
\newblock {\em Phys. Rev. D}, 103:042006, Feb 2021.

\bibitem{PhysRevD.104.044035}
Kallol Dey, Nikolaos Karnesis, Alexandre Toubiana, Enrico Barausse, Natalia Korsakova, Quentin Baghi, and Soumen Basak.
\newblock {Effect of data gaps on the detectability and parameter estimation of massive black hole binaries with LISA}.
\newblock {\em Phys. Rev. D}, 104:044035, Aug 2021.

\bibitem{PhysRevD.100.022003}
Quentin Baghi, James~Ira Thorpe, Jacob Slutsky, John Baker, Tito~Dal Canton, Natalia Korsakova, and Nikos Karnesis.
\newblock {Gravitational-wave parameter estimation with gaps in LISA: A Bayesian data augmentation method}.
\newblock {\em Phys. Rev. D}, 100:022003, Jul 2019.

\bibitem{castelli2024extractiongravitationalwavesignals}
Eleonora Castelli, Quentin Baghi, John~G. Baker, Jacob Slutsky, Jérôme Bobin, Nikolaos Karnesis, Antoine Petiteau, Orion Sauter, Peter Wass, and William~J. Weber.
\newblock {Extraction of gravitational wave signals in realistic LISA data}, 2024.
\newblock \url{https://arxiv.org/abs/2411.13402}.

\bibitem{Pollack2004}
Scott~E Pollack.
\newblock {LISA science results in the presence of data disturbances}.
\newblock {\em Classical and Quantum Gravity}, 21(14):3419–3432, June 2004.

\bibitem{carre2010}
Jérôme Carré and Edward~K. Porter.
\newblock {The Effect of Data Gaps on LISA Galactic Binary Parameter Estimation}, 2010.
\newblock \url{https://arxiv.org/abs/1010.1641}.

\bibitem{PhysRevD.103.082001}
Michele Vallisneri, Jean-Baptiste Bayle, Stanislav Babak, and Antoine Petiteau.
\newblock {Time-delay interferometry without delays}.
\newblock {\em {Phys. Rev. D}}, {103}:{082001}, {Apr} {2021}.

\bibitem{IMPRSStatsLecture8}
Jonathan~R. Gair.
\newblock {Statistics for Gravitational Wave Data Analysis - Lecture 8}.
\newblock \url{https://imprs-gw-lectures.aei.mpg.de/potsdam-2019/wp-content/uploads/sites/2/2020/01/IMPRSStatsforGWLecture8.pdf}, 2019.
\newblock Lecture notes, Accessed: 2024-09-17.

\bibitem{BabakLISAPresentation}
Stanislav Babak.
\newblock {LISA: Challenges for Data Analysis}.
\newblock \url{https://www.kiss.caltech.edu/workshops/LISA/presentations/Babak.pdf}, 2018.
\newblock Lecture notes, Accessed: 2024-09-17.

\bibitem{Otto_2012}
Markus Otto, Gerhard Heinzel, and Karsten Danzmann.
\newblock {TDI and clock noise removal for the split interferometry configuration of LISA}.
\newblock {\em Classical and Quantum Gravity}, 29(20):205003, aug 2012.

\bibitem{PhysRevD.99.084023}
Jean-Baptiste Bayle, Marc Lilley, Antoine Petiteau, and Hubert Halloin.
\newblock {Effect of filters on the time-delay interferometry residual laser noise for LISA}.
\newblock {\em Phys. Rev. D}, 99:084023, Apr 2019.

\bibitem{https://doi.org/10.15488/8545}
Markus Otto.
\newblock Time-delay interferometry simulations for the laser interferometer space antenna, 2015.
\newblock Gottfried Wilhelm Leibniz Universit\"{a}t Hannover, 10.15488/8545, \url{https://www.repo.uni-hannover.de/handle/123456789/8598}.

\bibitem{PhysRevD.107.083019}
Jean-Baptiste Bayle and Olaf Hartwig.
\newblock {Unified model for the LISA measurements and instrument simulations}.
\newblock {\em {Phys. Rev. D}}, {107}:{083019}, {Apr} {2023}.

\bibitem{PhysRevD.104.023006}
Jean-Baptiste Bayle, Olaf Hartwig, and Martin Staab.
\newblock {Adapting time-delay interferometry for LISA data in frequency}.
\newblock {\em {Phys. Rev. D}}, {104}:{023006}, {Jul} {2021}.

\bibitem{Houba2023}
Niklas Houba, Simon Delchambre, Gerald Hechenblaikner, Tobias Ziegler, and Walter Fichter.
\newblock {Time-delay interferometry infinity for tilt-to-length noise estimation in LISA}.
\newblock {\em {Classical and Quantum Gravity}}, {40}({10}):{107001}, April {2023}.

\bibitem{Tinto_Massimo_1999}
Massimo Tinto and John~W. Armstrong.
\newblock Cancellation of laser noise in an unequal-arm interferometer detector of gravitational radiation.
\newblock {\em Phys. Rev. D}, 59:102003, Apr 1999.

\bibitem{Armstrong_John_1999}
John~W. Armstrong, Frank~B. Estabrook, and Massimo Tinto.
\newblock {Time-Delay Interferometry for Space-based Gravitational Wave Searches}.
\newblock {\em The Astrophysical Journal}, 527(2):814--826, dec 1999.

\bibitem{Tinto_Massimo_2004}
Massimo Tinto, Frank~B. Estabrook, and John~W. Armstrong.
\newblock {Time delay interferometry with moving spacecraft arrays}.
\newblock {\em Physical Review D}, 69(8), Apr 2004.

\bibitem{PhysRevD.70.062002}
Massimo Tinto and Shane~L. Larson.
\newblock {LISA time-delay interferometry zero-signal solution: Geometrical properties}.
\newblock {\em Phys. Rev. D}, 70:062002, Sep 2004.

\bibitem{GeomTDIVallisneri2005}
Michele Vallisneri.
\newblock Geometric time delay interferometry.
\newblock {\em Physical Review D}, 72(4), August 2005.

\bibitem{SyntheticLISA}
Michele Vallisneri.
\newblock {Synthetic LISA: Simulating time delay interferometry in a model LISA}.
\newblock {\em Phys. Rev. D}, 71:022001, Jan 2005.

\bibitem{Tinto_Massimo_2021}
Massimo Tinto, Sanjeev Dhurandhar, and Prasanna Joshi.
\newblock Matrix representation of time-delay interferometry.
\newblock {\em Phys. Rev. D}, 104:044033, Aug 2021.

\bibitem{Bayle_Jean-Baptiste_2021}
Jean-Baptiste Bayle, Michele Vallisneri, Stanislav Babak, and Antoine Petiteau.
\newblock On the matrix formulation of time-delay interferometry, 2021.

\bibitem{Muratore_Martina_2020}
Martina Muratore, Daniele Vetrugno, and Stefano Vitale.
\newblock Revisitation of time delay interferometry combinations that suppress laser noise in lisa.
\newblock {\em Classical and Quantum Gravity}, 37(18):185019, aug 2020.

\bibitem{Page_Jessica_2021}
Jessica Page and Tyson~B. Littenberg.
\newblock Bayesian time delay interferometry.
\newblock {\em Physical Review D}, 104(8), Oct 2021.

\bibitem{de_Vine_Glenn_2010}
Glenn de~Vine, Brent Ware, Kirk McKenzie, Robert~E. Spero, William~M. Klipstein, and Daniel~A. Shaddock.
\newblock {Experimental Demonstration of Time-Delay Interferometry for the Laser Interferometer Space Antenna}.
\newblock {\em Phys. Rev. Lett.}, 104:211103, May 2010.

\bibitem{Mitryk_Shawn_2010}
Shawn Mitryk, Vinzenz Wand, and Guido Mueller.
\newblock {Verification of time-delay interferometry techniques using the University of Florida LISA interferometry simulator}.
\newblock {\em Classical and Quantum Gravity}, 27:084012, 2010.

\bibitem{Cruz_Rachel_2006}
Rachel~J. Cruz, James~I. Thorpe, Michael Hartman, and Guido Mueller.
\newblock {Time Delay Interferometry using the {UF} {LISA} Benchtop Simulator}.
\newblock In {\em {AIP} Conference Proceedings}. {AIP}, 2006.

\bibitem{2003TDI2ndGenMotiv}
Daniel~A. Shaddock, Massimo Tinto, Frank~B. Estabrook, and John~W. Armstrong.
\newblock {{{Data combinations accounting for LISA spacecraft motion}}}.
\newblock {\em {Physical Review D}}, {68}({6}), {Sep} {2003}.

\bibitem{PhysRevD.105.122008}
Olaf Hartwig, Jean-Baptiste Bayle, Martin Staab, Aur\'elien Hees, Marc Lilley, and Peter Wolf.
\newblock Time-delay interferometry without clock synchronization.
\newblock {\em Phys. Rev. D}, 105:122008, Jun 2022.

\bibitem{bayle:tel-03120731}
Jean-Baptiste Bayle.
\newblock {\em {Simulation and Data Analysis for LISA (Instrumental Modeling, Time-Delay Interferometry, Noise-Reduction Performance Study, and Discrimination of Transient Gravitational Signals)}}.
\newblock Theses, {Universit{\'e} de Paris ; Universit{\'e} Paris Diderot ; Laboratoire Astroparticules et Cosmologie}, October 2019.

\bibitem{PhysRevD.103.123027}
Olaf Hartwig and Jean-Baptiste Bayle.
\newblock {Clock-jitter reduction in LISA time-delay interferometry combinations}.
\newblock {\em Phys. Rev. D}, 103:123027, Jun 2021.

\bibitem{PhysRevD.106.042005}
Sarah Paczkowski, Roberta Giusteri, Martin Hewitson, Nikolaos Karnesis, Ewan~D. Fitzsimons, Gudrun Wanner, and Gerhard Heinzel.
\newblock {{{Postprocessing subtraction of tilt-to-length noise in LISA}}}.
\newblock {\em {Phys. Rev. D}}, {106}:{042005}, {Aug} {2022}.

\bibitem{Prince2002}
Thomas~A. Prince, Massimo Tinto, Shane~L. Larson, and John~W. Armstrong.
\newblock {LISA optimal sensitivity}.
\newblock {\em {Physical Review D}}, {66}({12}), December {2002}.

\bibitem{pfeiffer_turnbackLU}
{Kai Pfeiffer}.
\newblock {turnbackLU: Sparse matrix LU factorization}, {2024}.
\newblock {\url{{https://github.com/pfeiffer-kai/turnbackLU}}, accessed: 2024-09-12 }.

\bibitem{Moulin_Veeravalli_2018}
Pierre Moulin and Venugopal~V. Veeravalli.
\newblock {\em {Statistical Inference for Engineers and Data Scientists}}.
\newblock Cambridge University Press, 2018.

\bibitem{Bailer-Jones2017-iw}
Coryn A.~L. Bailer-Jones.
\newblock {\em {Practical Bayesian inference}}.
\newblock Cambridge University Press, Cambridge, England, July 2017.

\bibitem{Tang_2022}
Tang Niansheng.
\newblock {\em {Bayesian Inference}}.
\newblock IntechOpen, Rijeka, Nov 2022.
\newblock 10.5772/intechopen.97942.

\bibitem{Keskin2023}
Fesih Keskin.
\newblock {\em {Nested Sampling: A Case Study in Parameter Estimation}}, chapter~3.
\newblock IntechOpen, December 2023.

\bibitem{10.1093/biomet/45.3-4.296}
Thomas Bayes.
\newblock {An essay towards solving a problem in the doctrine of chances}.
\newblock {\em Biometrika}, 45(3-4):296--315, 12 1958.

\bibitem{noauthor_2023-fe}
Scott~M. Lynch.
\newblock {Markov chain Monte Carlo (MCMC) sampling methods}.
\newblock In {\em Applied Bayesian Statistics}, pages 62--110. SAGE Publications, Inc., 2455 Teller Road, Thousand Oaks California 91320, 2023.

\bibitem{vanOijen2024}
Marcel van Oijen.
\newblock {\em {Markov Chain Monte Carlo Sampling (MCMC)}}, page 35–40.
\newblock Springer International Publishing, 2024.

\bibitem{Spiegelhalter-95}
Walter~R. Gilks, Sylvia Richardson, and David Spiegelhalter.
\newblock {Introducing Markov chain Monte Carlo"}.
\newblock In {\em Markov Chain Monte Carlo in Practice}, pages 19--38. Chapman and Hall/CRC, December 1995.

\bibitem{Hanada2022}
Masanori Hanada and So~Matsuura.
\newblock {\em {Applications of Markov Chain Monte Carlo}}, page 113–168.
\newblock Springer Nature Singapore, 2022.

\bibitem{Hancock2004}
John~M. Hancock.
\newblock {Markov Chain Monte Carlo (MCMC, Metropolis‐Hastings, Gibbs Sampling)}, October 2004.
\newblock Dictionary of Bioinformatics and Computational Biology, ISBN 9780471650126.

\bibitem{Robert2015-pq}
Christian~P. Robert.
\newblock {The Metropolis-Hastings algorithm}.
\newblock {\em {arXiv e-prints}}, April 2015.
\newblock \url{http://arxiv.org/licenses/nonexclusive-distrib/1.0/}.

\bibitem{Berg2004-fr}
Bernd~A. Berg.
\newblock {\em Markov chain Monte Carlo simulations and their statistical analysis: With web-based {FORTRAN} code}.
\newblock World Scientific Publishing, Singapore, Singapore, December 2004.

\bibitem{Minh2015-ts}
David D.~L. Minh and Do~Le~(Paul) Minh.
\newblock {Understanding the Hastings algorithm}.
\newblock {\em Commun. Stat. Simul. Comput.}, 44(2):332--349, February 2015.

\bibitem{Gelman2011}
Andrew Gelman and Kenneth Shirley.
\newblock {\em Inference from Simulations and Monitoring Convergence}, page 162–174.
\newblock Chapman and Hall/CRC, May 2011.

\bibitem{https://doi.org/10.48550/arxiv.1812.09384}
Dootika Vats and Christina Knudson.
\newblock {Revisiting the Gelman-Rubin Diagnostic}, 2018.

\bibitem{lisainstrument22}
Jean-Baptiste Bayle, Olaf Hartwig, and Martin Staab.
\newblock {LISA Instrument}, November 2023.
\newblock \url{https://doi.org/10.5281/zenodo.10136689}.

\bibitem{https://doi.org/10.5281/zenodo.6351736}
Martin Staab, Jean-Baptiste Bayle, and Olaf Hartwig.
\newblock {PyTDI}, 2023.
\newblock \url{https://zenodo.org/record/6351736}.

\bibitem{Karnesis_Nikolaos_2023}
Nikolaos Karnesis, Michael~L. Katz, Natalia Korsakova, Jonathan~R. Gair, and Nikolaos Stergioulas.
\newblock {Eryn: a multipurpose sampler for Bayesian inference}.
\newblock {\em Monthly Notices of the Royal Astronomical Society}, 526(4):4814–4830, sep 2023.

\bibitem{Foreman-Mackey_Daniel_2013}
Daniel Foreman-Mackey, David~W. Hogg, Dustin Lang, and Jonathan Goodman.
\newblock {emcee: The MCMC Hammer}.
\newblock {\em {arXiv e-prints}}, 125(925):306, March 2013.

\bibitem{geyer1991markov}
Charles~J. Geyer.
\newblock {Markov Chain Monte Carlo Maximum Likelihood}.
\newblock In {\em Proceedings of the 23rd Symposium on the Interface}, pages 156--163. Interface Foundation of North America, 1991.

\bibitem{10.5555/2073946.2073948}
Christophe Andrieu, Nando de~Freitas, and Arnaud Doucet.
\newblock {Reversible jump MCMC simulated annealing for neural networks}.
\newblock In {\em Proceedings of the Sixteenth Conference on Uncertainty in Artificial Intelligence}, UAI'00, page 11–18, San Francisco, CA, USA, 2000. Morgan Kaufmann Publishers Inc.

\bibitem{PhysRevD.104.104054}
Neil~J. Cornish.
\newblock Heterodyned likelihood for rapid gravitational wave parameter inference.
\newblock {\em Phys. Rev. D}, 104:104054, Nov 2021.

\bibitem{doi:10.1137/0608037}
John~R. Gilbert and Michael~T. Heath.
\newblock Computing a sparse basis for the null space.
\newblock {\em SIAM Journal on Algebraic Discrete Methods}, 8(3):446--459, 1987.

\end{thebibliography}
\end{document}